\documentclass[aps,prc,showpacs,twocolumn,superscriptaddress]{revtex4}

\usepackage{graphicx}
\usepackage{longtable}
\usepackage{verbatim}
\usepackage{color}

\begin{document}

\title{$\alpha$-Cluster Structure of $^{18}$O}

\author{M. L. Avila}
\email{mavila@anl.gov}
\altaffiliation{Physics Division, Argonne National Laboratory, Argonne IL 60439, USA}
\affiliation{Department of Physics, Florida State University, Tallahassee, FL 
32306}

\author{G. V. Rogachev}
\email{rogachev@tamu.edu}
\affiliation{Department of Physics \& Astronomy and Cyclotron Institute, Texas 
A\&M University, College Station, TX 77843}

\author{V. Z. Goldberg}
\affiliation{Department of Physics \& Astronomy and Cyclotron Institute, Texas 
A\&M University, College Station, TX 77843}

\author{E. D. Johnson}
\affiliation{Department of Physics, Florida State University, Tallahassee, FL 
32306}

\author{K. W. Kemper}
\affiliation{Department of Physics, Florida State University, Tallahassee, FL 
32306}

\author{Yu. M. Tchuvil'sky}
\affiliation{Skobeltsyn Institute of Nuclear Physics, Lomonosov Moscow State University, 119991 Moscow, Russia}

\author{A. S. Volya}
\affiliation{Department of Physics, Florida State University, Tallahassee, FL 
32306}

\begin{abstract}
\begin{description}
\item[Background:] Clustering phenomena in $N \neq Z$ nuclei provide an 
opportunity to understand the interplay between cluster and nucleon degrees 
of freedom.
\item[Purpose:] To study resonances in the $^{18}$O spectrum, populated in $^{14}$C+$\alpha$ 
elastic scattering.
\item[Method:]  The Thick Target Inverse Kinematics (TTIK) technique was used to 
measure the excitation function for the $^{14}$C+$\alpha$ elastic scattering. A 
42 MeV $^{14}$C beam was used to populate states of excitation energy up to
14.9 MeV in $^{18}$O. The analysis was performed using a multi-level, multi-channel R-Matrix 
approach.  
\item[Results:] Detailed spectroscopic information, including 
spin-parities, partial $\alpha$- and neutron- decay widths and dimensionless reduced widths, 
was obtained for excited states in $^{18}$O between 8 and 14.9 MeV in excitation energy. 
Cluster-Nucleon Configuration Interaction Model calculations of the same quantities are performed and compared to the experimental results. 
\item[Conclusions:] Strong fragmentation of large $\alpha$-cluster strengths is observed 
in the spectrum of $^{18}$O making the $\alpha$-cluster structure of $^{18}$O 
quite different from the pattern of known quasi-rotational bands of alternating parity 
that are characteristic of $N=Z$, even-even nuclei like $^{16}$O and $^{20}$Ne. 
\end{description}
\end{abstract}


\maketitle

\section{Introduction}

\begin{figure}
\begin{center}\includegraphics[scale=0.31]{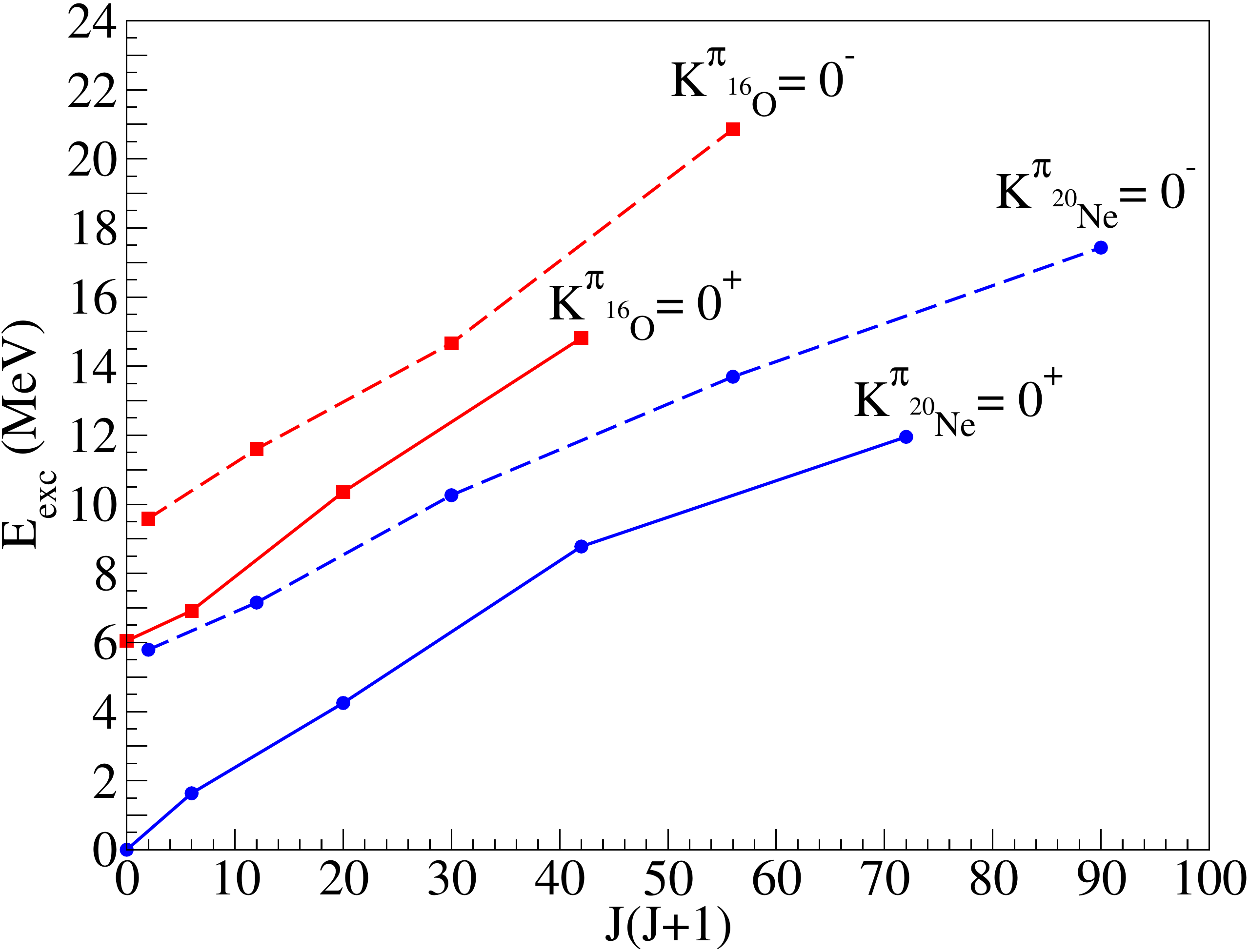}\end{center}
\caption{\label{fig:Bands} (Color online) Inversion doublet, $\alpha$-cluster quasi-rotational 
bands for $^{16}$O and $^{20}$Ne. All of the unbound states within the bands 
have large $\alpha$ reduced widths. The solid curves
connect the positive parity states, while the dashed lines connect the negative 
parity states. Blue and red color coding correspond to $^{20}$Ne and $^{16}$O 
respectively.}
\end{figure}

The concept of $\alpha$-clustering has been successfully applied to explain 
multiple features in the nuclear spectrum. In particular, a number of known 
structure peculiarities in light $N=Z$ even-even nuclei such as $^{8}$Be, $^{12}$C, 
$^{16}$O, and $^{20}$Ne, is associated with clustering. The most striking are the 
inversion doublet quasi-rotational $\alpha$-cluster bands \cite{Hori68}, as shown in 
Fig. \ref{fig:Bands}. All members of these bands that have excitation energies 
above the $\alpha$-decay threshold, posses $\alpha$-reduced widths close to the 
single particle limit, indicating their extreme $\alpha$-cluster character. 
Extensive experimental and theoretical studies 
\cite{Ames82,Bill79,John69,Ried84,Free07,Hori68,Suzu76,Kana95} have suggested an interpretation
of these bands as well developed $\alpha$+core structures. 

It has proven to be far more difficult to study clustering phenomena in 
non-self-conjugate, $N\neq Z$ nuclei because  the ``extra'' nucleons 
introduce additional degrees of freedom which may modify, create, enhance or destroy 
cluster structures. In addition
experimental studies require a more complicated analysis due to the 
presence of low-lying nucleon decay channels and a higher level density than in 
$N=Z$, even-even nuclei.

In $N=Z$, even-even nuclei the $\alpha$-decay threshold is usually lower in energy than
the nucleon decay threshold. However, for $N \ne Z$ nuclei like $^{18}$O the energy thresholds for neutron and $\alpha$-decay are close (in the mirror nucleus, 
$^{18}$Ne, both the proton and the two-proton thresholds  are below the 
$\alpha$-threshold), so one can expect that the decay properties of the 
states with both large and small $\alpha$-widths in $N\ne Z$ nuclei also contain 
information on the nucleon widths. The closeness of the decay thresholds 
for $N \ne Z$ nuclei allows one to explore the interplay 
between the single nucleon and cluster degrees of freedom.

Several different theoretical approaches were developed to describe the cluster and single 
particle phenomena simultaneously. The Antisymmetrized Molecular Dynamics (AMD) 
\cite{Kana01} and Fermionic Molecular Dynamics \cite{Neff08} approaches were 
particularly successful. Within these frameworks clustering emerges from 
nucleon-nucleon interactions without the need to introduce clusters {\it a priori}. 
The {\it ab initio}, Green's Function Monte Carlo (GFMC) calculations were also 
successful in reproducing clustering for the ground state of $^8$Be(g.s.) 
\cite{Wiri00}.

An approach to clustering exploiting the Elliott SU(3) model \cite{Ell58} 
under the assumption that the wave functions of clustered states possess and 
unmixed SU(3) symmetry has been particularly successful in studies of 
multi-cluster systems such as $^8$Be,$^{10}$Be,$^{12}$C,$^{16}$O,$^{32}$S
\cite{GKT06,GKT07,GKT08,GKT10,GKT13}. However, purely algebraic models lack 
configuration mixing and are not expected to describe complex spectroscopy of 
states such as those in $^{18}$O with two valence neutrons.

The emergence of clustering in large scale shell model calculations is also of
interest. Ideally, the complete shell model 
basis with the inclusion of the reaction continuum is sufficient for the description 
of the cluster structures in a nucleus. However, practically, it is often 
necessary to restrict the basis of the shell model wave functions, thus losing a
large fraction of the $\alpha$-cluster components. For example, recent {\it ab 
initio} calculations \cite{Maris2009} show difficulties in obtaining a correct excitation 
energy for the well known $\alpha$-cluster second 0$^+$ state in $^{12}$C.
Comparison of clustering observables to the shell model predictions addresses the emergence 
of cluster configurations as basis increase in size and highlights the interplay between 
single-nucleon and cluster degrees of freedom in N$\ne$Z nuclei.

The goal of this paper is to examine the structure of $^{18}$O using detailed 
R-matrix analysis of the $^{14}$C+$\alpha$ elastic scattering excitation 
functions and to examine the results using shell model. In our work we use  
Cluster-Nucleon Configuration Interaction Model  (CNCIM) \cite{CNCIM}, which 
represents the latest developments of the shell model approach to clustering. 

There have been many experimental efforts to study the $\alpha$-cluster 
structure of $^{18}$O using different approaches 
\cite{Zhao89,Buch07,Fort78,Oert10,Cuns81,Cuns82,Cuns83,Smit85,Smit88,San56,Bair66,Wag02,Cobe80,Yild06,Curt02,Jahn78,Wood78,Morg70,Wein58,Gold04,John09}. 
These experiments are sensitive to cluster states of different energy, spin, width 
and configuration to various degree, thus they contributed valuable complimentary 
experimental information. The detailed R-matrix analysis of the $^{14}$C+$\alpha$ 
elastic scattering excitation functions over a wide range of energies is performed here for the 
first time, and summarizes several years of activity. The same 
experimental setup and analysis techniques are used in this work as in \cite{John09}. The beam energy 
used here is higher than in \cite{John09} in order to study the 
excitation function for $^{14}$C+$\alpha$ at higher energies.

\section{Experiment and Analysis}
The experiment was performed at the Florida State University, John D.\ Fox 
Superconducting Linear Accelerator facility. The Thick Target Inverse Kinematics  
(TTIK) technique was used to measure the $^{14}$C+$\alpha$ elastic scattering 
excitation function. The technique was first suggested by \cite{Arte90,Gold93}. 
More details about the technique can be found in \cite{Mark00,Lonn10}. 
In this approach helium gas is used as 
the target and the $^{14}$C ions as the beam. The pressure of the helium gas in 
the chamber was adjusted for the beam to stop completely inside the chamber 
before reaching the detectors. When an interaction between the beam and the gas 
target occurs, the $\alpha$ particle gains kinetic energy from the 
projectile and propagates forward. Specific energy loss of the 
$\alpha$-particle is much smaller than that of $^{14}$C, which allows the 
$\alpha$-particle to emerge from the target with little energy loss. The energy 
spectrum of the $\alpha$-particles measured by the detectors, also placed in the 
target gas, would then reflect the $^{14}$C+$\alpha$ excitation function. This 
technique allows one to measure a large range of excitation energies without the 
need to change the initial energy of the beam, making the experiment more 
efficient and less time consuming. There is also the additional benefit of not 
having to use a radioactive $^{14}$C target. The $^{14}$C beam was produced by 
an FN Tandem Van de Graaff accelerator using a special $^{14}$C SNICS-II 
cesium-sputter ion source. The $^{14}$C beam of 42 MeV energy was sent into a 
chamber filled with 99.9\% pure helium gas ($^4$He).

\begin{figure}
\begin{center}\includegraphics[scale=0.28]{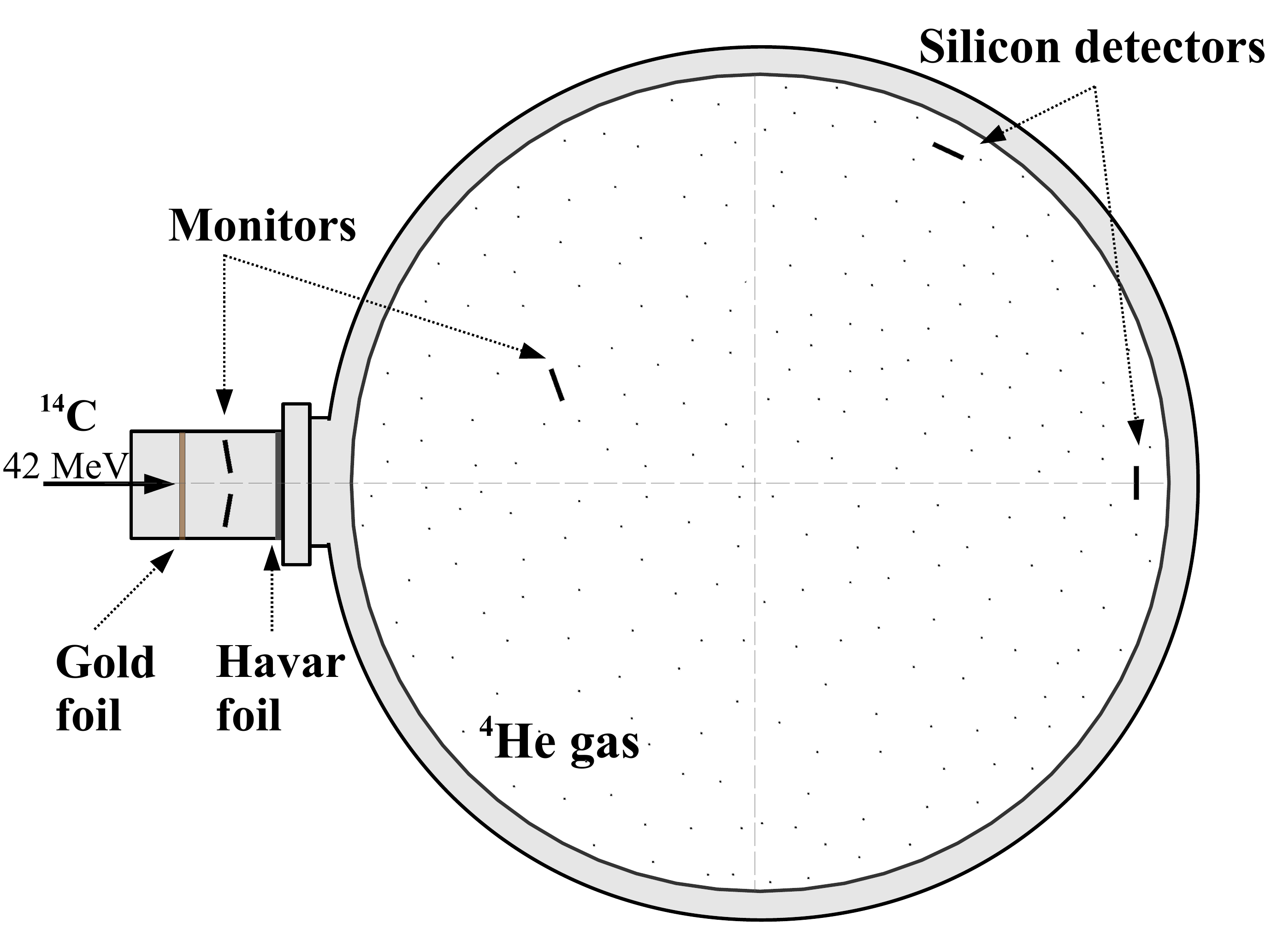}\end{center}
\caption{\label{fig:Chamber} Experimental setup for measurements of the elastic 
scattering of $^{14}$C by $\alpha$ particles using the Thick Target Inverse 
Kinematics technique. The silicon detectors were placed at angles
ranging from 0$^{\circ}$ to 60$^{\circ}$ in steps of 5$^{\circ}$. Only the extreme
angles are shown.}
\end{figure}

To monitor the beam quality and alignment during the run a gold foil was used 
before the entrance window, where elastic scattering was measured by silicon 
detectors arranged symmetrically with respect to the beam axis. The entrance 
window of the chamber was covered with a 1.27 $\mu$m Havar foil. In a 
conventional experiment (with thin target) it is usually easy to measure the 
intensity of the beam using a Faraday cup. In the thick target approach it is 
not possible, since the beam ions stop inside the target. 
Therefore, the intensity 
of the incoming beam was determined using elastic scattering of the beam ions from 
the Havar entrance window as measured by a monitor detector taking into account 
each of the components of the Havar foil. The monitor detector was placed at 15 
degrees, 22 cm away from the entrance window. 
To calculate the elastic scattering each of the components of the Havar foil are taken 
into account. It was verified using the optical model that the cross section at this angle is 
mostly Rutherford for most of the components of the Havar foil. The contribution of each
component is weighted according to the percentage 
of Havar chemical composition.
The experimental setup is shown in 
Fig. \ref{fig:Chamber}.  The accuracy of the absolute normalization is 15\%. An 
array of silicon detectors was placed inside the chamber at angles ranging from 
0$^{\circ}$ to 60$^{\circ}$ in steps of 5$^{\circ}$ to detect the recoiling $\alpha$-particles.
In Fig \ref{fig:Chamber} only the extreme angles for the detector positions are shown.
Excitation functions 
covering the excitation energy region of 8-14.9 MeV were measured at 13 different 
angles. 

The spectra of $\alpha$-particles measured in the laboratory frame have to be 
converted into c.m. excitation functions for further analysis. Because of the extended gas target 
the scattering angle is not fixed and has to be calculated from the energy of 
the recoil $\alpha$-particle and the known location of the detectors. The 
energy loss of an $\alpha$ particle and the solid angle also depend on the 
location of interaction point and have to be calculated for each energy bin in 
the measured spectrum. This is done using a code which takes into account the 
relevant experimental conditions. Details of the procedure can be found in 
\cite{SantaTecla10}. 

Monte Carlo simulations based on the GEANT 3.21 library were performed in order 
to evaluate the dependence of the experimental energy resolution on the c.m. 
energy and scattering angle, and also to correct for the detector mount shadow 
effects in the calculation of the absolute cross section.This information was 
used in a convolution of the R-Matrix calculations.

The excitation functions were analyzed using a multi-level, multi-channel 
R-Matrix approach \cite{Lane58}. The details of the R-Matrix analysis can be found in 
Refs. \cite{EDJ-Thesis,Avila13}. The results are presented by quoting the 
quantum numbers and energies of the resonances together with 
the partial decay width $\Gamma_c$ for every open channel. The width is also 
expressed via the dimensionless reduced width parameter $\theta^2$, which 
represents the ratio of the observed decay width relative to the single-particle 
limit. This dimensionless reduced width was compared with spectroscopic factors 
from theoretical calculations. The $\alpha$ and neutron decay channels of 
$^{18}$O were included in the fit. For the $^{14}$C($\alpha$,n) reaction the 
decay channels to the ground state and to the first excited state of $^{17}$O 
were included. At $^{14}$C beam energies above 27.4 MeV the inelastic channel 
$^{14}$C($\alpha$,$\alpha^{\prime}$) is also open. It is not possible to 
distinguish between the $\alpha$-particles coming from elastic and inelastic 
reactions in this specific realization of the TTIK approach. Due to the negative Q-value 
for the inelastic scattering (-6.1 MeV) the recoiled $\alpha$-particles from 
inelastic events would have significantly smaller energy than the elastically 
scattered $\alpha$-particles from the same location in the target, and would 
show up in the elastic scattering spectrum as background at low energies. A 
direct comparison with the low energy data from the previous experiment 
\cite{John09}, performed at 25 MeV, where the inelastic channel was not open, 
showed no evidence of inelastic contribution.

\begin{figure}
  \begin{center}\includegraphics[scale=0.3]{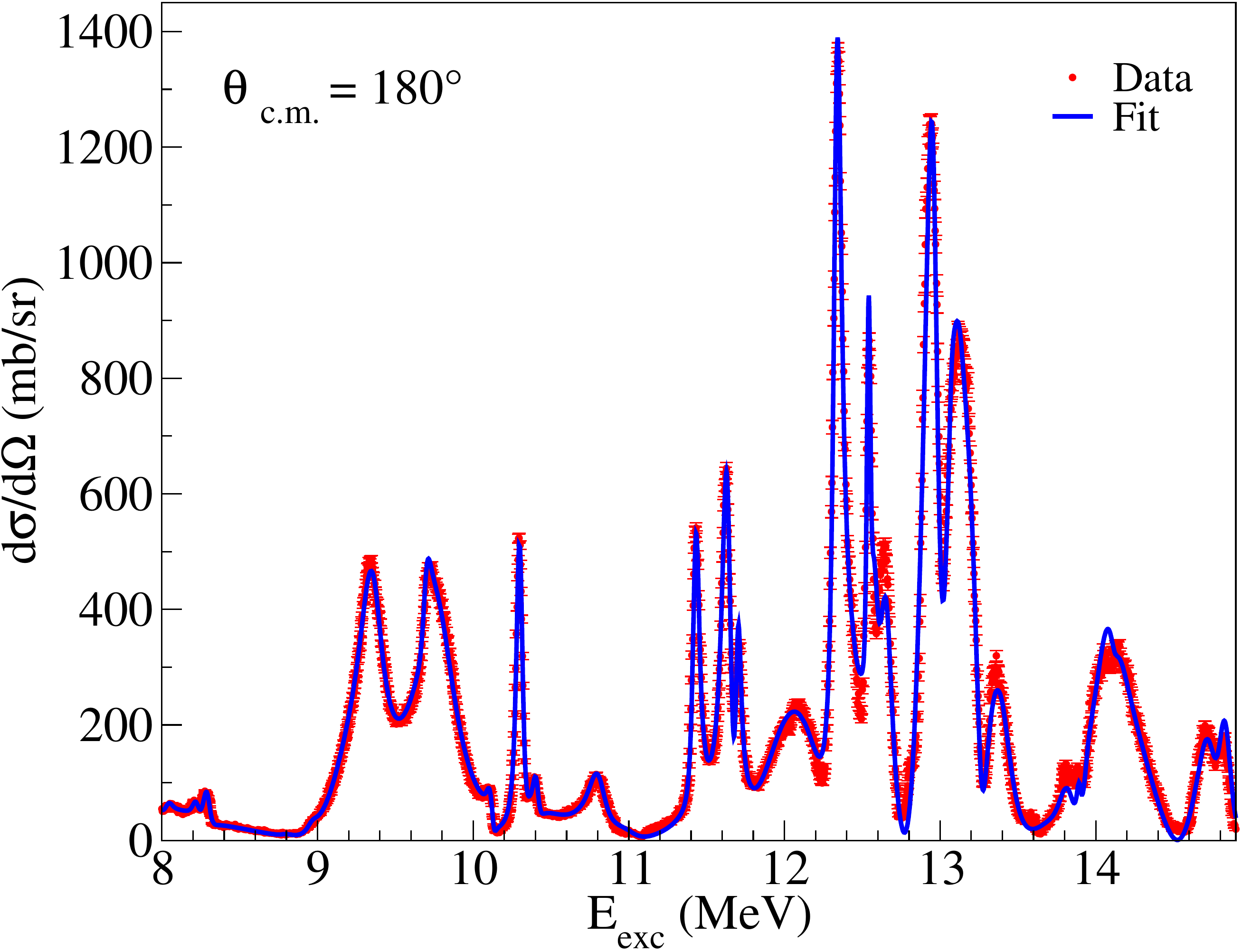}\end{center}
   \caption{\label{fig:Spectrum} (Color online) Excitation function for $^{14}$C+$\alpha$ 
elastic scattering at 180$^{\circ}$ in the c.m. frame measured with a 42 MeV $^{14}$C beam. The solid curve is the best R-matrix fit.}
\end{figure}

A total of fifty-four resonances were used to fit the data. The excitation 
function at 180$^{\circ}$ in the c.m. for the entire energy range measured in this experiment is shown in 
Fig. \ref{fig:Spectrum}. 
The uncertainties of the best fit parameters were determined using a Monte Carlo procedure. The parameter
values were varied randomly, but only values that produced no more than one standard deviation from the best fit $\chi^2$ values
where excepted. The resulting distribution of parameter values was used to determine the uncertainty. 
This was done state by state, not taking into account correlation between different states. Therefore the provided uncertainties
give good indication on the sensitivity of the fit to the specific parameters, but the actual uncertainty values may be enhanced 
if correlation between the states is taken into account.

Detailed description of the observed spectra is given 
in the next section.

\section{Discussion of the $^{18}$O spectrum at excitation energies between 8.0 MeV 
and 14.9 MeV\label{sec:18OResults}}

This section contains a detailed discussion of the properties of the states in 
$^{18}$O extracted from the experimental data of this work. Also, the results are 
compared to the available data from previous experimental studies. The discussion 
is structured according to the excitation energies of the resonances, which are 
grouped into 1 MeV intervals.The experimental data are used for calculating 
the dimensionless reduced $\alpha$ and nucleon widths, which are defined as
ratios of the corresponding reduced widths to the corresponding single particle limit 
($\gamma^2/({\hbar}^2/{\mu}R^2$)). Here $\mu$ and $R$ are, respectively, the 
reduced mass and channel radius for the corresponding decay channel (5.2 fm for $\alpha$ decay 
and 4.6 fm for nucleon decay).

\subsection{Resonances in the excitation energy range between 8.0 MeV and 9.0 
MeV}

The lowest excitation energy state that is clearly visible in the measured 
excitation function is the known 1$^-$ state at 8.0378(7) MeV \cite{Till95}. 
This is a narrow state with a width significantly smaller than our experimental 
resolution ($\approx$40 keV at 1.8 MeV in the c.m.), and therefore a direct 
measurement of the width was not possible. However, this state is below the 
neutron decay threshold (8.044 MeV), and only one decay channel is open, 
$\alpha$ decay to the ground state of $^{14}$C. If the effects related to 
interference with other states are neglected, then the width of the state can be determined from the 
measured experimental cross section 
($\Gamma_{nat}=\sigma_{ex}/\sigma_{th}\times\Delta$, where $\Delta$ is the 
experimental energy resolution calculated using Monte Carlo simulation). The  
width of this state was determined as 2.0(7) keV.  The only definitive width 
measurement for this state, made prior to this experiment, was reported in 
\cite{Buch07} where this state was populated in the $\beta$-decay of $^{18}$N, and 
a width of 0.95$^{+0.4}_{-0.9}$ keV was determined. The result of this work is in 
fair agreement with that of \cite{Buch07} and the upper limit 
for the width given in \cite{Till95}. This width corresponds to the $\alpha$ 
dimensionless reduced width of only $\theta^2=0.02$, so
that this state does not have a strong overlap with the $^{14}$C(g.s)+$\alpha$ 
configuration. This finding contradicts previous suggestions made in 
\cite{Ashw06,Oert10} that this state might be the band head of the negative 
parity, inversion doublet, $\alpha$-cluster quasi-rotational band.
 
\begin{figure}[ht!]
\begin{center}
\includegraphics[scale=0.35]{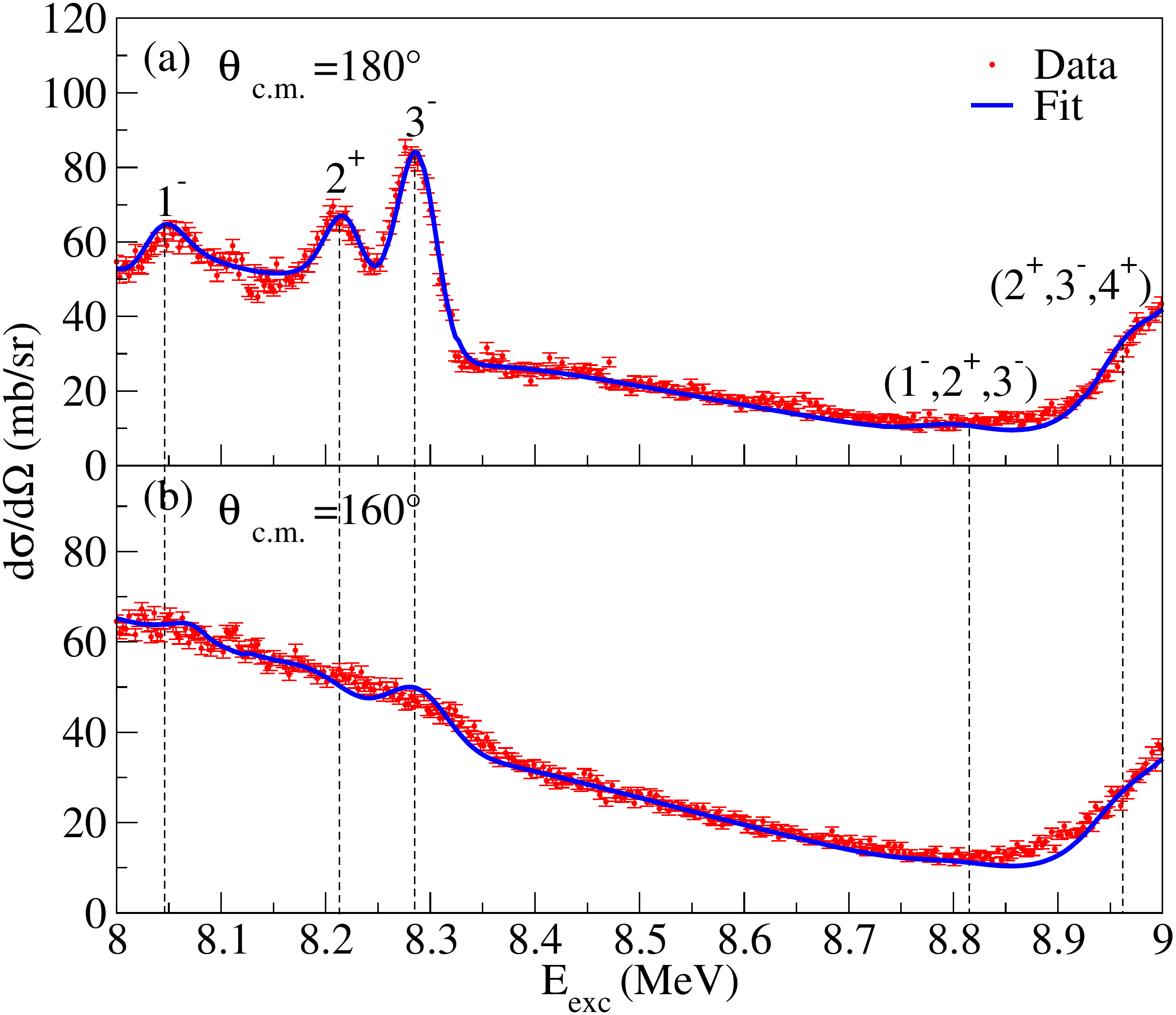}
\end{center}
\caption{\label{fig:8-9} (Color online) Excitation functions for the elastic scattering of 
$\alpha$ particles
from $^{14}$C from 8 to 9 MeV with the best $ $\textbf{R}-Matrix fit(solid curve). 
A 3$^-$ spin-parity
assignment for the state at 8.82 MeV and a 2$^+$ for the state at 8.96 MeV were 
used for this diagram.}
\end{figure}

Four more states are observed in the 8 to 9 MeV excitation region (Fig. 
\ref{fig:8-9}). The 8.213(4) MeV 2$^+$ state is a well known narrow state 
\cite{Till95}that is above the neutron decay threshold and has been 
observed in the $^{14}$C($\alpha$,n) reaction \cite{San56,Bair66}. In \cite{San56} 
the state was observed at 8.223 MeV with a width of 1.6(10) keV, and in 
\cite{Bair66} the state was located at 8.217 MeV with a width of 1(1) keV, but 
spin and party were not assigned to it. The inverse reaction $^{17}$O($n,\alpha$) has 
also been studied in \cite{Wag02}, where the 2$^+$ state was found at 8.213(4) 
MeV with width of 2.26(14) keV.  In the $^{14}$C($\alpha$,$\alpha$) elastic 
scattering studied by \cite{Wein58} the 2$^+$ state was observed at 8.222 MeV 
and its width was determined to be 1.2(8) keV. In this work the combined analysis 
of the $^{14}$C($\alpha$,$\alpha$) and $^{14}$C($\alpha$,n) data constrains the 
properties of this state rather well. The 2$^+$ state is 
observed at 8.22(1) MeV and its dimensionless $\alpha$ reduced width  
is 0.03 and the total width is 1.9(2) keV.

A 3$^-$ state is observed at 8.290(6) MeV with a width of 8.5(9) keV. This is 
the dominant feature in the $^{14}$C($\alpha$,n) spectrum and also is prominent 
in the $^{14}$C($\alpha$,$\alpha$) excitation function. The properties of this 
state can be constrained reasonably well from these two data sets. We determine 
that this state has a significant dimensionless $\alpha$ reduced width of 0.18 and 
is the only $^{14}$C(g.s.)+$\alpha$ cluster state in the 8-9 MeV region. It was 
observed earlier in \cite{San56,Wein58} at 8.293 MeV with a width of 10(1) keV and 
7.7(9) keV respectively. In \cite{Bair66} the narrow state was found at energy 
of 8.287 MeV but the spin assignment was not made.  The 3$^-$ state was also observed 
in \cite{Wag02} at 8.282(3) MeV with a width of 14.74(59) keV. The results of the present work are in 
good agreement with the parameters found in \cite{Wein58,San56} and in fair 
agreement with \cite{Bair66}, but disagree with the neutron and $\alpha$-partial widths 
determined in \cite{Wag02}. The partial $\alpha$ width 
is 2.9(2) keV and the neutron width is 5.6(7) keV here, whereas in 
\cite{Wag02} the partial $\alpha$ width and neutron width were 
13.661(416) keV and 1.08(2) keV, respectively. When the parameters from 
\cite{Wag02} are used in the analysis, the $^{14}$C($\alpha$,n) spectrum is still well reproduced but 
the $^{14}$C($\alpha$,$\alpha$) spectrum is not.

The state at 8.82(3) MeV with width of 60(10) keV is clearly visible in the 
$^{14}$C($\alpha$,n) spectrum (Fig. \ref{fig:neutron}), but its evidence in the 
$^{14}$C($\alpha$,$\alpha$) spectrum is weak. This state has been observed earlier in 
\cite{San56} at 8.832 MeV with a width of 100(20) keV and in \cite{Bair66} at 
8.809 MeV and with a width of 80(20) keV, both from the $^{14}$C($\alpha$,n) reaction. 
However, spin and parity assignments were not made. A state at 8.82 MeV with 
a width of 70(12) keV was observed in \cite{Sell95} using $^{18}$O$(e,e')$ and a
tentative (1$^+$) spin-parity assignment was made. If this unnatural parity 
assignment is correct then the state cannot be observed in the $^{14}$C($\alpha$,n) 
or $^{14}$C($\alpha$,$\alpha$) reactions.  Therefore these results
should correspond to two different states, or the assignment of unnatural parity 
is not correct. It is not possible to determine the spin of the state from our 
data since it is barely seen in the $^{14}$C($\alpha$,$\alpha$) spectrum, and 
different spin assignments (1$^-$,2$^+$,3$^-$) fit the $^{14}$C($\alpha$,n) and 
the $^{14}$C($\alpha$,$\alpha$) spectra fairly well.

The 8.96(1) MeV state with a width of 70(30) keV shows up as a small bump in the 
$^{14}$C($\alpha$,$\alpha$) spectrum, and it is a relatively strong state in the 
$^{14}$C($\alpha$,n) excitation function. It was observed in \cite{San56} at 
8.966 MeV with a width of 54(3) keV, and in \cite{Bair66} at 8.956 MeV with a width of 
65(10) keV. The 4$^+$ assignment was made for this state in \cite{Sell95}. A 
pair of states around 9.0 MeV was observed in the $^{16}$O(t,p) reaction in 
\cite{Cobe80}. Indirect arguments were given supporting the 4$^+$ assignment for 
at least one of them \cite{Fortune1985} and it was argued in \cite{Fortune1985} 
that a 4$^+$ state at 9.0 MeV probably has $(1d_{5/2})(1d_{3/2})$ configuration. Our 
data are consistent with a 4$^+$ assignment for the 8.96 MeV, however 2$^+$ and 
3$^-$ assignments cannot be excluded. If the 4$^+$ assignment is correct then 
this state has a substantial $\alpha$-cluster component (see discussion in section 
\ref{sec:CNCIM}).

\subsection{Resonances in the excitation energy range between 9.0 MeV and 10.0 
MeV}

This energy region is defined by the double peak structure with a very large 
$^{14}$C($\alpha$,$\alpha$) cross section at c.m. angles close to 180$^\circ$.  
This structure is the result of interference between several broad overlapping 
$\alpha$-cluster resonances, which makes the analysis very difficult. Six 
resonances were used to fit this energy range with five of them having very large 
dimensionless reduced $\alpha$-widths.

\begin{figure}[ht!]
\begin{center}\includegraphics[scale=0.39]{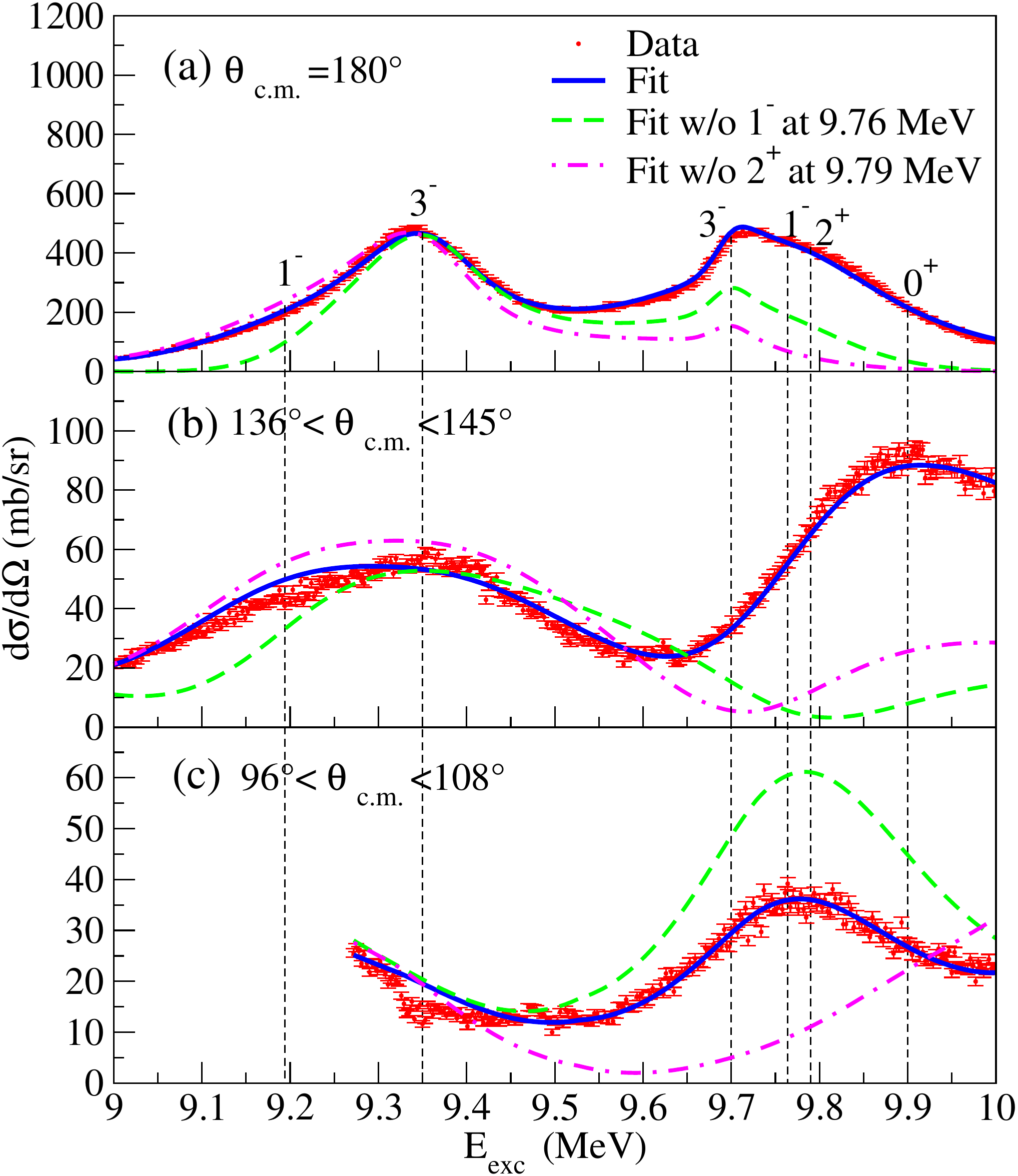}
 % 9-10MeV.eps: 0x0 pixel, 300dpi, 0.00x0.00 cm, bb=
\end{center}
\caption{\label{fig:9-10} (Color online) Excitation functions for the elastic scattering of 
$\alpha$ particles from $^{14}$C with the best R-Matrix fit (solid curve) for the 
energy range of 9-10 MeV. The dashed and the dash-dotted curves represent the 
best R-Matrix fit without the inclusion of  the 1$^-$ and 2$^+$ resonances at 
9.76 and 9.79 MeV, respectively}
\end{figure}

Two broad 1$^-$ states are observed. The first one at 9.19(2) MeV shows up as 
the tail on the left side of the first broad peak (mainly a 3$^-$ state) at 
180$^{\circ}$ shown in Fig. \ref{fig:9-10} (a). This state can also be seen as 
the first peak shown in Fig. \ref{fig:9-10} (b) at 140$^\circ$. At 
$\theta\approx$ 140$^\circ$ the contribution from the 3$^-$ state  disappears, 
which makes the 1$^-$ state very obvious. This state was first observed in 
\cite{Zhao89} at about 9.2 MeV and with a partial $\alpha$-width close to 500 keV 
(although, it was interpreted as the interference of six unresolved states), and 
then in \cite{Buch07} using $\beta$-delayed $\alpha$ emission of $^{18}$N.  In 
\cite{Buch07} the excitation energy and partial $\alpha$-width of this 1$^-$ 
state are reported to be 9.16(10) MeV and 420(200) keV respectively. The 
1$^-$ state was also suggested in \cite{Gold04} at 9.027$^{+.15}_{-.03}$  MeV 
with a width of 500$^{+150}_{-50}$ keV. Our analysis indicates that the width of this state is 
220(30) keV, which is significantly smaller, but still within the large error bars 
of \cite{Buch07}. This state also has a smaller width when compared to \cite{Gold04}. 
However, the analysis in \cite{Gold04} does not take into account interference 
with other states, which can lead to overestimation of the width.
The width of the state is dominated by the partial 
$\alpha$-width and it has a dimensionless $\alpha$ reduced width is 
$\theta_{\alpha}^2$=0.20. While it is a factor of two smaller than in 
\cite{Buch07}, it is still large enough for the state to be considered as a strong 
cluster state with $^{14}$C(g.s)+$\alpha$ configuration. 

The other 1$^-$ state is at 9.76(2) MeV with a width of 700(120) keV. The 
parameters of this state are in good agreement with \cite{Buch07} where it 
was observed at 9.85(50) MeV and with a
partial $\alpha$-width of 560(200) keV. Constructive interference of this state 
and a 2$^+$ state at 9.79 MeV makes a strong contribution to the  second peak at 
180$^\circ$ and 140$^\circ$, as can be seen in Fig. \ref{fig:9-10} (a,b).  
The dashed curve in Fig. \ref{fig:9-10} shows the R-Matrix fit without the 
inclusion of this 1$^-$ state.

Two 3$^-$ states are observed in this energy region. The first one at 9.35(2) 
MeV is a dominant $\alpha$-cluster resonance and makes a dominant contribution 
to the first broad peak at 180$^{\circ}$ in Fig. \ref{fig:9-10}(a). This state produces 
a very prominent peak in the $^{14}$C($\alpha,n$) reaction also, as can be seen 
in Fig. \ref{fig:neutron}. The state was observed in \cite{Curt02} at 9.35 MeV, 
and a suggestion of a 2$^+$ or 3$^-$ spin-parity assignment was made. In 
\cite{Man90} a state at 9.36 MeV with a 2$^+$ or 3$^-$ spin parity assignment was 
observed but its small width of $<$20 keV indicates that this may not be the 
same state. A 3$^-$ state was also suggested in \cite{Gold04} at
9.39(2) MeV with a width of 200(20) keV. Our best fit for the width is 180(30) 
keV, with a dimensionless reduced width of 0.48. This is the strongest  
$\alpha$-cluster state in this energy range.

The other 3$^-$ state is at 9.70(1) MeV and has a width of 140(10) keV. This 
state has a small dimensionless reduced $\alpha$ width, and therefore little 
influence on the elastic cross section. It is needed, however, to reproduce 
the neutron spectrum as is shown in Fig. \ref{fig:neutron}. This state gives little 
contribution to the shape of the second peak at 180$^\circ$  (Fig. 
\ref{fig:9-10}(a)), but no influence at 140$^\circ$ (Fig. \ref{fig:9-10}(b)) 
indicating a 3$^-$ spin-parity assignment. A state was observed in \cite{Sell95} 
at 9.71(1) MeV and it was identified as a tentative (5$^-$) state. It was also 
seen in \cite{Curt02} at 9.70 MeV where a tentative (1$^-$,2$^+$,3$^-$) 
assignment was made. A 3$^-$ state was observed by \cite{Gold04} at 9.711(15) MeV with 
a width of 75(15) keV. In reference \cite{Oert10} a 3$^-$ state was suggested at similar 
energy (9.715(5) MeV) but with a much smaller width.

A 2$^+$ state was found at 9.79(6) MeV with a width of 170(80) keV and 
dimensionless reduced width of 0.1. This state strongly interferes with 
the 1$^-$ state at 9.76(2) MeV, which makes a large contribution to the second peak 
at all angles shown in Fig. \ref{fig:9-10}. At angles close to 90$^{\circ}$  the 
cross section for any negative parity states vanish, which makes the existence of 
this 2$^+$ state very evident as shown in Fig. \ref{fig:9-10} (c). The 
dash-dotted curve in Fig. \ref{fig:9-10} shows the R-Matrix fit without this 
2$^+$ state.

The last state in this energy range is a very broad 0$^+$ state at 9.9(1) MeV 
with partial $\alpha$ width of 3.2(8) MeV. A more detailed discussion of this 
state will be given in Section \ref{Sec:PM}.

\subsection{Resonances in the excitation energy range between 10.0 MeV and 11.0 
MeV}

\begin{figure}[t!]
\begin{center}
 \includegraphics[scale=0.4]{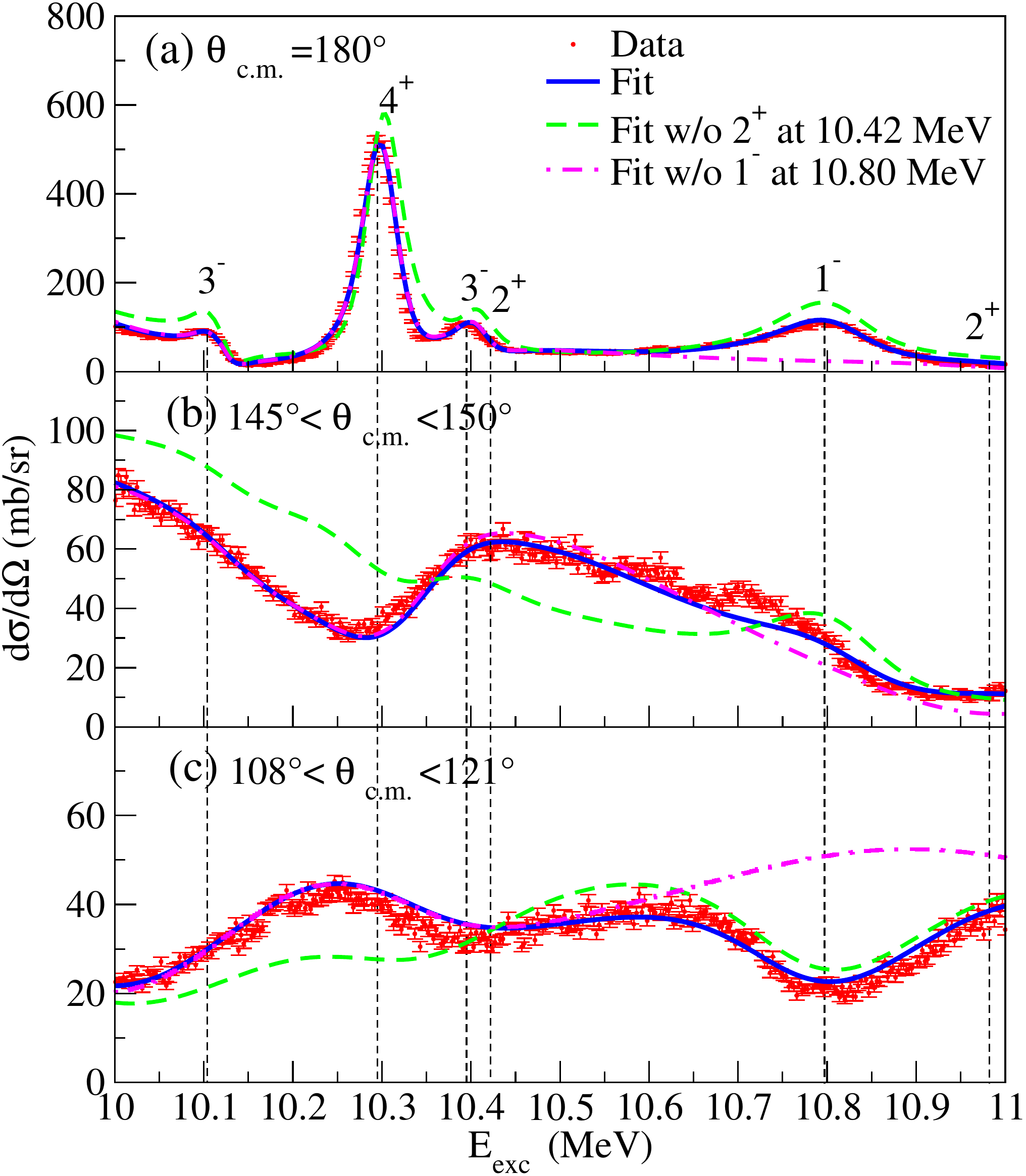}
 % 10-11.eps: 0x0 pixel, 300dpi, 0.00x0.00 cm, bb=
\end{center}
\caption{\label{fig:10-11} (Color online) Excitation functions for elastic scattering of 
$\alpha$-particles from $^{14}$C at 180$^\circ$, 147$^\circ$ and 115$^\circ$ with 
the best R-Matrix fit(solid curve) for the energy range of 10-11 MeV. The dashed 
line and the dash-dotted curve represents the best R-Matrix fit without the 
inclusion of the 2$^+$ and 1$^-$ states at 10.42 and 10.80 MeV, respectively.}
\end{figure}

Six resonances were used to fit this energy region and only one resonance has 
a dimensionless reduced $\alpha$-width of more than 10\% of an $\alpha$ single 
particle width. This region has well defined resonances and most of them can be 
seen as distinct peaks at 180$^\circ$ in Fig \ref{fig:10-11}(a).

Two 3$^-$ states were observed in this region and they are fairly obvious 
narrow peaks in the large c.m. angle data shown in Fig. \ref{fig:10-11}(a), at 
10.11(1) and 10.395(9) MeV. They are also important resonances in the 
$^{14}$C($\alpha$,n) spectra as can be seen in Fig. \ref{fig:neutron}. The 
widths of these states are 16(5) and 70(20) keV, respectively. These states have 
also been identified in previous works \cite{Till95,Gold04,Oert10}. In 
\cite{Gold04} the width for both states was determined to be 45(8) keV at 
10.10(1) and 10.365(10) MeV. In \cite{Oert10} these states were at 10.111(5) 
and 10.400(7) MeV with widths of 12 keV and 30 keV, respectively.

Two 2$^+$ states were observed. The first 2$^+$ is at 10.42(1) MeV and has a width 
of 180(40) keV and it strongly interferes with its neighboring 
states. Its presence is needed to reproduce the minimum at 10.3 MeV at 147$^{\circ}$ (Fig. 
\ref{fig:10-11}(b)). At 110$^\circ$ this state contributes to the first broad 
peak shown in Fig. \ref{fig:10-11}(c). The dashed line in Fig. \ref{fig:10-11} 
shows the fit without the inclusion of this 2$^+$ state. A 2$^+$ state at 10.43(15) 
MeV was suggested earlier in \cite{Gold04} with the somewhat larger width of 500(150) 
keV.

The second 2$^+$ state is at 10.98(4) MeV and has a width of 280(130) keV. This state is 
weak in the $^{14}$C($\alpha$,$\alpha$) channel. However, due to its interference 
with a broad 2$^+$ state at higher energy it helps to shape the cross section at 
150$^\circ$ at around 11 MeV.

A sharp 4$^+$ state is observed at 10.290(4) MeV with a width of 29(4) keV. This well known
state is seen as the peak with a large cross section at 180$^\circ$ (Fig. 
\ref{fig:10-11}(a)) and has been observed previously in 
\cite{Oert10,Morg70,Cuns81,Curt02,Gold04,Ashw06,Sell95,Cobe80,Smit88,Jahn78,Yild06}.

The last state in this energy region is a $1^-$ state at 10.80(3) MeV with a width 
of 690(110) keV. This state is seen as the broad peak at 180$^\circ$ shown in 
Fig. \ref{fig:10-11} (a). At angles close to 118$^{\circ}$ this state is needed 
to reproduce the minimum in the cross section shown in Fig. \ref{fig:10-11} (c). 
In Fig \ref{fig:10-11} the dash-dotted line shows the R-matrix fit without this 
1$^-$ state. A 1$^-$ state was previously seen by \cite{Buch07} at 10.89(10) MeV 
with a partial $\alpha$-width of 300(100) keV. The partial $\alpha$-width in our 
best fit is 630(90) keV.

\subsection{Resonances in the excitation energy range between 11.0 MeV and 12.0 
MeV}

Seven resonances were used to fit this energy region. Three of these resonances 
have $\alpha$ dimensionless reduced widths of more than 10\%.

\begin{figure}[t!]
\begin{center}
 \includegraphics[scale=0.4]{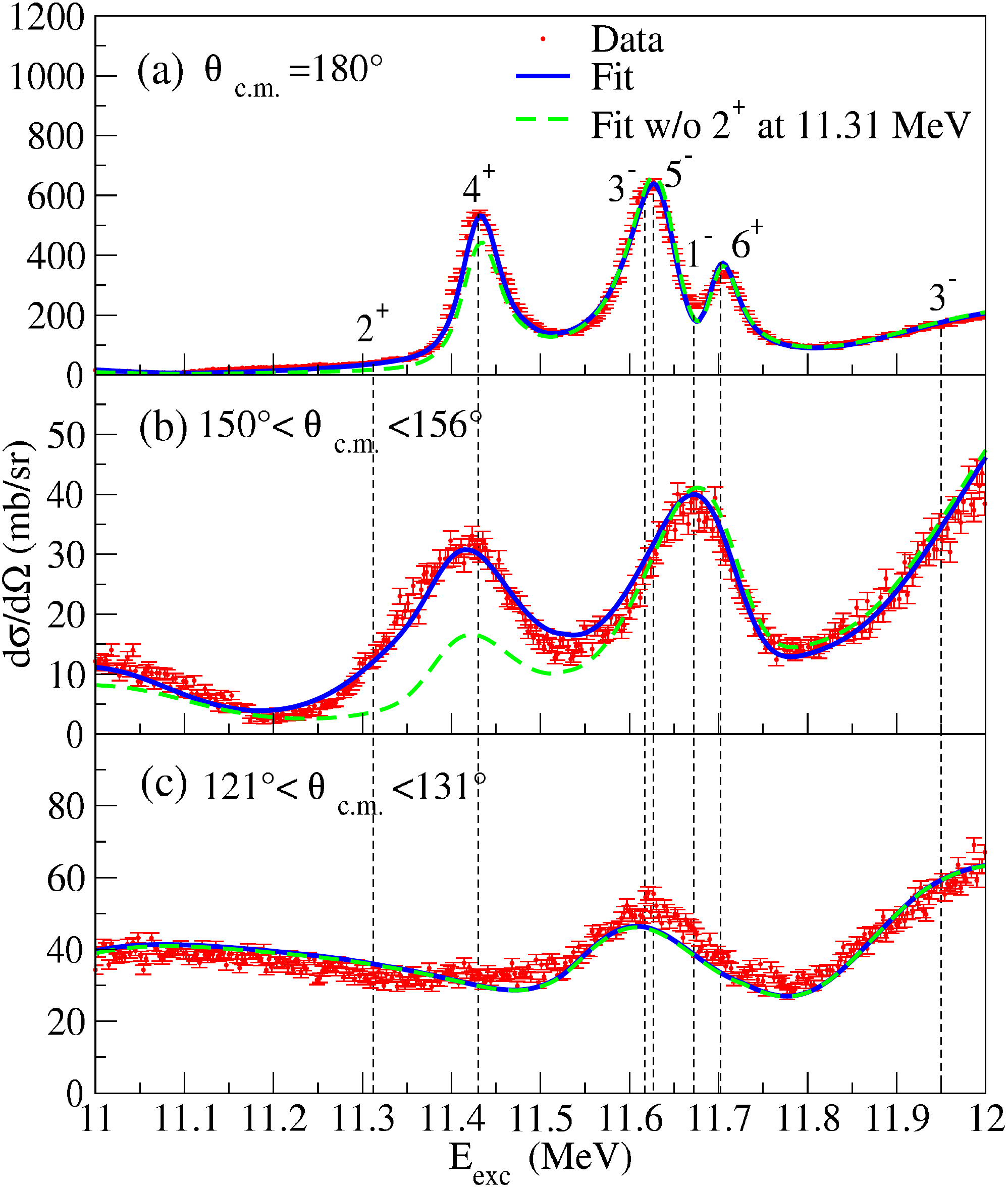}
 % 10-11.eps: 0x0 pixel, 300dpi, 0.00x0.00 cm, bb=
\end{center}
\caption{\label{fig:11-12} (Color online) Excitation functions for elastic scattering of 
$\alpha$ particles from $^{14}$C at 180$^\circ$, 150$^\circ$ and 120$^\circ$ with 
the best R-Matrix fit (solid curve) for the energy range of 11-12 MeV. The 
dashed line represents the R-matrix fit without the inclusion of a 2$^+$ state 
at 11.31 MeV}
\end{figure}

A 2$^+$ state is observed at 11.31(8) MeV with a width of 250(100) keV. This state 
contributes to the first peak shown in Fig. \ref{fig:11-12} (a,b) mostly due to 
interference with a broad 2$^+$ state at higher excitation energy. At angles close to 125$^\circ$ 
the cross section for a 2$^+$ state vanish having no effect on the cross section 
in Fig. \ref{fig:11-12} (c). The dashed curve in Fig. \ref{fig:11-12} represents 
the R-matrix fit without the inclusion of this state. A broad 2$^+$ state at 
11.39 MeV was suggested in \cite{Morg70}.

A 4$^+$ state is observed at 11.43(1) MeV with a width of 40(10) keV. This state 
is the main contributor to the peak shown in Fig. \ref{fig:11-12}(a,b) and was 
previously seen in \cite{Morg70,Cuns81,Wood78,Ashw06,Gold04,Smit88,Oert10}. It 
was suggested as a (4$^+$) by \cite{Morg70} and later by \cite{Cuns81}; 
however, the width was not measured. In \cite{Gold04} it was found at 11.415(5) 
MeV with a width of 45 keV, which is in good agreement with our findings. In 
\cite{Oert10} this state was at 11.423(5) MeV with width of a 35 keV. This state 
is also visible in the $^{14}$C($\alpha$,n) spectrum (Fig. \ref{fig:neutron}).

Two 3$^-$ states were observed in this energy range. The first one is at 
11.62(3) MeV and has a width of 150(20) keV. This state is needed to reproduce the 
second peak in Fig. \ref{fig:11-12} (b). In \cite{Sell95} a state at 11.67(2) 
MeV with a width of 112(7) keV  was identified as a possible 3$^-$ state, which 
agrees with our values. The second 3$^-$ state is at 11.95(1) MeV. It is a 
broad state in both the $^{14}$C($\alpha$,$\alpha$) and the $^{14}$C($\alpha$,n) 
spectra with a total width of 560(70) keV. It has a strong influence at all 
angles except at angles close to 140$^\circ$, making this a good indicator for a 
3$^-$ state. This state brings the cross section up at energies around 12 MeV at 
all the angles shown in Fig. \ref{fig:11-12} and is one of the states with a 
larger degree of clustering in this energy range with dimensionless reduced 
$\alpha$-width of 0.17. In the $^{14}$C($\alpha$,n) spectrum this state is seen 
as a broad peak (Fig. \ref{fig:neutron}).

There is the well known 5$^-$ state observed at 11.627(4) MeV with a width of 
40(5) keV. It can be seen as the second peak in Fig. \ref{fig:11-12}(a), and it 
also contributes to the peak in Fig. \ref{fig:11-12} (c).  It has no 
contribution to Fig. \ref{fig:11-12}(b) because at 155$^\circ$ the cross section 
for a 5$^-$ state becomes zero. It has a significant dimensionless reduced $\alpha$ 
width of 0.13. This state was previously identified as 5$^-$ in 
\cite{Oert10,Curt02,Ashw06,Rae84,Cuns81,Yild06,Gold04,Smit88,Morg70} with a width 
of 60(5) keV as measured in \cite{Gold04} and 25 keV in \cite{Oert10}.

A 1$^-$ state is observed at 11.67(2) MeV and has a width of 200(90) keV. It shows 
up at all angles contributing to the shape of the second peak in Fig. 
\ref{fig:11-12} (a,b) and to the only peak seen in Fig. \ref{fig:11-12}(c).  In 
\cite{Buch07} a 1$^-$ state with $\alpha$ partial width of 220(100) keV was 
found at 11.56(10) MeV.

There is one 6$^+$ state at 11.699(5) MeV with a width of 23(2) keV. This state is the 
last peak seen in Fig. \ref{fig:11-12} (a). It has a dimensionless reduced 
$\alpha$-width of 0.23, and is recognized as the first strong 6$^+$ 
$\alpha$-cluster state. There is a strong interference between this state and 
another 6$^+$ state found at a higher energy that was previously observed and 
identified as 6$^+$ in \cite{Oert10,Morg70,Cuns81,Smit88,Gold04}. The width of 
this state was determined to be 35(5) keV in \cite{Gold04} and 27 keV in 
\cite{Oert10}.

\subsection{Resonances in the excitation energy range between 12.0 MeV and 13.0 
MeV}

This is an energy range with a higher density of states and strong interference 
among the states made this energy interval very challenging to fit. Twelve 
resonances were found and four of them have large dimensionless reduced $\alpha$ 
-widths. 

\begin{figure}[t!]
\begin{center}
 \includegraphics[scale=0.4]{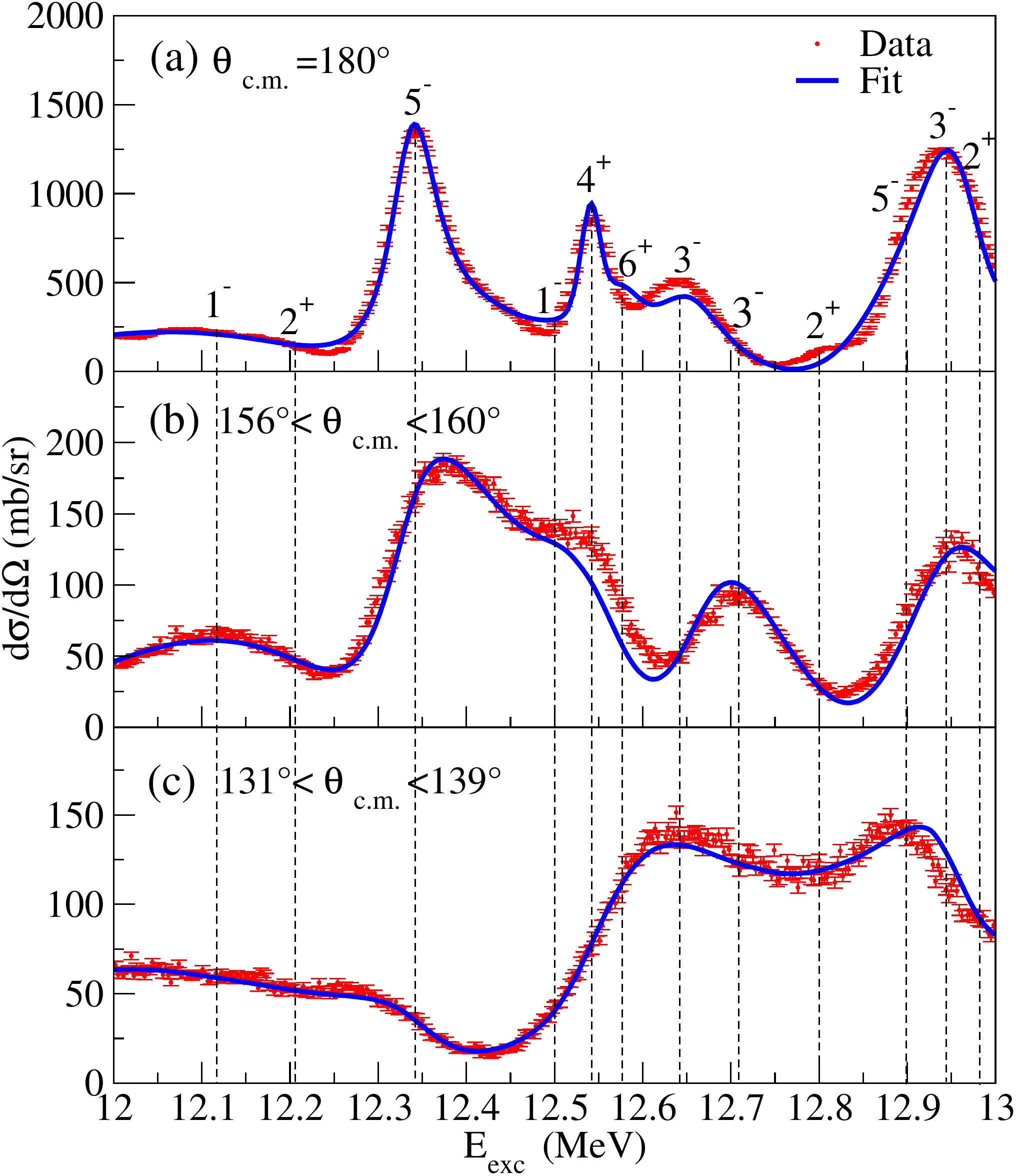}
 % 10-11.eps: 0x0 pixel, 300dpi, 0.00x0.00 cm, bb=
\end{center}
\caption{\label{fig:12-13} (Color online) Excitation functions for elastic scattering of 
$\alpha$ particles from $^{14}$C at 180$^\circ$, 156$^\circ$ and 131$^\circ$ with 
the best R-Matrix fit (solid curve) for the energy range of 12-13 MeV.}
\end{figure}

Two 1$^-$ states were observed. The first one is at 12.12(1) MeV and has a width 
of 410(120) keV. It contributes to the cross section of the first small bump 
seen in Fig. \ref{fig:12-13}(a,b). A 1$^-$ or 2$^+$ was suggested by 
\cite{Sell95} for a state at 12.09(2) MeV. The width of this state was not 
determined in \cite{Sell95}. It was also seen and identified as 1$^-$ by 
\cite{Buch07} at 12.12(10) MeV with a partial $\alpha$-width of 22(7) keV. The 
partial $\alpha$-width that we observed for this state is 50(10) keV in fair 
agreement with \cite{Buch07}. The other 1$^-$ state is at 12.5(1) MeV and has 
a width of 900(400) keV. It is important at large c.m. angles to reproduce the 
right side of the first peak shown in Fig. \ref{fig:12-13} (a,b). At smaller 
angles the interference of this state with other states brings the cross section 
down, making it possible to fit the rise of the cross section between 12.5 and 
12.6 MeV (Fig. \ref{fig:12-13} (c)). A 1$^-$ state at 12.95(50) MeV was also 
observed in \cite{Buch07} with a partial $\alpha$-width of 210(100) keV. Even 
though this state was found at a higher energy, our state is still within the 
error bars of \cite{Buch07}. The partial $\alpha$-width determined in this work, 
300(100) keV, is in good agreement with that determined in \cite{Buch07}. 

Three 2$^+$ resonances were observed. Two of them are strong $\alpha$-cluster 
states. The first one is at 12.21(8) MeV and has a width of 1100(300) keV and a
dimensionless reduced $\alpha$-width of 0.37. The second 2$+$ 
state is at 12.90(3) MeV and has a width of 310(30) keV. It contributes to 
the highest energy peak seen in Fig. \ref{fig:12-13}. The other 2$^+$ state is a very 
broad state one 12.8(3) MeV with a width of 4.8(4) MeV. More detailed discussion 
about this broad state will appear in section \ref{Sec:PM}. 

Two 5$^-$ states were observed. The first one is at 12.339(4) MeV and has a width 
of 39(4) keV. This state was suggested before in 
\cite{Oert10,Morg70,Cuns81,Smit88,Gold04}. In \cite{Gold04} it was found at 
12.317(10) MeV with a width of 80(10) keV and in \cite{Oert10} at 12.327(9) MeV 
with a width of 45 keV. It is evident at large c.m. angles and can be seen as the 
peak with larger cross section in Fig. \ref{fig:12-13} (a,b). At 130$^{\circ}$ 
this state shows up as the right part of the first bump in Fig. \ref{fig:12-13} 
(c). The other 5$^-$ state is at 12.94(1) MeV and has a width of 40(10) keV. This state 
can be seen as the last peak at all angles shown in Fig. \ref{fig:12-13}. The 
state has a small reduced $\alpha$-width of ${\theta}^2_{\alpha}$=0.02, and its decay 
is dominated by neutron emission. It was probably observed  before in the
$^{14}$C($^6$Li,d) reaction \cite{Cuns83} at 12.9 MeV, where a tentative (5$^-$) 
spin-parity assignment was made.

A 4$^+$ state was found at 12.542(4) MeV with a width of 6(3) keV. It is the 
second peak in Fig. \ref{fig:12-13} (a,b). It helps to produce the dip and the 
left side of the first peak at 135$^{\circ}$ (Fig. \ref{fig:12-13}(c)). The 
4$^+$ strength was suggested before in \cite{Morg70} at 12.5 MeV but the width 
of this state was not specified.

A 6$^+$ state was observed at 12.576(9) MeV with a width of 70(20) 
keV. It is a strong $\alpha$-cluster state with a dimensionless reduced 
$\alpha$-width of 0.38. At an angle close to 160$^\circ$ the 6$^+$ state does not 
Contribute and therefore it is not visible in Fig. \ref{fig:12-13} (b). However, 
this state becomes important at angles between 135 and 150$^\circ$, determining the 
shape of the cross section in Fig. \ref{fig:12-13}(c).  It was suggested before 
in \cite{Curt02,Cuns81,Smit88,Ashw06,Morg70,Gold04}. In \cite{Gold04} it was 
found at 12.527(10) MeV with a width of 32(5) keV. This state was also observed in 
\cite{Oert10} but its width was determined to be only 24 keV. We have observed 
that the interference of this state with the 6$^+$ state at 11.7 MeV significantly 
modifies the R-matrix parameters for both states and may be the main reason 
why the best fit width and excitation energy of this state in this work are 
different from those in \cite{Gold04,Oert10}.

Three 3$^-$ states were observed. The first one is at 12.642(4) MeV and has a width 
of 110(40) keV. The second 3$^-$ state is at 12.71(2) MeV with a width of 300(30) 
keV. The interference of these two states appears at higher c.m. angles as the 
third peak in Fig. \ref{fig:12-13} (a). The last 3$^-$ state is a very broad 
state and it interferes with neighboring states, making the inclusion of this 
state necessary for reproducing the shape of the cross section. It is at 
12.98(4) MeV and has a width of 1040(200) keV and
a dimensionless reduced $\alpha$-width of 0.32 making it a cluster state. 
\subsection{Resonances in the excitation energy range between 13.0 MeV and 14.0 
MeV}

Nine resonances were observed in this energy range. Two of them have large 
dimensionless reduced widths.

\begin{figure}[ht!]
\begin{center}
 \includegraphics[scale=0.4]{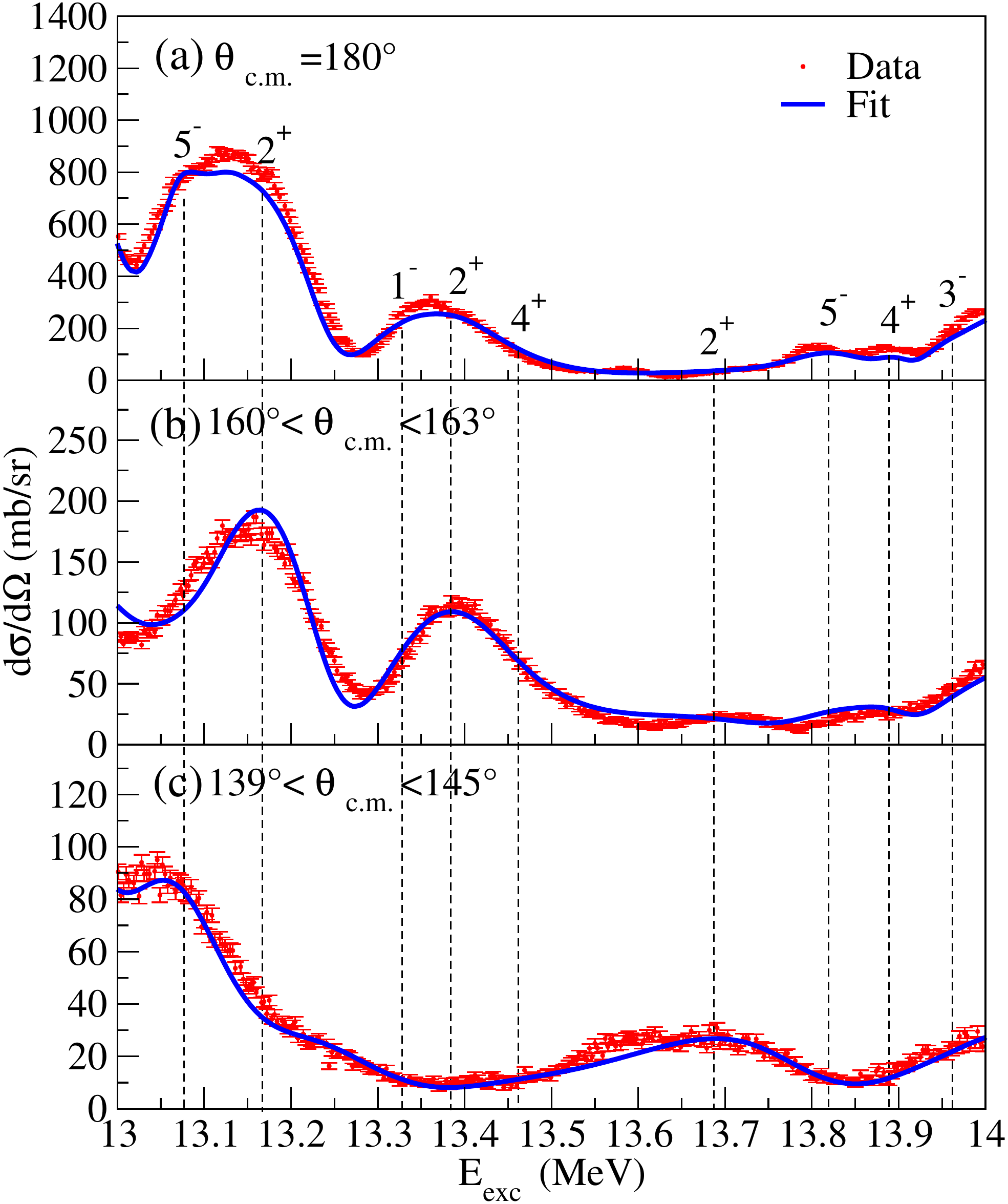}
 % 10-11.eps: 0x0 pixel, 300dpi, 0.00x0.00 cm, bb=
\end{center}
\caption{\label{fig:13-14} (Color online) Excitation functions for elastic scattering of 
$\alpha$-particles from $^{14}$C at 180$^\circ$, 160$^\circ$ and 140$^\circ$ with 
the best R-Matrix fit(solid curve) for the energy range of 13-14 MeV.}
\end{figure}

There is one 5$^-$ state found at 13.08(1) MeV with a width of 180(20) keV. It is 
a cluster state with a dimensionless reduced $\alpha$-width of 0.17. It is needed 
to reproduce the peak at 13.08 MeV near 140$^{\circ}$ in the c.m. (Fig. 
\ref{fig:13-14}).
No other solution produced a good fit. The 5$^-$ state has not 
been observed before at this energy. However, a 5$^-$ at 13.3 MeV, which does 
not show up in our analysis, is suggested in the $^{14}$C($^6$Li,d) reaction 
\cite{Cuns83}. Also, states at 13.1 MeV and 13.26 MeV were observed in 
\cite{Oert10}. The data indicates 5$^-$ strength in the region, but a very 
high level density (8 states per 400 keV interval) as well. It is possible that, due to 
the complexities of the spectrum and the interference effects, the excitation energy of 
the 5$^-$ state is shifted with respect to the transfer reaction data of 
\cite{Cuns83,Oert10}.

Three 2$^+$ states were observed. One is at 13.17(3) MeV and has a width of 
150(50) keV. This state was introduced to fit the shape of the cross section for 
the first peak seen at all angles in Fig. \ref{fig:13-14}. The second 2$^+$ 
state is at 13.38(2) MeV and has a width of 250(40) keV. It can be seen as a peak 
shown in Fig. \ref{fig:13-14}(a,b). The last 2$^+$ state is at 13.69(1) MeV with a
width of 530(120) keV. This state shows up as the second bump in Fig. 
\ref{fig:13-14}(c).

One 1$^-$ state is observed at 13.33(2) MeV with a width of 300(130) keV. This is  
a very weak state in the $^{14}$C($\alpha$,$\alpha$) channel having a
dimensionless reduced $\alpha$ width of less than 0.01. However, it 
improves the fit near 13.3 MeV.

There are two 4$^+$ states in this energy range. The first one is at 13.46(2) 
MeV and has a width of 540(80) keV.  This state contributes to the second peak 
seen on Fig. \ref{fig:13-14} (a,b). It also interferes with other states 
bringing the cross section down at higher energies. The second 4$^+$ state is at 
13.89(1) MeV with a width of 24(10) keV. It is a very weak state in both the  
$^{14}$C($\alpha$,$\alpha$) and the $^{14}$C($\alpha$,$n$) channels. It is 
introduced to reproduce the small bump seen in Fig. \ref{fig:13-14}(a).

A 5$^-$ state was observed at 13.82(2) MeV. It is a weak state with a width of 
25(6) keV. It can be seen at 180$^\circ$ as a small bump (Fig. 
\ref{fig:13-14}(a)). Interference of this state with another 5$^-$ state at a
higher energy helps to reproduce the shape of the cross section to the right 
side of this peak at all angles. In reference \cite{Oert10} a 5$^-$ state was 
seen at 13.82(2) with a width of 28 keV.

A 3$^-$ state is found at 13.96(2) MeV with a width of 150(50) keV. This state is 
responsible for the increase in cross section toward 14 MeV.

\subsection{Resonances in the excitation energy range between 14.0 MeV and 14.9 
MeV}

The R-matrix analysis of this last energy interval is not very reliable because 
of the featureless behavior of the excitation function and because of the influence 
of higher-lying, unknown states that are not included in the R-matrix fit. 
Nevertheless, it appears that the region is dominated by $5^{-}$ strengths 
which is consistent with the predictions of the cluster-nucleon configuration 
interaction model (see Chapter \ref{sec:CNCIM}).

Nine resonances have been introduced. Four of these states have large 
dimensionless reduced $\alpha$  widths. A very broad 3$^-$ state with a
dimensionless reduced $\alpha$ width of 0.7 was introduced at 14.0(2) MeV with 
a width of 2.6(5) MeV. The inclusion of this broad state was necessary to bring 
down the cross section near 14 MeV.

\begin{figure}[ht]
\begin{center}
 \includegraphics[scale=0.4]{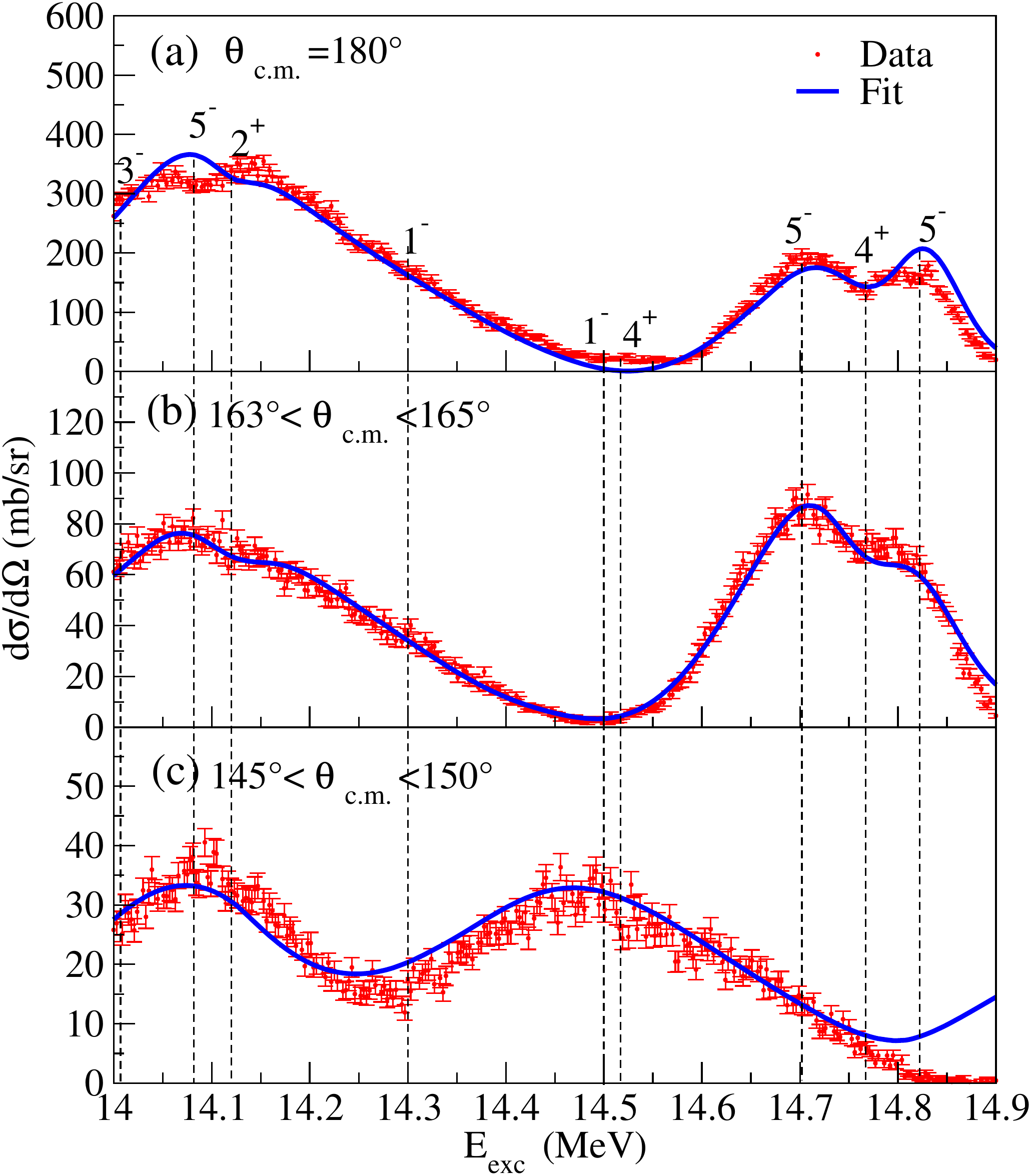}
 % 10-11.eps: 0x0 pixel, 300dpi, 0.00x0.00 cm, bb=
\end{center}
\caption{\label{fig:14-15} (Color online) Excitation function for elastic scattering of 
$\alpha$ particles from $^{14}$C at 180$^\circ$, 163$^\circ$ and 145$^\circ$ with 
the best R-Matrix fit(solid curve) for the energy range of 14-14.9 MeV.}
\end{figure}

Three 5$^-$ states were observed. The first one is at 14.1(1) MeV and has a width 
of 560(70) keV and a dimensionless reduced $\alpha$-width of 0.23. Its presence can be seen as the first bump in Fig. 
\ref{fig:14-15}.The second 
state is at 14.7(1) MeV with a width of 280(100) keV. This is a broad state 
with a dimensionless reduced $\alpha$-width of 0.16. This state appears on the 
left side of the last peak in Fig. \ref{fig:14-15}(a,b).  We used a 5$^-$ state 
at 14.82(7) MeV with a width of 140(60) keV and a dimensionless reduced width of 
0.07 to reproduces the right side of the last peak in Fig. \ref{fig:14-15} 
(a,b). However, since this state is at the edge of the measured excitation 
energy range its parameters are very unreliable and should only be considered as 
an indication of 5$^-$ strength at that energy.       

A 2$^+$ state was observed at 14.12(7) MeV with a width of 160(60) keV. This state 
shapes the cross section for the first peak in Fig. \ref{fig:14-15}.

There are two 1$^-$ states. One at 14.3(3) MeV and the second one at 14.5(2) MeV 
with widths of 900(300) and 450(220) keV, respectively. These states are more 
obvious at and around 150$^\circ$, where the state at 14.5 MeV is seen as the 
last peak in Fig. \ref{fig:14-15}(c) and the state at 14.3 MeV is used to shape 
the cross section of the same peak.

A 4$^+$ state was observed at 14.52(1) MeV with a width of 250(29) keV. It is 
needed to reproduce the near zero cross section at 14.5 MeV (Fig. 
\ref{fig:14-15}(a,b)). Another strong 4$^+$ state is at 14.77(5) MeV having a dimensionless reduced 
$\alpha$-width of 0.28. This state corresponds to the last peak in Fig. 
\ref{fig:14-15}(a,b). These 4$^+$ states make no contribution in Fig 
\ref{fig:14-15}(c) since the cross section for a 4$^+$ state at 150$^\circ$ is 
zero.

\subsection{Neutron Excitation Function\label{sub:Neutron-Excitation-Function}}

\begin{figure}[ht!]
\begin{center}
\includegraphics[scale=0.35]{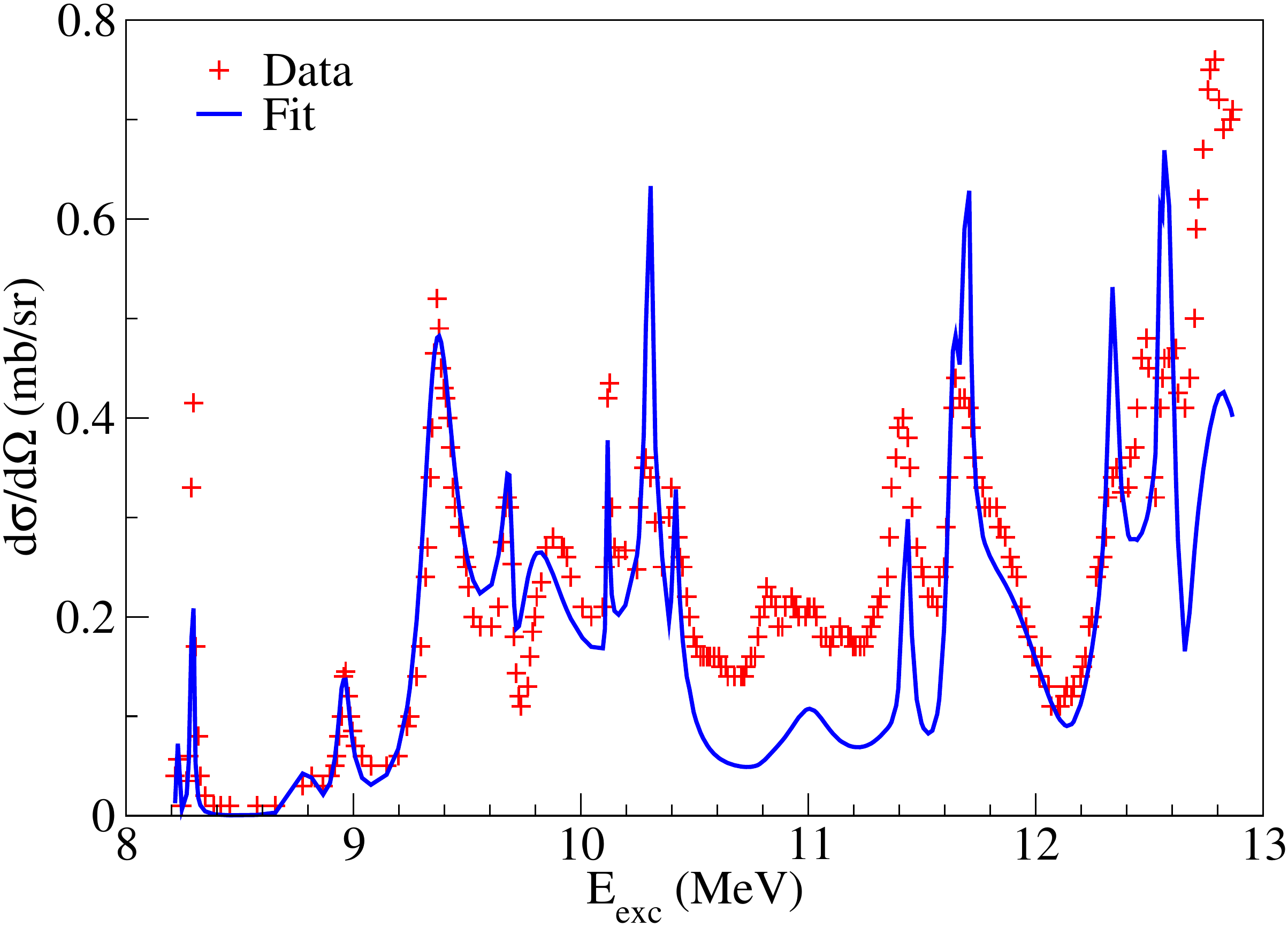}
\end{center}
\caption{\label{fig:neutron} (Color online) Total cross section for the $^{14}$C($\alpha$,n) 
reaction taken from \cite{Bair66} and \cite{Morg70} where absolute normalization was 
not performed. The solid curve corresponds to the R-matrix calculation with resonance 
parameters from the fit to the $^{14}$C($\alpha$,$\alpha$) data. The R-matrix 
prediction was used to normalize the data here.}
\end{figure}

The $^{14}$C($\alpha,n$) total reaction cross section data \cite{Bair66,Morg70} 
was used in the manual R-matrix analysis but not in the automated fit procedure. 
The total neutron cross section data for the excitation energy range of 8.1 MeV 
to 10.2 MeV was taken from \cite{Bair66}, and corresponding excitation energies above
10.2 MeV from \cite{Morg70} (see Fig. \ref{fig:neutron}). 

The absolute cross section was not measured in \cite{Bair66,Morg70}, and, in 
addition, the data from \cite{Morg70} was not corrected for energy variation of 
the detector efficiency and for neutron decays to the excited states of $^{17}$O 
which also contribute to the total cross section. Therefore, we do not expect 
the fit to match the data perfectly and we only focused on the resonance 
structures and their relative strengths. In order to compare the neutron data with 
our fit the data points were normalized to the fit curve. The data and the 
R-matrix fit are shown in Fig \ref{fig:neutron}.  The resonance structure for 
the ($\alpha$,n) excitation function is reasonably well reproduced with  most of the 
discrepancies seen at higher excitation energies. These discrepancies are 
either due to the fact that some of the resonances important for the neutron channel 
 may be too weak in the $^{14}$C($\alpha$,$\alpha$) channel to be observed, or may be related to the neutron decay to the excited states of 
$^{17}$O. This decay channel is not included in the curve shown in Fig. 
\ref{fig:neutron}. We calculated the total cross section for neutron decay to 
the first excited state of $^{17}$O and verified that the gap between the R-matrix 
fit and the experimental data at 11 MeV can be completely eliminated by this 
channel.

% \newcolumntype{C}[1]{>{\centering\arraybackslash}m{#1}}
% \renewcommand{\arraystretch}{1.01}

\begin{table*}
\begin{center}
\caption{\label{tab:ResPars1}Summary of the parameters of resonances observed in 
the $^{14}$C+$\alpha$ elastic scattering excitation function. $E_{exc}$ is the 
excitation energy, $J^{\pi}$ is the spin-parity, $\Gamma_{tot}$ is the total 
width of the state, $\Gamma_{\alpha}$ is the partial alpha width, $\Gamma_{n}$ 
is the partial neutron width and $\theta^2_{\alpha}$ is the $\alpha$ 
dimensionless reduced $\alpha$-width. The states observed in this work are shown in the left 
and the states from previous experimental studies are shown on the right side of the table.}
\begin{tabular}{cc}
\hline 
\hline 
\hspace{4.17cm}This work \hspace{4.13cm} & \hspace{2.9cm}Previous data \hspace{2.83cm} \tabularnewline
\end{tabular}
\begin{tabular}{ccccccccccc}
\hline
\hspace{0.45cm}E$_{exc}$ \hspace{0.45cm}&  \hspace{0.45cm}$J^{\pi}$\hspace{0.45cm} 
&\hspace{0.47cm}$\Gamma_{tot}$\hspace{0.45cm} 
&\hspace{0.47cm}$\Gamma_{\alpha}$\hspace{0.45cm} 
&\hspace{0.47cm}$\Gamma_{n}$\hspace{0.45cm} 
&\hspace{0.47cm}$\theta_{\alpha}^{2}$\hspace{0.45cm} & \hspace{0.45cm}E$_{exc}$ 
\hspace{0.75cm}& \hspace{0.45cm}$J^{\pi}$ \hspace{0.45cm}	& 
\hspace{0.4cm}$\Gamma_{tot}$ \hspace{0.4cm}	
&\hspace{1.25cm}Ref.\hspace{1.25cm}
\tabularnewline
(MeV) &  &(keV)&(keV)   &(keV)  &  &(MeV) &  & (keV) & \tabularnewline
\hline
8.04(2)	  &  $1^{-}$	&  2.0(7)	&  2.0(7)	&    -		&  0.02	
&  8.038	&  $1^-$	&  0.95$^{+0.4}_{-0.9}$	&  \cite{Buch07} 
\tabularnewline
	  &		&            	&	      	& 	  	&      	
&  8.0378(7) 	&  $1^-$	& \textless 2.5	& \cite{Till95} \tabularnewline
8.22(1)   &  $2^{+}$	&  1.9(2)   	&  1.7(1)  	& 0.2(1)    	&  0.03	
&  8.223       	&		& 1.6(10)     	& \cite{San56} \tabularnewline
	  &		&           	&	      	& 	  	&      	
&  8.217      	&		&  1(1)  	& \cite{Bair66} \tabularnewline
	  &		&           	&	      	& 	  	&      	
&  8.222      	&  $2^{+}$	&  1.2(8)  	& \cite{Wein58} \tabularnewline
	  &		&           	&	      	& 	  	&      	
&  8.213      	&  $2^{+}$	&  2.26(14)  	& \cite{Wag02} \tabularnewline
8.290(6)   &  $3^{-}$	&  8.5(9)   	&  2.9(2)  	& 5.6(7)    	&  0.18	
&  8.293      	&($1^{-}$,$3^{-}$) & 10(1)     & \cite{San56} \tabularnewline
	  &		&           	&	      	& 	  	&      	
&  8.287      	&		&  17(5)  	& \cite{Bair66} \tabularnewline
	  &		&           	&	      	& 	  	&      	
&  8.293      	&  $3^{-}$	&  7.7(9)  	& \cite{Wein58} \tabularnewline
	  &		&           	&	      	& 	  	&      	
&  8.282(3)   	&  $3^{-}$	&  14.74(59)  	& \cite{Wag02} \tabularnewline  
8.82(3)   &\footnotesize{($1^{-}$,$2^{+}$,$3^{-}$)}	& 60(10)  &0.3(2)	& 60(10)     	
& \footnotesize{\textless0.01 (if 2$^+$)} &  8.832	&    		& 100(20)       & 
\cite{San56} \tabularnewline
	  &		&          	&	   	& 	  	&      	
&  8.809     	&		&  80(20)  	& \cite{Bair66} \tabularnewline
8.96(1)   &\footnotesize{($2^{+}$,$3^{-}$,$4^{+}$)} & 70(30) &  5(1)  	& 65(30)   	
& 0.2 (if 4$^+$)	&  8.966    	&       	&  54(3)      	& 
\cite{San56} \tabularnewline
	  &		&           	&	      	& 	  	&      	
&  8.956   	&		&  65(10)     	& \cite{Bair66} \tabularnewline
	  &	      	&           	&	   	& 	  	&      	
&  8.96    	&   ($4^{+}$)	&  43(4)      	& \cite{Sell95} \tabularnewline
9.19(2)   &  $1^{-}$  	&  220(30)   	&  200(10)  	& 20(10)    	& 0.20 	
&  9.2     	&  $1^-$	&  *500      	& \cite{Zhao89} \tabularnewline
	  &		&		&		&		&	
& 9.16(100)   	&  $1^-$       	& $^*$420(200)  & \cite{Buch07} \tabularnewline
	  &		&		&		&		&	
&9.027$^{+.15}_{-.03}$& $1^-$  &550$^{+150}_{-50}$& \cite{Gold04} 
\tabularnewline
9.35(2)   &  $3^{-}$  	&  180(30)   	&  110(30)  	& 70(5)     	& 0.48 	
& 9.36		&  $2^+$	&\textless20 	& \cite{Man90} \tabularnewline
	  &		&		&		&		&	
& 9.35   	&($2^+$,$3^-$)	&       	& \cite{Curt02} \tabularnewline
	  &		&		&		&		&	
& 9.39(2)   	&   $3^-$	&  200(20)	& \cite{Gold04} \tabularnewline
9.70(1)   &  $3^{-}$  	&  140(10)   	&  15(2)   	& 125(10)    	& 0.04 	
&  9.71(1)     	&    ($5^-$)  	&        	& \cite{Sell95}\tabularnewline
	  &		&		&		&		&	
& 9.70(2)   	&\footnotesize{($1^{-}$,$2^{+}$,$3^{-}$)}&  	& \cite{Curt02} 
\tabularnewline
  &		&		&		&		&	
& 9.711(15)   	& $3^{-}$	&  75(15)	& \cite{Gold04} \tabularnewline
	  &		&		&		&		&	
& 9.715(5)   	& $3^{-}$	&  15		& \cite{Oert10} \tabularnewline

9.76(2)   &  $1^{-}$  	&  700(120)  	&  630(60) 	& 70(50)     	& 0.46 	
&   9.85(50)   	&  $1^{-}$      &  $^*$560(200)	& \cite{Buch07} \tabularnewline
9.79(6)   &  $2^{+}$  	& 170(80) 	& 90(30) 	& 80(50)  	& 0.10  
&         	&         	&        	& \tabularnewline
9.9(1)   &  $0^{+}$  	& 3200(800)  	& 3200(800) 	&    -      	& 1.85	
&        9.9(3)   	&         $0^{+}$ 	&    2100(500) 	& \cite{John09} \tabularnewline

10.11(1) & $3^{-}$  	&  16(5)	& 7(2)    	& 9(3)   	& 0.01	
&  10.10(1)	&  $3^{-}$	& 45(8)  	& \cite{Gold04} \tabularnewline
	  &		&		&		&		&	
&  10.111(5)   	&       	&  12		& \cite{Oert10} \tabularnewline
10.290(4) & $4^{+}$	& 29(4)  	& 19(2)   	& 10(2)   	& 0.09 	
&  10.287(10)	&  $4^{+}$	& 30(7)		& \cite{Gold04}\tabularnewline
	  &		&		&		&		&	
&  10.293(6)   	&       	&  28		& \cite{Oert10} \tabularnewline
	  &             &             &               &               &      &   
10.29          &               &              & 
\cite{Morg70,Cuns81,Curt02,Ashw06,Sell95}  \tabularnewline
	  &             &             &               &               &      &   
10.29          &               &              & 
\cite{Cobe80,Smit88,Jahn78,Yild06} \tabularnewline   
10.395(9) & $3^{-}$  	
& 70(20) 	& 50(5)   	& 50(20) 	& 0.03  &  10.365(10)	& 
$3^{-}$ 	&  45(8)	& \cite{Gold04}\tabularnewline
	  &		&		&		&		&	
&  10.400(7)   	&        	&  30		& \cite{Oert10} \tabularnewline
10.42(1) & $2^{+}$	& 180(40)	&  40(10)		& 140(40) 	
& 0.03  &  10.43(15)	& $2^{+}$ 	& 500(100)	& \cite{Gold04} 
\tabularnewline
10.80(3) & $1^{-}$	& 690(110) 	& 630(90) 	& 60(30)  	& 0.29 	
&  10.89(10)	&  $1^{-}$	& $^*$300(100)	& \cite{Buch07} \tabularnewline
10.98(4) & $2^{+}$  	& 280(130) 	& 20(10)   	& 260(120) 	& 0.01  
&  		&  		&  		& \tabularnewline
11.31(8) & $2^{+}$  	& 250(100) 	& 90(30)   	& 160(80) 	& 0.02  
& 11.39(2) 	& ($2^{+}$) 		&       		& \cite{Till95} \tabularnewline
11.43(1) & $4^{+}$  	& 40(10)  	& 30(10)   	& 10(5)   	& 0.05 	
&  11.415(5)	&  $4^{+}$	&  45 		& \cite{Gold04} \tabularnewline
	 &		&		&		&		&	
&  11.423(5)   	&       	&  35		& \cite{Oert10} \tabularnewline
11.62(3)  & $3^{-}$  	& 150(20) 	& 30(5)  	& 120(20)  	& 0.01  
&  11.67(2)	&  ($3^{-}$) 	& 112(7)	& \cite{Sell95} \tabularnewline
11.627(4)  & $5^{-}$  	& 40(5) 	& 30(3)   	& 10(3)	 	& 0.13  
&  11.609(10)	&  $5^{-}$	& 60(5)		& \cite{Gold04} \tabularnewline
&		&		&		&		&	
&  11.616(8)   	&       	&  25		& \cite{Oert10} \tabularnewline

	  &             &              &               &               &      &      
11.62        &             &              &  
\footnotesize{\cite{Curt02,Ashw06,Rae84,Cuns81,Yild06,Smit88,Morg70}} \tabularnewline
11.67(2)  & $1^{-}$  	& 200(90) 	& 120(40)  	& 80(50)  	& 0.04  
&  11.56(10)	&  $1^{-}$	& *220(100)	& \cite{Buch07} \tabularnewline
\hline
\end{tabular}
\end{center}
\end{table*}

\begin{table*}
\begin{center}
% \caption*{Table \ref{tab:ResPars1}: Continued}
%\footnotesize
\begin{tabular}{cc}
\hline 
\hline 
\hspace{4.17cm}This work \hspace{4.13cm} & \hspace{2.9cm}Previous data \hspace{2.83cm} \tabularnewline
\end{tabular}
\begin{tabular}{ccccccccccc}
\hline
\hspace{0.55cm}E$_{exc}$ \hspace{0.55cm}&  \hspace{0.5cm}$J^{\pi}$\hspace{0.5cm} 
&\hspace{0.55cm}$\Gamma_{tot}$\hspace{0.55cm} 
&\hspace{0.55cm}$\Gamma_{\alpha}$\hspace{0.55cm} 
&\hspace{0.55cm}$\Gamma_{n}$\hspace{0.55cm} 
&\hspace{0.55cm}$\theta_{\alpha}^{2}$\hspace{0.55cm} & \hspace{0.5cm}E$_{exc}$ 
\hspace{0.75cm}& \hspace{0.5cm}$J^{\pi}$ \hspace{0.5cm}	& 
\hspace{0.5cm}$\Gamma_{tot}$ \hspace{0.5cm}	
&\hspace{1.25cm}Ref.\hspace{1.25cm}
\tabularnewline
(MeV) &  &(keV)&(keV)   &(keV)  &  &(MeV) &  & (keV) & \tabularnewline
\hline

11.699(5)   & $6^{+}$  & 23(2)  	& 12(1)   	& 11(1)   	& 0.23 	
&  11.695(10)	&   $6^{+}$	& 35(5)		& \cite{Gold04} \tabularnewline
	  &		&		&		&		&	
&  11.702(6)   	&       $6^{+}$	&  27		& \cite{Oert10} \tabularnewline
	  &		&		&		&		&	
&  11.69     	&   $6^{+}$    	&        	& \cite{Morg70,Cuns81,Smit88} 
\tabularnewline	 
11.95(1)  & $3^{-}$  	& 560(70) 	& 300(30)  	& 260(40)  	& 0.17  
&      11.82(2)         &   ($3^-$)    	&       	&  \cite{Till95} \tabularnewline
12.12(1)  & $1^{-}$  	& 410(120) 	& 50(10)  	& 360(110)  	& 0.02	
&  12.12(10)	&  $1^-$	& *22(7)	& \cite{Buch07} \tabularnewline
12.21(8)  & $2^{+}$  	& 1100(300) 	& 1000(250)   	& 100(50) 	& 0.37  
&  	12.04(2)	&  	($2^+$)	&  		& \cite{Till95} \tabularnewline
12.339(4)  & $5^{-}$  & 39(4) 	& 26(2)   	& 13(2) 	& 0.06  &  
12.327(9)	&        	&  45		& \cite{Oert10} \tabularnewline
	   &  	 	& 	 	& 	   	& 	 	& 	 
&  12.317(10)	&   $5^-$     	&  80(10)	& \cite{Gold04} \tabularnewline
	   &  	 	& 	 	& 	   	& 	 	& 	 
&  12.32    	&   $5^-$     	&       	& \cite{Morg70,Cuns81,Smit88} 
\tabularnewline
12.5(1)  & $1^{-}$ 	& 900(400) 	& 300(100)  	& 600(300)  	& 0.08  
&  12.95(50)	&  1$^-$		& *210(100)  		& \cite{Buch07} 
\tabularnewline
12.542(4)  & $4^{+}$	& 6(3) 		& 5(2)	 	& 1(1)		
&\textless 0.01  & 12.5 &  &  & \cite{Morg70} \tabularnewline
12.576(9)   & $6^{+}$	& 70(20)	& 50(10)		& 20(10)		
& 0.38  &  12.557(7)   &         $6^+$         &  24  & \cite{Oert10} 
\tabularnewline
	    & 		& 		& 		& 		& 	 
&  12.527(10)   &  $6^+$      &  32(5)  & \cite{Gold04} \tabularnewline
	    & 		& 		& 		& 		& 	 
&  12.53           &  $6^+$      &  32(5)  &  
\cite{Curt02,Cuns81,Smit88,Ashw06,Morg70} \tabularnewline 
12.642(4)  & $3^{-}$	& 110(40)	& 10(5)		& 100(40)	& \textless0.01  &  &  &  & 
\tabularnewline

12.71(2)  & $3^{-}$	& 300(30)	& 120(10)	& 180(30)	& 0.05  
&  &  &  & \tabularnewline
12.8(3)  & $2^{+}$   & 4800(400)& 4800(400)	& - 		& 1.56  &  &  &  
& \tabularnewline
12.90(3)  & $2^{+}$	& 310(30)	& 285(30)	& 25(5)		& 0.09  
&  &  &  & \tabularnewline
12.94(1)  & $5^{-}$	& 40(10)	& 15(2)		& 25(10)		
& 0.02  & 12.9 & (5$^-$)  &  & \cite{Cuns83} \tabularnewline

12.98(4)  & $3^{-}$	& 1040(200) &  770(120)   	& 270(100) 	& 0.32  
&  	&  		&  		& \tabularnewline
13.08(1)   & $5^{-}$   & 180(20)	& 120(10)       & 60(15)		
& 0.17  &  &  &  & \tabularnewline
13.17(3)  & $2^{+}$   & 150(50)		& 130(40)	& 20(10)		
& 0.04  &  &  &  & \tabularnewline
13.33(2)  & $1^{-}$  	& 300(130) 	& 30(15)  	& 270(120)  	& 
\textless0.01	&  		&  		&  		& 
\tabularnewline
13.38(2)  & $2^{+}$	& 250(40)	& 220(30)	&  40(15)	& 0.07  
&  &  &  & \tabularnewline
13.46(2)   & $4^{+}$   & 540(80)	& 210(10)	& 330(70)	& 0.12  
&  &  &  & \tabularnewline

13.69(1)  & $2^{+}$	& 530(120)	& 40(20)	& 490(100)	& 0.01  
&  &  &  & \tabularnewline
13.82(1)  & $5^{-}$   & 25(6)		& 3(1)          & 22(5)		
&\textless 0.01  & 13.82(2) & $5^{-}$ & 28 & \cite{Oert10} \tabularnewline
13.89(1)  & $4^{+}$    & 24(10)	        & 14(6)	       & 10(4)		& 0.01  
&  &  &  & \tabularnewline
13.96(2)  & $3^{-}$	& 150(50) 	& 80(10)   	& 70(50) 	& 0.03  
&  	&  		&  		& \tabularnewline
14.0(2)  & $3^{-}$	& 2600(500) 	& 2100(300)   	& 500(200) 	& 0.70  
&  	&  		&  		& \tabularnewline
14.1(1)   & $5^{-}$   & 560(70)	& 260(20)	& 300(40)	& 0.23  &  &  &  
& \tabularnewline
14.12(7)   & $2^{+}$	& 160(60)	& 100(30)	& 60(30)	        
& 0.03  &  &  &  & \tabularnewline
14.3(3)    & $1^{-}$  	& 900(300) 	& 400(150)  	& 500(150)  	& 0.10	
&  	14.45(5)	&  	   &  1070	& \cite{Till95} \tabularnewline
14.5(2)  & $1^{-}$  	& 450(220) 	& 230(100)  	& 220(130)  	& 0.05	
&  14.7	&  	$1^-$    &  	800	& \cite{Till95} \tabularnewline
14.52(1)   & $4^{+}$   & 250(30)	& 80(10) 	& 170(20)	& 0.03  
&  &  &  & \tabularnewline
14.7(1)   & $5^{-}$    & 280(100)	& 230(80)	& 50(25)	& 0.16  
&  &  &  & \tabularnewline
14.77(5)  & $4^{+}$    & 680(50)	& 680(50)	& 2(1)  	& 0.28  
&  &  &  & \tabularnewline
14.82(7)   & $5^{-}$   & 140(60)	       & 100(40)	& 40(20)	
& 0.07 &  &  &  & \tabularnewline
\hline
\hline
\end{tabular}
\begin{tabular}{c}
\footnotesize{*Partial $\alpha$ width.}
\end{tabular}
\end{center}
\end{table*}

\subsection{Summary to section \ref{sec:18OResults}.}
Table \ref{tab:ResPars1} summarizes the data for 54 states in $^{18}$O in the 8.0 MeV - 14.9 MeV 
excitation energy region that were observed in the present work. This number (of states) is double 
the number of natural parity levels with known quantum characteristics (including tentative ones) 
given in compilation \cite{Till95}. The number of investigated levels in this work is large,
but this is not so surprising because resonance studies with the TTIK method (see, for example,
\cite{Norrby2011a,Norrby2011b}) as this techniques allows a broad range of 
excitation energies to be covered in a single run. However, 
this is the first time that data has been obtained in a broad angular interval and a 
complete R-matrix analysis performed for this large quantity of TTIK data. The reliability of the experimental 
information extracted from the data is demonstrated by a detailed comparison with the previously 
known results and by a fair simulation of the ($\alpha$,n) spectrum on the basis of the excitation functions
for the elastic scattering. These data will be most useful in the development of theoretical tools that are 
capable of giving a microscopic description of clustering in non-self-conjugate nuclei. 
One example of such theoretical approaches is discussed in section \ref{sec:CNCIM}. 
In the following section (Sec. \ref{sec::rotational}) we concentrate only on the states that 
have the highest degree of clustering, in an attempt to identify the members of 
$\alpha$-cluster, inversion doublet, quasi-rotational bands.

\section{Rotational Bands in $^{18}$O\label{sec::rotational}}

One of the most striking manifestations of $\alpha$-clustering in light nuclei is 
the appearance of a sequence of highly clustered states that form rotational 
bands of alternating parities. The positive parity $\alpha$-cluster rotational band 
is found at a lower energy than the corresponding negative parity band  by several MeV. These bands are called 
inversion doublets and have been conclusively identified in $^{16}$O and 
$^{20}$Ne (see Fig. \ref{fig:Bands}) 
\cite{Ames82,Bill79,John69,Ried84,Free07,Hori68,Suzu76,Kana95}. There have been 
numerous attempts to find the members of the inversion doublet rotational bands 
in $^{18}$O that correspond to an $^{14}$C(g.s.)+$\alpha$ configuration, in 
analogy to the well known rotational bands in $^{16}$O and $^{20}$Ne, but their the assignment into band members remains controversial.

Predictions for $\alpha$-cluster rotational bands in $^{18}$O have been made in 
\cite{Desc85,Furu08}.The Generator Coordinate Method (GCM) was used in  
\cite{Desc85} to investigate clustering in $^{18}$O \cite{Desc85}. This is done 
by calculating the quadrupole moments, rms radii, and reduced $\alpha$ widths of 
the resonances using the antisymmetric $^{14}$C$_{g.s.}$+$\alpha$ and 
$^{14}$C(2$^{+}$,7.01 MeV)+$\alpha$ wave functions. Furutachi $et\: al.$ 
studied the $\alpha$-cluster structure of $^{18}$O using AMD+GCM 
(antisymmetrized molecular dynamics plus the generator coordinate method) 
\cite{Furu08}. The main difference between the two calculations was that in 
\cite{Furu08} clusters emerged as a result of nucleon-nucleon and three-nucleon 
interaction while in \cite{Desc85} cluster configurations were assumed {\it a 
priori}. Three positive parity rotational bands are predicted in \cite{Desc85}, 
but only one of them has a distinct $\alpha$+$^{14}$C(g.s.) configuration, 
which this experiment is particularly sensitive to. Similar predictions were 
made in \cite{Furu08}. 

Three members of the $^{14}$C(g.s.)+$\alpha$ rotational band are below the 
energy range studied in this experiment as two of them are bound. 
A 6$^+$ state predicted at 
an excitation energy of 11.6 MeV with a dimensionless reduced $\alpha$-width of 0.15 
\cite{Desc85}is observed in this work, in good agreement with the value of 0.23 determined for the 
6$^+$ state at 11.7 MeV. However, the situation is complicated by the fact that 
there is another 6$^+$ state at 12.58 MeV that has an even larger dimensionless 
reduced $\alpha$-width of 0.38. This almost equal splitting of $\alpha$-strength 
between the two 6$^+$ states is not predicted in \cite{Desc85,Furu08}. Another 
important discrepancy between the GCM calculations and the experimental data is 
the relatively large dimensionless reduced $\alpha$-width for the 4$^+$ state at 
10.29 MeV (9\%) which is at least one order of magnitude larger than that 
assigned to this state in \cite{Desc85}. This state was suggested to belong to 
the $^{14}$C(2$^+$)+$\alpha$ rotational band and its $^{14}$C(g.s.)+$\alpha$ 
dimensionless reduced $\alpha$-width is predicted to be below one percent. As in 
the case of the 6$^+$ state, we see that the $\alpha$-strength is spread more evenly 
between several 4$^+$ states, rather than concentrated in just one state. 

Three negative parity rotational bands are suggested in \cite{Desc85}, but only 
one has a distinct $^{14}$C(g.s.)+$\alpha$ configuration. We propose that 
the $1^{-}$ and $3^{-}$ states predicted at 9.6 and 9.8 MeV excitation energies, 
respectively, are associated with the strongly $\alpha$-clustered states we 
observe at 9.16 and 9.39 MeV. This assignment of states is different from that
proposed in \cite{Desc85}, but due to its large dimensionless reduced 
$\alpha$-width, the $1^{-}$ state at 9.16 MeV should be considered as the band 
head of the $0^{-}$ $\alpha$-cluster rotational band. However, just as in the 
positive parity band, the existence of a strongly 
$\alpha$-clustered $1^{-}$ state at 9.76 MeV makes the situation complicated.
There is also another strong 3$^-$ 
state ($\theta^2=0.18$) at excitation energy of 8.28 MeV 1 MeV below 
the 3$^-$ state with the largest clustering at 9.36 MeV ($\theta^2$=0.48). Again, there is a 
strong splitting of $\alpha$-strength between these two states.

The 5$^-$ state is predicted at 13 MeV as the most clustered state in the band 
with $\theta^2=0.6$ \cite{Desc85} and would be the most dominant one in our 
spectrum. No such state was observed. Instead, there are several 5$^-$ states 
with substantial $\alpha$-strength spread out over a 3 MeV energy interval 
between 12 and 15 MeV. Their combined $\alpha$-strengths add up to 0.8. Again, 
this splitting of $\alpha$-strength is not predicted by the GCM. Obviously, 
while some general properties of the cluster states are reproduced in 
\cite{Desc85}, the model missed the physics that determines the splitting of the
$\alpha$-strength. In the calculations done in 
\cite{Furu08} for the negative parity rotational band there is a large 
difference in the excitation energies of the $1^{-}$ and $3^{-}$ states. It 
appears that there is a systematic shift in the location of the states. A splitting 
of $\alpha$-cluster strength is predicted in \cite{Furu08} where it is related to proton excitation 
of the $^{14}$C core, but only one of the states for each spin-parity should 
have a dominant $\alpha$ width. This is only partially correct. We do observe 
the splitting of the $\alpha$-cluster states, and an obviously dominant state 
does not exist for $1^{-}$, rather, there are two equally strong 
$\alpha$-cluster states separated by only 600 keV. While some of the features 
predicted in \cite{Desc85,Furu08} agree with what is experimentally observed, 
the nearly equal splitting of $\alpha$ strength among several states observed 
experimentally is not reproduced.

Assignments of the $^{14}$C$_{g.s.}$+$\alpha$ rotational bands have been 
suggested in the experimental works \cite{Morg70,Cuns81,Cuns82,Oert10}. The states 
0$^+$(3.63 MeV), 2$^+$(5.24 MeV), 4$^+$(7.11 MeV) and 6$^+$(11.69 MeV) have been 
considered as members of the positive parity rotational band, which resembles the 
$^{20}$Ne g.s. rotational band. In \cite{Morg70} the 4$^+$ state at 10.29 MeV was 
suggested as a member of this rotational band instead of the 4$^+$ state at 7.11 MeV. 
An 8$^+$ state at 17.6 MeV and 18.06 MeV was suggested by \cite{Cuns82} and 
\cite{Oert10} respectively as the fifth member of this rotational band. Only the 
6$^+$ state at 11.7 MeV is within the energy range measured in this work and we 
confirm that this is a highly clustered state ($\theta^2=0.23$), but the existence 
of the second strong 6$^+$ state at 12.58 MeV ($\theta^2=0.38$) forces us to conclude 
that clustering in $^{18}$O is more complicated and cannot be described by a 
single pair of inversion doublet rotational bands. Our experimental results, and 
also hints from the Cluster-Nucleon Configuration Interaction Model Calculations 
discussed in the next section, point to the importance of the $(1s0d)^4$ and  
$(0p)^2(1s0d)^2$ configuration mixing for the positive parity $\alpha$-cluster 
states.

We now focus on the discussion of the locations of the band head and other 
members of the negative parity inversion doublet, $\alpha$-cluster, quasi-rotational 
band in $^{18}$O. This question has a rich history and has been discussed in 
many theoretical and experimental papers (see \cite{Oert10} and refs. therein). 
The 1$^-$ state at 8.035 MeV was proposed as the band head for this band 
in recent work of W. von Oertzen, at el. \cite{Oert10} and the same 
suggestion was made in \cite{Curt02}. Our result excludes this state as a 
member of the 0$^-$ band due to its small dimensionless reduced  $\alpha$-width 
($\theta^2=0.02$). In fact, none of the states identified in \cite{Oert10} as 
members of the negative parity inversion doublet (1$^-$ (8.04 MeV), 3$^-$ (9.7 
MeV ), 5$^-$ (13.6 MeV), 7$^-$ (18.63 MeV)) can belong to this band except 
maybe a 7$^-$ state that lies beyond the energy region studied in this work. 
The 3$^-$ state at 9.7 MeV has $\theta^2$ of only 0.04,
an order of magnitude less than the 
3$^-$ at 9.3 MeV and the 5$^-$ state at 13.6 MeV is not observed at all in this work, 
which rules out the assignment made in \cite{Oert10}.

It appears that the situation for the negative parity inversion doublet 
rotational band is similar to that for the positive parity inversion doublet. 
The $\alpha$-strength is split among several states and it is not possible to 
identify a single, dominant $\alpha$-cluster rotational band. Configuration 
mixing is probably at work here as well. We discuss this in more detail in the 
next section.

As a short summary of this section, we note that the previous calculations 
for the $\alpha$ structure in $^{18}$O were only partially successful. No 
clear evidence for inversion doublet rotational bands in $^{18}$O was observed.
Unlike in neighboring $N=Z$, even-even nuclei, $^{16}$O and $^{20}$Ne, 
the $\alpha$-cluster strength is split among several states of the same spin-parity.

\section{Cluster-Nucleon Configuration Interaction Model Calculations 
\label{sec:CNCIM}}

In order to gain further understanding of the structure of many-body states in 
$^{18}$O and to examine the distribution of the $\alpha$-cluster strength in 
$^{18}$O we performed Cluster-Nucleon Configuration Interaction Model (CNCIM) 
calculations \cite{CNCIM}, the approach can be summarized as follows.

First, the structure of the states of the $^{18}$O nucleus is treated in the unrestricted
$p-sd$ configuration space with the effective interaction Hamiltonian from \cite{Utsu11}.
The $p-sd$ shell gap is slightly adjusted by $100$ keV. This small adjustment assures the
best reproduction of nuclear spectra in this mass region. The matrix dimension for 
positive parity J$_z$=0 magnetic projection is 42269424. Other dimensions are of similar order.
The same approach is applied to obtain the wave function (WF) of the ground state of daughter 
nucleus $^{14}$C. This WF is used to construct the $^{14}$C(g.s.)+$\alpha$ channel.

Second, the WF of the $\alpha$-particle is considered to be the lowest (0s)$^4$ 
translationally-invariant four-nucleon oscillator function. Taking into account that the WF
of the relative $^{14}$C+$\alpha$ motion is related to a simple SU(3) representation
($\lambda$,0) one only needs to project the overlap of the WFs of mother and daughter
nuclei onto the scalar superposition of four-nucleon configurations possessing
a required symmetry. This procedure is carried out by diagonalization of the proper Casimir
operators. Thus the large scale shell model WFs were used to obtain four-nucleon structures and
to calculate corresponding fractional parentage coefficients. The next step
is the projection of the four-nucleon WF resulting from the discussed procedure onto the
$\alpha$-particle WF and the WF of its relative motion. It is performed by use of
the so-called cluster coefficients, defined and expressed in \cite{Sm77}. Naturally the
requirement of translational invariance is rigorously met for the WFs of the
$\alpha$-cluster channels. The relevant SU(3)-classified four-nucleon 
configurations include: $(0p)^4\, (4,0)$;  $(0p)^3 (1s0d)^1\, (5,0) $; 
$(0p)^2(1s0d)^2\, (6,0) $; $(0p)^1(1s0d)^3\, (7,0);$ and $(1s0d)^4\, (8,0).$ 
Here the nucleon configurations are listed together with the corresponding 
$(\lambda, \mu)$ quantum numbers of the SU(3) symmetry. 
The permutational symmetry is fixed as $[f]=[4]$. 

Third, the channels were orthogonalized and normalized by direct diagonalization of the 
norm-kernel in harmonic oscillator basis.
 
Formal details of the CNCIM can be found in Ref. \cite{CNCIM}.
The results of the calculations related to the energy range under study 
are summarized in Table \ref{tab:SU3}. Subsequent discussion is arranged in the following way.
Positive parity states are discussed first going from the lowest spin to the highest. 
Then the negative parity states are discussed in the same order. The two very broad 
resonances that have extremely large dimensionless reduced $\alpha$-widths are 
discussed in section \ref{Sec:PM}.
\begin{figure}[h!]
\begin{center}
 \includegraphics[scale=0.63]{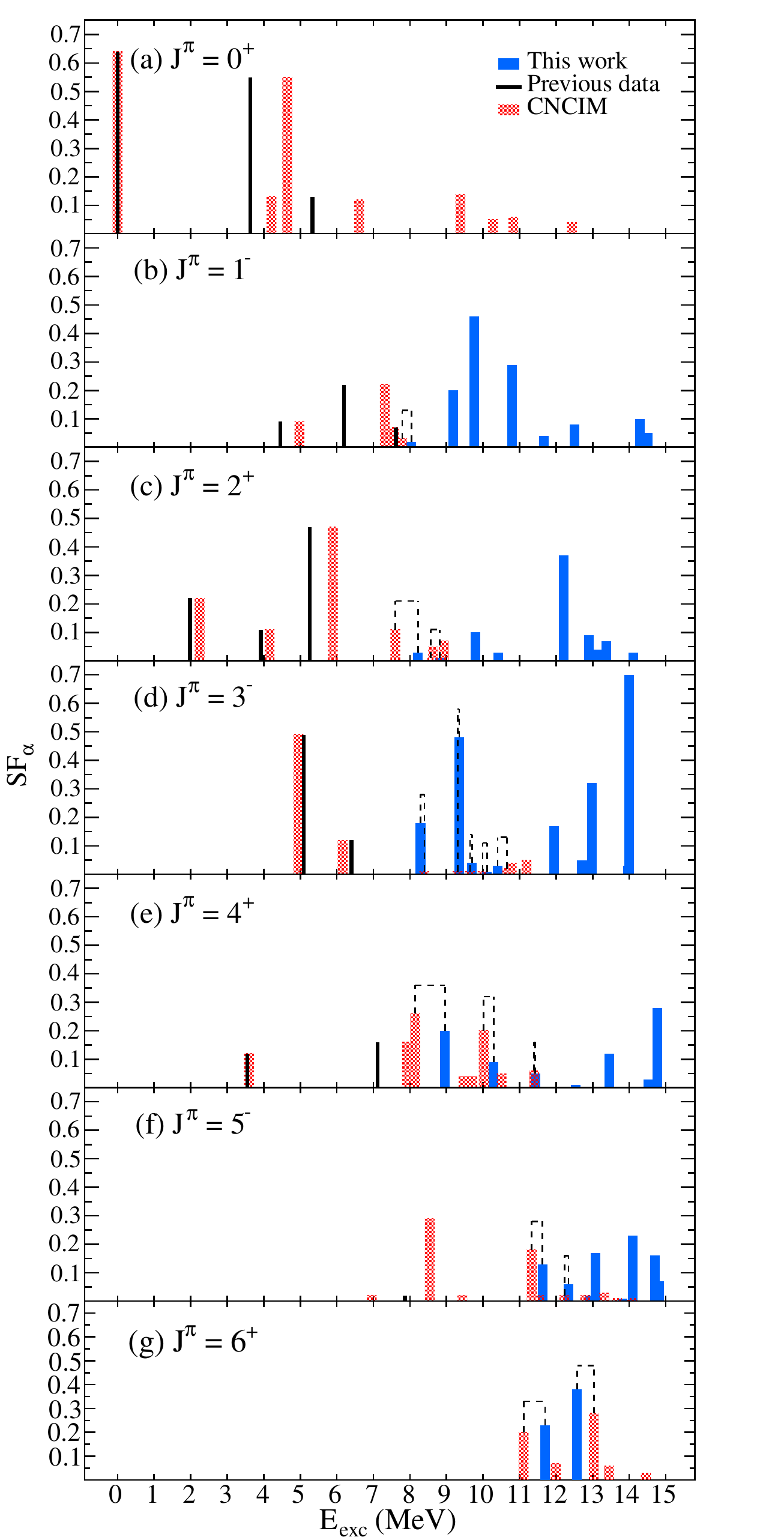}
 % 10-11.eps: 0x0 pixel, 300dpi, 0.00x0.00 cm, bb=
\end{center}
\caption{\label{fig:Alpha_SF} (Color online) Distribution of $\alpha$-strength by spin-parity.
Solid blue bars are the states observed in this work and the hatched red bars are the predictions of the CNCIM.
The length of the bars corresponds to the experimental dimensionless reduced $\alpha$ (solid blue bars) or 
the calculated SF$_{\alpha}$ (hatched red bars). The black lines represent known bound or very narrow near threshold 
states. The length of the black line is set to the calculated SF$_\alpha$ of the corresponding CNCIM state.
The suggested association between the states observed in this work and those predicted by CNCIM calculations
is indicated by connecting lines. States with $\alpha$-strength below 2\% are ignored, except for few cases for
which association between the CNCIM and the experimental state is suggested.}
\end{figure} 

\subsubsection{0$^+$ states}

The only 0$^+$ level which was observed experimentally in this work is
a very broad ($\Gamma$=3200 keV) state at 9.9 MeV. 
There are reasons to assume that the dominant nucleon
configuration of this and the broad 2$^+$ state at 12.9 MeV is $(1p0f)^2(1s0d)^2 (10,0)$. 
These states are not predicted by the CNCIM calculations restricted to the {\it p-sd} 
configuration space. The nature of these states and their properties are analyzed in 
section \ref{Sec:PM}.

The CNCIM predicts most $\alpha$-strength to be concentrated in two $0^+$ states: the
ground state, and a state predicted at 4.64 MeV. This splitting of $\alpha$-strength is
due to the strong mixture of $(1s0d)^4\, (8,0)[4]$ and $(0p)^2(1s0d)^2\, (6,0)[4]$
configurations. In the excitation energy region between 8 and 13 MeV, six 0$^+$ states 
restricted to {\it p-sd}-shell configuration are predicted. All of them
have small to moderate (0.02 $\leq SF_\alpha \leq$ 0.14) $\alpha$-strength.
Taking into account that the cross section is proportional to (2J+1) and
the fact that the angular distribution is isotropic, the weak 0$^+$ states are easy 
to miss experimentally. Only three 0$^+$ states were known experimentally before this 
work. All are bound. We believe that the highly clustered 0$^+$ state, predicted by the 
CNCIM calculations corresponds to the 3.634 MeV state, that is known to be strongly 
populated in the ($^7$Li,t) reaction \cite{Morgan1970}, unlike the 5.336 MeV state. 

\subsubsection{2$^+$ states}

Twelve 2$^+$ states were observed in the present experiment. If one neglects the 
broad state at 12.9 MeV the sum of the $\alpha$-strength is equal to $\approx$0.80.

The two strongest $\alpha$-cluster 2$^+$ states are predicted at 2.25 and 5.89 
MeV (SF$_\alpha$=0.22,0.47). This prediction does not contradict the 
assignment of the third 2$^+$ state at 5.26 MeV to the positive parity 
$\alpha$-cluster inversion doublet rotational band (see previous section). 
However, appreciable $\alpha$-clustering predicted for the first 2$^+$ state in 
$^{18}$O again indicates strong spread of the cluster strength in $^{18}$O 
due to configuration mixing. No other strong 2$^+$ $\alpha$-cluster states below 
10.5 MeV are predicted, and none are observed experimentally. Generally, the CNCIM 
predictions of the global properties of the $\alpha$-widths are reasonable, although
in some cases it is hard to establish a direct correspondence between the 
experimental and the theoretical results. The CNCIM predicts a 
higher density of 2$^+$ states than we observe experimentally which may be expected, because most of the predicted 2$^+$s states have very small 
SF$_\alpha$ with decay dominated by $\ell=0$ neutron emission.

\subsubsection{4$^+$ states}

Seven 4$^+$ states were observed in this experiment. The state at 8.96 MeV
may also have 4$^+$ spin-parity assignment. The sum of the $\alpha$-strength is 
$\approx$0.80.

The first three 4$^+$ states were predicted to have significant $\alpha$-cluster 
components according to the CNCIM calculations. The second 4$^+$ state, predicted at 
7.92 MeV and observed in previous experiments at 7.117 MeV, is considered to be a member 
of the $\alpha$-cluster positive parity rotational band mentioned in the 
previous section. However, it is not the strongest cluster 4$^+$ state. The 
third 4$^+$ state (predicted at 8.14 MeV) has the largest S$_{\alpha}$=0.26. The only 
possible candidate that may correspond to this state is the one at 8.96 MeV. We could 
not fix the spin-parity of this state, but 4$^+$ assignment is possible. If we 
assume 4$^+$ for this state then indeed its $\theta^2_{\alpha}$ is large 
(${\approx}0.2$). Moreover, in spite of its cluster nature the width of this 
state is dominated by neutron decay due to the large experimental value of 
the dimensionless reduced neutron width for $\ell=2$ decay, 
$\theta^2_n{\approx}$0.3. This is in-line with the CNCIM which predicts this 
state to neutron decay to $^{17}$O g.s. with $\ell=2$ and SF$_n$=0.6. 

As seen in Table \ref{tab:SU3} the only strong $\alpha$-cluster 4$^+$ state other than
8.96 MeV below 12 MeV is predicted at 10.02 MeV. This corresponds to the well known
10.29 MeV state, which has appreciable experimental $\theta^2_\alpha$ of 0.09(1). In
addition, four 4$^+$ states which have moderate $\alpha$-widths and two 
4$^+$ states with small $\alpha$-widths are predicted. A good correspondence of the 
parameters is found for the state at 11.43 MeV. While this is not a strong cluster state with 
$\theta^2_\alpha$=0.05, it corresponds to a prominent feature in the $^{14}$C($\alpha$,$\alpha$) excitation 
function because its total width is dominated by the partial 
$\alpha$-width, with neutron decay being negligible, and because of favorable 
$\ell=4$ $\alpha$-penetrability factors at this excitation energy.

It may not be surprising that the predicted 4$^+$ states at 9.47, 9.7, 10.5, 11.1 MeV and 12.01 MeV 
are not observed experimentally as these are not cluster states and have 
small neutron $\ell=2$ SF's. These states are probably too narrow to be observed 
in this experiment.

It appears that not only global properties of the spectra of 4$^+$ states but the characteristics of
the individual levels are reproduced in the CNCIM calculations reasonably well.

\subsubsection{6$^+$ states}

Two 6$^+$ states were observed in this experiment. Both of them have large values of the
dimensionless reduced widths, which is expected because the high potential barrier
makes the $\alpha$-decay widths small and therefore only the levels with 
substantial reduced $\alpha$-width tend to be visible in the $\alpha$+$^{14}$C 
elastic scattering excitation function.

The lowest 6$^+$ state is predicted at 11.11 MeV, with an SF$_\alpha$ of 0.2. This is close to the 
experimental 6$^+$ state at 11.7 MeV with $\theta^2_\alpha$=0.23(2). More important is 
the prediction of the $\alpha$-strength splitting. Another cluster 6$^+$ state 
is predicted at 13.03 MeV with SF$_\alpha$=0.28. We observe the strong 6$^+$ state
at 12.58 MeV with $\theta^2_\alpha$=0.38. Theoretical results indicate that 
the large value of SF$_\alpha$ for these states are due to the significant $(1s0d)^4\, 
(8,0)[4]$ component.  It appears that the exact splitting is also a consequence 
of $(1s0d)^4 (8,0)[4]$ orthogonality with the $(0p)^2(1s0d)^2\, (6,0)[4] $ channel.  Eight 6$^+$ 
states are predicted by CNCIM below 15 MeV, but only two of them have large 
SF$_\alpha$'s. Others are located in the same energy region but possess SF$_\alpha$'s 
that are at least four times smaller. Consequently, it is most likely that
the other 6$^+$ states are too weak to be observed in this experiment, so the results of the
CNCIM shell model calculations for the 6$^+$ spin-parity are in reasonable 
agreement with the experimental data.

\subsubsection{1$^-$ and 3$^-$ states}

As can be seen in the Table \ref{tab:ResPars1} a large number of 1$^-$ and 3$^-$
levels with various widths are observed in the experiment. The strongest 1$^-$
$\alpha$-cluster states are located at 9.19, 9.76 and 10.8 MeV. The sum rule of the
{1$^-$} $\alpha$-strength is equal to 1.26. The strongest 3$^-$ $\alpha$-cluster
states are observed at 8.29, 9.35, 11.95, 12.98 and 14.0 MeV. The sum of the
{3$^-$} $\alpha$-strength is equal to 1.92.

Predictions of the CNCIM with respect to the strong states are difficult to match
with the experimental data. Indeed, the only relatively strong $\alpha$-cluster 1$^-$
state predicted by the CNCIM is the state at 7.3 MeV, which probably corresponds to the
experimental 6.2 MeV state. The first 3$^-$ at 4.95 MeV is predicted to be the strongest
$\alpha$-cluster state (SF$_\alpha$=0.5) and the second 3$^-$ predicted at 6.16 MeV 
has the second largest SF (0.12). All of the remaining 1$^-$ and 3$^-$ states have small 
SF$_\alpha$'s with a sum rule  $\alpha$-strength an order of 
magnitude smaller than the experimentally observed values. It is obvious that the pattern of the
calculated spectra is very different from the experimental observations. One can
conclude that major features in the 1$^-$ and 3$^-$ $\alpha$-cluster spectrum are not
reproduced by the CNCIM in the measured energy range. Possible basic causes for 
this difference are discussed below.

\subsubsection{5$^-$ states}

Eight 5$^-$ states were observed in the present experiment with the  cluster states
located at 11.63, 13.08, 14.1 and 14.7 MeV. There are four levels which have 
moderate dimensionless reduced $\alpha$-widths. The sum of the $\alpha$-strength is 
$\approx$0.80.

The lowest cluster 5$^-$ state (SF$_\alpha$=0.29) appears at 8.54 MeV in the 
CNCIM. Though it is within the energy range of this experiment, it was not 
observed because a 5$^-$ state at this energy is too narrow due to its low 
penetrability factor. There is a known 5$^-$ state at 8.125 MeV, but the 
partial $\alpha$-width for this state is not known. It would be very interesting 
to see if this state indeed corresponds to a cluster configuration, which would 
be surprising, since we normally do not expect to find cluster states below the 
corresponding state in the inversion doublet band. Another cluster 5$^-$ state 
with SF$_\alpha$=0.18 is predicted at 11.33 MeV. This clearly corresponds to the 
known 11.62 MeV resonance with the $\theta^2_\alpha$=0.13 determined in this 
work. A detailed comparison shown in Table \ref{tab:SU3} reveals that the prediction 
of the CNCIM in the energy region below 13 MeV is reasonable. In principle, 
all of the SM 5$^-$ states could be identified with the experimental states, 
except for the state at 11.534 MeV. However, the SF$_{\alpha}$ predicted for 
all 5$^-$ states, other than those at 8.54 and 11.33 MeV, are small ($<$0.03), but this is 
not what was found experimentally. The experimental dimensionless reduced 
$\alpha$-widths for the 5$^-$ states at 13.08 and 14.07 MeV are much larger 
than the predicted values. The sum rule of the SFs also confirms this discrepancy. 
Indeed, if one neglects the state at 8.54 MeV the sum is equal to 0.3, which is 
far below the experimental value. The basic causes of the discrepancies are 
probably the same as for the 1$^-$ and 3$^-$ states.

\begin{table*}[]
\begin{center}
\caption{\label{tab:SU3}Comparison of the states predicted by SU(3) shell model 
calculations and the experimental values reported in TUNL and the present work.}
\begin{tabular}{cccccc}
\hline
\hline
\hspace{0.9cm}$J^{\pi}$ \hspace{1.5cm}&  \hspace{0.94cm}Energy\hspace{0.94cm} 
&\hspace{1cm}Exp. energy \hspace{1cm} &\hspace{0.cm}Exp. energy \hspace{1cm}&\hspace{1cm}SF from SM\hspace{1cm} 
&\hspace{1cm}$\theta_{\alpha}^{2}$ from\hspace{1cm}
\tabularnewline
& from SM & from TUNL & from this work & & this work  \tabularnewline
& (MeV)  & (MeV) & (MeV)   &  &   \tabularnewline
\hline
0$^+$	  &  0		&  0	 &  	 &    0.64		&  
\tabularnewline
	  &  4.212	&  5.336 &  	 &    0.13		&  
\tabularnewline
	  &  4.642	&  3.634 &  	 &    0.55		&  
\tabularnewline
	  &  6.609	&  	 &  	 &    0.12		&  
\tabularnewline
	  &  9.382	&  	 &  	 &    0.14		&  
\tabularnewline
	  &  10.274	&  	 &  	 &    0.05		&  
\tabularnewline
	  &  10.830	&  	 &  	 &    0.06		&  
\tabularnewline
	  &  11.416	&  	 &  	 &    \textless0.01	&  
\tabularnewline
	  &  12.433	&  	 &  	 &    0.04		&  
\tabularnewline
	  &  12.940	&  	 &  	 &    0.02		&  
\tabularnewline

1$^-$	  &  4.975	&  4.456 &  	 &    0.09		&  
\tabularnewline
	  &  7.312	&  6.198 &  	 &    0.22		&  
\tabularnewline
	  &  7.564	&  7.616 &  	 &    0.07		&  
\tabularnewline
	  &  7.790	&  8.038 &  8.04 &    0.03		&  0.02 
\tabularnewline
	  &  9.561	&  	 &  	 &    0.01		& 
\tabularnewline
	  &  10.454	& 	 &  	 &    	\textless0.01	&   
\tabularnewline
	  &  10.662	& 	 &  	 &    	\textless0.01	&   
\tabularnewline
	  &  10.782	& 	 &  	 &    0.01	&   \tabularnewline
	  &  11.250	& 	 &  	 &    	0.01	&   \tabularnewline
	  &  11.488	& 	 &  	 &    	\textless0.01	&   
\tabularnewline

2$^+$	  &  2.246	&  1.982 &  	 &    0.22		&  
\tabularnewline
	  &  4.161	&  3.920 &  	 &    0.11		&  
\tabularnewline
	  &  5.893	&  5.255 &  	 &    0.47		&  
\tabularnewline
	  &  7.601	&  8.213 &  8.22 &    0.11    &   0.03  
\tabularnewline
	  &  8.569	& 	 &  8.82  &    0.01	& $<$0.01  
\tabularnewline
	  &  8.633	&  	 &  	 &    0.05		&  
\tabularnewline
		  &  8.931	&  	 &  	 &    0.07		&  
\tabularnewline
	  &  9.397	&  9.361	 &  	 &    \textless0.01	&  
\tabularnewline
	  &  9.687	&  	 &  	 &    \textless0.01	&  
\tabularnewline
	  &  10.424	&  	 &  	 &    \textless0.01	&  
\tabularnewline  

3$^-$	  &  4.950	&  5.098 &  	 &    0.49		&  
\tabularnewline
	  &  6.160	&  6.404 &  	 &   0.12	&  \tabularnewline
	  &  8.394	&  8.282 &  8.290 &    \textless0.01	&  0.18 
\tabularnewline

	  &  9.057	&  	 &  	 &    \textless0.01	&  
\tabularnewline
  &  9.306	&  	 & 9.35	 &    \textless0.01	&  0.48 
\tabularnewline
	  &  9.644	&  9.67   &  9.70 	 &    0.01	& 0.04  
\tabularnewline
&  9.990	&  10.11	 & 10.11 	 &    0.01	&  0.01 \tabularnewline
 &  10.655  & 10.40 	 & 10.40 	 &    0.02	&  0.03 \tabularnewline
	    &  10.789	&  	 &  	 &    0.04	&  \tabularnewline
		  &  11.192	&  	 &  	 &    0.05	&  \tabularnewline

4$^+$	  &  3.597	&  3.555 &  	 &    0.12		&  
\tabularnewline
	  &  7.921	&  7.117 &  	 &    0.16		&  \tabularnewline
	    &  8.141	&  8.955 &  8.96 &    0.26		&  ($\approx 
0.2$) \tabularnewline
	  &  9.468	&  	 &  	 &    0.04		&  
\tabularnewline

	  &  9.706	&  	 &  	 &    0.04		&  
\tabularnewline

	  &  10.020	&  10.295& 10.29 &    0.20		& 0.09 
\tabularnewline
	  &  10.510	&  	 &  	 &    0.05		&  
\tabularnewline
  &  11.106	&  	 &  	 &    \textless0.01	&  
\tabularnewline
\hline
\end{tabular}
\end{center}
\end{table*}

\begin{table*}
\begin{center}
\begin{tabular}{cccccc}
\hline
\hline
\hspace{0.9cm}$J^{\pi}$ \hspace{1.5cm}&  \hspace{0.94cm}Energy\hspace{0.94cm} 
&\hspace{1cm}Exp. energy \hspace{1cm} &\hspace{0.cm}Exp. energy \hspace{1cm}&\hspace{1cm}SF from SM\hspace{1cm} 
&\hspace{1cm}$\theta_{\alpha}^{2}$ from\hspace{1cm}
\tabularnewline
& from SM & from TUNL & from this work & & this work  \tabularnewline
& (MeV)  & (MeV) & (MeV)   &  &   \tabularnewline
\hline
	  &  11.396	&  11.41 &  11.43 &    0.06		& 0.05 
\tabularnewline
	  &  12.011	&  	 &  	 &    \textless0.01	&  
\tabularnewline

5$^-$	  &  6.958	&  7.864 &  	 &    0.02		&  
\tabularnewline
	  &  8.546	&  8.125 &  	 &    0.29		&  
\tabularnewline
	  &  9.432	&  9.713 &  	 &    0.02		&  
\tabularnewline
	  &  11.332	&  11.62 & 11.62 &    0.18	& 0.13  \tabularnewline
	  &  11.534	&  	 &  	 &    0.02		&  
\tabularnewline
	  &  12.229	& 12.33	 &  12.34 &    0.02		& 0.06 
\tabularnewline
	  &  12.801	&  	 & 12.94 &    0.02		& 0.02 
\tabularnewline
	  &  13.316	&  	 & 13.08 &    0.03		& 0.18 
\tabularnewline
	  &  13.665	&  	 & 13.82 &    \textless0.01	& \textless0.01 
\tabularnewline
	  &  14.080	&  	 & 14.07 &    \textless0.01	& 0.23 
\tabularnewline

6$^+$	  &  11.112	&  11.690 & 11.70 &    0.20		& 0.23 
\tabularnewline
	  &  11.991	&  	  &       &    0.07		&  
\tabularnewline
	  &  13.035	&  12.530 & 12.58 &    0.28		& 0.38 
\tabularnewline
	  &  13.445	&  	  &  	 &    0.06		&  
\tabularnewline
	  &  13.806	&  	  &  	 &  	\textless0.01	&  
\tabularnewline
	  &  14.263	&  	  &  	 &    0.01		&  
\tabularnewline
	  &  14.455	&  	  &  	 &    0.03		&  
\tabularnewline
	  &  14.784	&  	  &  	 &    	\textless0.01	&  
\tabularnewline
\hline
\hline
\end{tabular}
\end{center}
\end{table*}

\subsubsection{Overview of CNCIM}

The distribution of the $\alpha$-strength is shown in Fig. \ref{fig:Alpha_SF} for each spin-parity. 
The solid blue bars correspond to the experimental data from this work and the hatched red bars are CNCIM calculation.
The bar length reflects the dimensionless reduced $\alpha$-width for the experimental states 
or the SF$_{\alpha}$ for the calculated state. The black lines are known bound states and their 
length is set to be that of the associated CNCIM calculated state. We use connecting lines to indicate the 
suggested link between the states observed in this work and those predicted by CNCIM calculations. 
States with SF$_{\alpha}$ below 2\% are not shown in Fig. \ref{fig:Alpha_SF} except for cases for which
association between CNCIM and experimental states are suggested in \ref{tab:SU3}.

Overall, we can conclude that the CNCIM as it stands now provides an adequate description of 
$\alpha$-clustering for many low-lying states. Unfortunately, many of these states, especially 
those with low spin are below the $\alpha$-decay threshold, hence not accessible in this 
experiment. In the limits where comparison is possible, the results appear to be encouraging.
The total number of the calculated and observed positive parity states with large 
and moderate $\alpha$-strengths are in reasonable agreement as well as some details of the positive 
parity spectrum. It shows that the recent advances in configuration 
interaction techniques, which includes expanded computational capabilities, 
better theoretical understanding of phenomenological interactions, and a closer 
link to fundamental {\it ab-initio} and no-core approaches, make some treatment 
of clustering feasible.

The experimental results, however, also point to noticeable discrepancies, 
especially for the negative parity states. Truncation of the configuration space is 
the most likely reason for that. The {\it p-sd} model space turns out to be more or less adequate
for the description of an essential part of positive parity states in the energy domain
under study. It is not the case for the negative parity states. It is possible that the majority of 
the negative parity states above 8 MeV of excitation energy contain a significant component of particle excitations 
related to the {\it sd} $\to$ {\it pf} nucleon transfer which is not contained in the basis 
that was used for the calculations. Future model with extended valence space Hamiltonian should be 
able to overcome this limitation. Nevertheless, numerous improvements in the 
existing approach, including basis expansion, are possible (see discussion in Ref. \cite{CNCIM}).

A similar interpretation appears to be natural for the extremely broad positive parity
states. The dominate  component of these states is most likely the
$(1p0f)^2(1s0d)^2 (10,0)$ configuration which is not present in the
employed basis. The structure of these states, discussed in detail in the next section, 
may also be influenced by their strong continuum coupling through a so-called superradiance 
mechanism \cite{Auerbach:2011}. It is known that the structure of states that are strongly coupled
to decay channels gets modified so that almost all the decay strength is concentrated in a
broad super-radiant state, while other states become narrow
\cite{Volya:2003PRC,Volya:2009}. Due to the centrifugal barrier the effect is most
noticeable for channels with low angular momentum partial waves.

\section{Broad $0^+$ and $2^+$ States \label{Sec:PM}}
 
The R-matrix fit discussed in Sec. \ref{sec:18OResults} shows that it is 
necessary to include two very broad, purely $\alpha$-cluster low spin states, namely 
the 0$^+$ state at 9.9 MeV and the 2$^+$ state at 12.9 MeV. Due to the large widths of these 
0$^+$ and 2$^+$ states, they are not seen as narrow peaks, instead they 
influence the excitation function over a broad energy range.
 These levels were not predicted by any of the 
models considered in previous sections. 

Broad 0$^+$ and 2$^+$ states with very large partial $\alpha$-widths are known in $^{20}$Ne 
at excitation energies of $\sim$8.7 MeV and 8.9 MeV \cite{Till95}. Observation 
of the broad 0$^+$ and 2$^+$ states in $^{18}$O at 9.9 MeV and 12.9 MeV provides evidence 
that these purely cluster states are common features in light nuclei. Similar structures were observed 
recently in $^{12}$C \cite{Fynbo2005} and suggested in $^{10}$B(T=1) \cite{Kuchera2011}.   
It is worthwhile to note that reliable identification of broad 0$^+$ states is difficult, because 
the cross sections are small (due to the 2J+1 factor) and does not vary with angle. 
It is easy to attribute the contribution of the level to background. A considerable effort has been spent to find a direct signature 
for the presence of this broad level. In $^{18}$O \cite{John09}, it was found that the interference 
of this resonance with the Rutherford amplitude near 90$^{\circ}$ in c.m. (where only positive 
parity levels contribute) presented an unambiguous identification. If the excitation energy 
of the 0$^+$ state would have been several MeV higher, then the Rutherford amplitude would be 
smaller and identification would become even more difficult.

Here we investigate further the nature of these states. The reduced $\alpha$-widths 
of these states are so large that we apply a simple $\alpha$-cluster potential model 
that is known to work well for neighboring $^{16}$O and $^{20}$Ne \cite{Goldberg1974,Buck1975}.
In this model the $\alpha$-particle can be seen as a cluster orbiting a closed core  
with quantum numbers N (number of nodes) and L (orbital angular momentum). 
Following the Talmi-Moshinksy relations these quantum numbers are found in 
terms of the corresponding quantum numbers of individual nucleons $n_i$ and $l_i$ 
that make up the $\alpha$-cluster
\begin{equation}\label{nnodes}
\lambda=2N+L=\sum_{i}^{4}(2n_i+l_i).
\end{equation}
The prevailing 4-nucleon structures of the CNCIM with the corresponding SU(3) 
label $(\lambda\, 0)$ provide a guidance to the selection of these parameters. 

To construct an effective interaction between the core and the cluster one could double 
fold the cluster and core mass densities, or use a parametrized potential shape. 
It is important to recognize that the Pauli exclusion principle does not 
generally allow for the cluster scattering to be represented as  potential 
scattering. The corresponding Schr\"odinger equation must be generalized to 
include a norm operator. For our purposes, where the width is so large that the 
$\alpha$ is nearly completely excluded from the internal region the exclusion can 
be modeled by limiting the configuration space to a correct minimum number of nodes 
($N$) or by introducing a repulsive core which effectively blocks the excluded 
spatial region. The results were very similar for the potential 
with a core and without one. The details on the potential parameters
are given in  \cite{Avila13}.
\begin{figure}[ht]
 \begin{center}
 \includegraphics[scale=0.35]{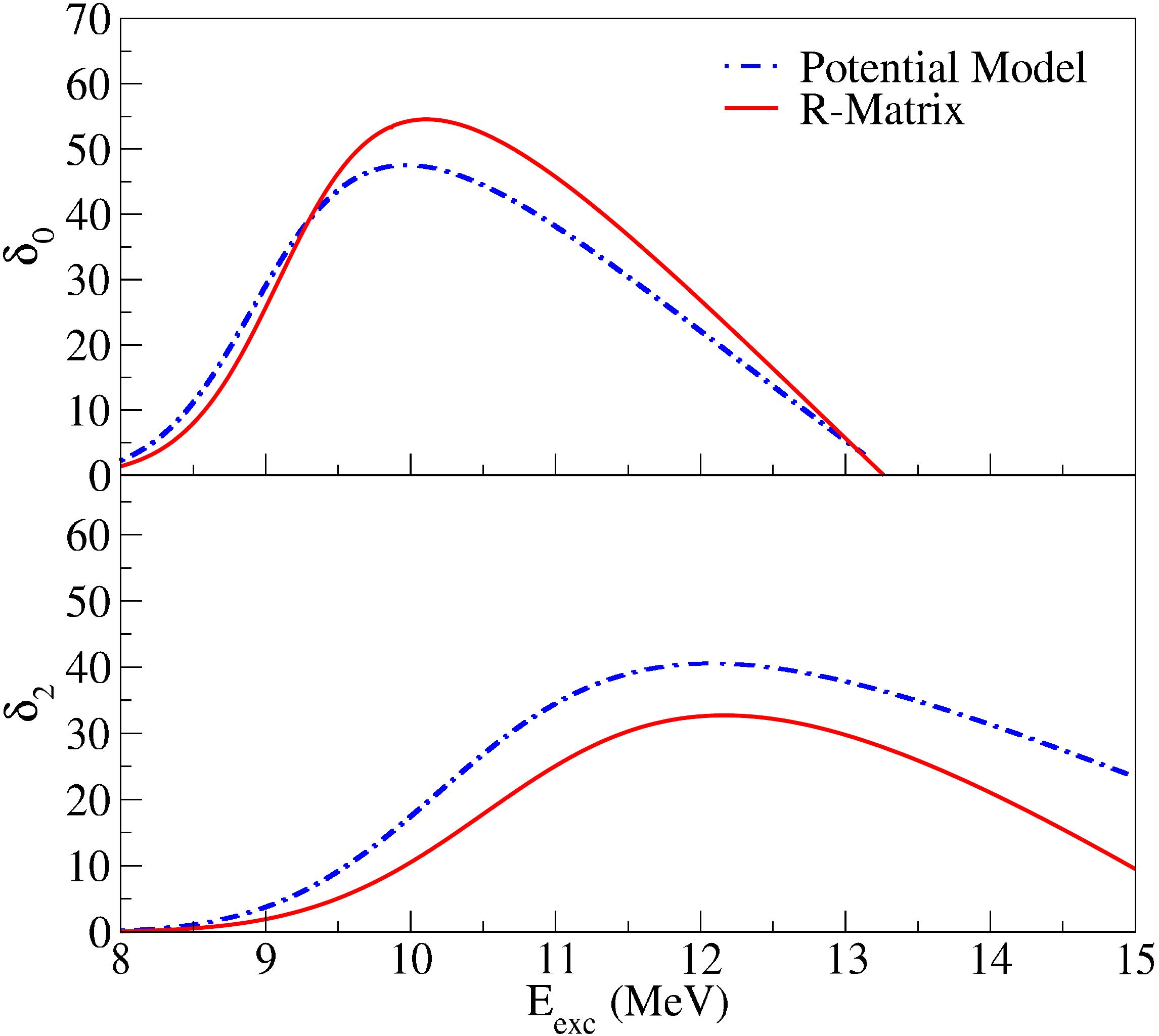}
 % comparison.pdf: 360x277 pixel, 72dpi, 12.70x9.77 cm, bb=0 0 360 277
\end{center}
\caption{\label{fig:phase_shift} (Color online) Phase shifts for $0^+$ (top) and $2^+$ 
(bottom) states in $^{18}$O from Potential Model (dash-dotted 
curve) and R-Matrix (solid line) calculations}
\end{figure}

Fig. \ref{fig:phase_shift} presents a comparison of $\ell=0$ and $\ell=2$ phase shifts for 
the broad 0$^+$ and 2$^+$ resonances calculated in the potential model approach 
and from the R-matrix fit. The potential model phase shift, representing the extreme 
$\alpha$-cluster model agrees very well with the experimental one from the R-matrix fit and can be considered as one more piece of evidence for the pure $\alpha$-cluster structure 
of the 0$^+$ and 2$^+$ states in question. The pure structure of these states 
leads to expected observation of other features of the cluster bands, like 
higher spin members of the positive parity band and the observation of a negative 
parity band (considerations of this kind led H.T. Fortune to suggest the existence of the broad
2+ state \cite{Fort12}, after he obtained knowledge on the properties of the broad 0$^+$ 
resonance at 9.9 MeV from \cite{John09}). Evidently, more experimental effort 
is needed to verify or reject these 
speculations. It seems that the even more interesting question is the origin of these 
new and extreme $\alpha$-cluster states.

Why does the $\alpha$-cluster structure split at low excitation energies in $^{18}$O, but 
appears to be pushed into one very highly clustered state in the region of a higher density 
of states? It could be related to the super-radiance phenomenon, described in 
\cite{Auerbach:2011,Volya:2003PRC,Volya:2009}, or
it can be because these states are so broad and their lifetime is so short 
that they decay before mixing with the nearby states can occur.
Indeed, we are discussing properties of broad 0$^+$ states close to 
the $\alpha$-particle threshold. Realizing that the presence of the states in 
question could be a common feature in light nuclei, it is worthwhile to note that 
there has been no observations of  broad cluster $\ell=0$ levels in nuclei with 
an odd number of nucleons.

\section{Conclusions\label{sec:18OConclusions}}

Excitation functions for $^{14}$C+$\alpha$ elastic scattering in the 
excitation energy range between 8 and 14.9 MeV were measured using the Thick Target 
Inverse Kinematics technique over a wide angular range. The results of a detailed R-matrix analysis that also included 
available data for the $^{14}$C($\alpha$,n) reaction yielded spin-parity 
assignments, excitation energies and partial widths for 54 
excited states in $^{18}$O. The $^{14}$C($\alpha$,$\alpha$) elastic scattering 
is particularly sensitive to the states that have a $^{14}$C(g.s.)+$\alpha$ 
configuration, and the completeness of the experimental data complemented by the detailed 
R-matrix analysis allowed for an accurate determination of $\alpha$-clustering in $^{18}$O.

The search for $\alpha$-cluster inversion doublet rotational bands in $^{18}$O has 
been the subject of many experimental and theoretical studies 
(see for example \cite{Desc85,Curt02,Furu08,Oert10} and references therein), but the corresponding 
assignments remain controversial. We conclude that unlike for the 
$N=Z$, $^{16}$O and $^{20}$Ne nuclei, the 
$\alpha$-strength is split about evenly between two or more states for each 
spin-parity and it is not possible to define inversion doublet rotational 
bands in the same sense as for $^{16}$O and $^{20}$Ne. The Cluster-Nucleon Configuration 
Interaction Model (CNCIM) calculations presented in section \ref{sec:CNCIM} 
indicate that splitting of the $\alpha$-strength for the positive parity band is a 
result of $(1s0d)^4\, (8,0)[4]$ and $(0p)^2(1s0d)^2\, (6,0)[4] $ configuration 
mixing. For the negative parity states and the very broad positive parity states 
the (1p0f) shell (not included in the CNCIM) probably plays an important role. 
Comparison of the results of this experiment to the 
predictions of the GCM cluster model \cite{Desc85} and the AMD+RGM \cite{Furu08} model show
significant discrepancies as they underestimate the  splitting and predict that only one state 
has dominant $^{14}$C(g.s.)+$\alpha$ configuration, which is not the case 
experimentally. These findings highlight the need for sophisticated microscopic 
analyses if we are to understand $N\ne Z$ nuclei.

Assignment of the $\alpha$-cluster rotational bands without any knowledge of the partial 
$\alpha$-widths is dangerous. The most striking example is the assignment of the $0^-$ inversion doublet 
rotational band in \cite{Oert10}. The present work shows that all states in the 0$^-$ rotational band suggested in \cite{Oert10} have 
$\alpha$-strengths that are at least a factor of 10 smaller than the $\alpha$-strength of 
the strongest cluster state with the corresponding spin-parity (except, maybe the for 
7$^-$ state that has an excitation energy too high to be populated in this study), 
which obviously excludes them from being members of the 0$^-$ inversion doublet rotational band.

Broad, purely $\alpha$-cluster $0^{+}$ and 2$^+$ states at 9.9 MeV and 12.9 
MeV were observed in $^{18}$O. While these states were not predicted by any microscopic 
calculations, their origin is probably similar to the well known 0$^+$ and 
2$^+$ broad states in $^{20}$Ne at 8.7 and 8.9 MeV. The presence of broad, very 
clustered states could be a common feature in light nuclei.

Detailed spectroscopic information, including spin-parities, partial $\alpha$- and neutron- decay
widths and dimensionless reduced widths, was obtained for excited states in $^{18}$O between 
8.0 to 14.9 MeV in excitation energy. Experimental results are compared with existing theoretical
models. While some features of the $^{18}$O spectrum are reproduced by the models, it appears
that the complete theoretical description of clustering phenomena in non-self-conjugate nuclei 
is still out of rich.
 
\begin{acknowledgments}
The authors would like to acknowledge the financial support provided
by the National Science Foundation under grant No. PHY-456463,
the U.S. Department of Energy under contracts No. DE-SC0009883 
and No. DE-FG03-93ER40773.
\end{acknowledgments}

\bibliography{myrefs}

\begin{thebibliography}{69}
\expandafter\ifx\csname natexlab\endcsname\relax\def\natexlab#1{#1}\fi
\expandafter\ifx\csname bibnamefont\endcsname\relax
  \def\bibnamefont#1{#1}\fi
\expandafter\ifx\csname bibfnamefont\endcsname\relax
  \def\bibfnamefont#1{#1}\fi
\expandafter\ifx\csname citenamefont\endcsname\relax
  \def\citenamefont#1{#1}\fi
\expandafter\ifx\csname url\endcsname\relax
  \def\url#1{\texttt{#1}}\fi
\expandafter\ifx\csname urlprefix\endcsname\relax\def\urlprefix{URL }\fi
\providecommand{\bibinfo}[2]{#2}
\providecommand{\eprint}[2][]{\url{#2}}

\bibitem[{\citenamefont{Horiuchi and Ikeda}(1968)}]{Hori68}
\bibinfo{author}{\bibfnamefont{H.}~\bibnamefont{Horiuchi}} \bibnamefont{and}
  \bibinfo{author}{\bibfnamefont{K.}~\bibnamefont{Ikeda}},
  \bibinfo{journal}{Progress of Theoretical Physics}
  \textbf{\bibinfo{volume}{40}}, \bibinfo{pages}{277} (\bibinfo{year}{1968}),
  \urlprefix\url{http://ptp.oxfordjournals.org/content/40/2/277.abstract}.

\bibitem[{\citenamefont{Ames}(1982)}]{Ames82}
\bibinfo{author}{\bibfnamefont{L.~L.} \bibnamefont{Ames}},
  \bibinfo{journal}{Phys. Rev. C} \textbf{\bibinfo{volume}{25}},
  \bibinfo{pages}{729} (\bibinfo{year}{1982}),
  \urlprefix\url{http://link.aps.org/doi/10.1103/PhysRevC.25.729}.

\bibitem[{\citenamefont{Billen}(1979)}]{Bill79}
\bibinfo{author}{\bibfnamefont{J.~H.} \bibnamefont{Billen}},
  \bibinfo{journal}{Phys. Rev. C} \textbf{\bibinfo{volume}{20}},
  \bibinfo{pages}{1648} (\bibinfo{year}{1979}),
  \urlprefix\url{http://link.aps.org/doi/10.1103/PhysRevC.20.1648}.

\bibitem[{\citenamefont{John et~al.}(1969)\citenamefont{John, Aldridge, and
  Davis}}]{John69}
\bibinfo{author}{\bibfnamefont{J.}~\bibnamefont{John}},
  \bibinfo{author}{\bibfnamefont{J.~P.} \bibnamefont{Aldridge}},
  \bibnamefont{and} \bibinfo{author}{\bibfnamefont{R.~H.} \bibnamefont{Davis}},
  \bibinfo{journal}{Phys. Rev.} \textbf{\bibinfo{volume}{181}},
  \bibinfo{pages}{1455} (\bibinfo{year}{1969}),
  \urlprefix\url{http://link.aps.org/doi/10.1103/PhysRev.181.1455}.

\bibitem[{\citenamefont{Riedhauser}(1984)}]{Ried84}
\bibinfo{author}{\bibfnamefont{S.~R.} \bibnamefont{Riedhauser}},
  \bibinfo{journal}{Phys. Rev. C} \textbf{\bibinfo{volume}{29}},
  \bibinfo{pages}{1961} (\bibinfo{year}{1984}),
  \urlprefix\url{http://link.aps.org/doi/10.1103/PhysRevC.29.1961}.

\bibitem[{\citenamefont{Freer}(2007)}]{Free07}
\bibinfo{author}{\bibfnamefont{M.}~\bibnamefont{Freer}}, \bibinfo{journal}{Rep.
  Prog. Phys.} \textbf{\bibinfo{volume}{70}}, \bibinfo{pages}{2149}
  (\bibinfo{year}{2007}),
  \urlprefix\url{http://stacks.iop.org/0034-4885/70/i=12/a=R03}.

\bibitem[{\citenamefont{Suzuki}(1976)}]{Suzu76}
\bibinfo{author}{\bibfnamefont{Y.}~\bibnamefont{Suzuki}},
  \bibinfo{journal}{Progress of Theoretical Physics}
  \textbf{\bibinfo{volume}{55}}, \bibinfo{pages}{1751} (\bibinfo{year}{1976}),
  \urlprefix\url{http://ptp.oxfordjournals.org/content/55/6/1751.abstract}.

\bibitem[{\citenamefont{Kanada-En'yo and Horiuchi}(1995)}]{Kana95}
\bibinfo{author}{\bibfnamefont{Y.}~\bibnamefont{Kanada-En'yo}}
  \bibnamefont{and} \bibinfo{author}{\bibfnamefont{H.}~\bibnamefont{Horiuchi}},
  \bibinfo{journal}{Progress of Theoretical Physics}
  \textbf{\bibinfo{volume}{93}}, \bibinfo{pages}{115} (\bibinfo{year}{1995}),
  \urlprefix\url{"http://ptp.oxfordjournals.org/content/93/1/115.abstract"}.

\bibitem[{\citenamefont{Kanada-En'yo and Horiuchi}(2001)}]{Kana01}
\bibinfo{author}{\bibfnamefont{Y.}~\bibnamefont{Kanada-En'yo}}
  \bibnamefont{and} \bibinfo{author}{\bibfnamefont{H.}~\bibnamefont{Horiuchi}},
  \bibinfo{journal}{Progress of Theoretical Physics Supplement}
  \textbf{\bibinfo{volume}{142}}, \bibinfo{pages}{205} (\bibinfo{year}{2001}),
  \urlprefix\url{http://ptps.oxfordjournals.org/content/142/205.abstract}.

\bibitem[{\citenamefont{Neff and Feldmeier}(2008)}]{Neff08}
\bibinfo{author}{\bibfnamefont{T.}~\bibnamefont{Neff}} \bibnamefont{and}
  \bibinfo{author}{\bibfnamefont{H.}~\bibnamefont{Feldmeier}},
  \bibinfo{journal}{International Journal of Modern Physics E}
  \textbf{\bibinfo{volume}{17}}, \bibinfo{pages}{2005} (\bibinfo{year}{2008}),
  \urlprefix\url{http://www.worldscientific.com/doi/abs/10.1142/S0218301308010994}.

\bibitem[{\citenamefont{Wiringa et~al.}(2000)\citenamefont{Wiringa, Pieper,
  Carlson, and Pandharipande}}]{Wiri00}
\bibinfo{author}{\bibfnamefont{R.~B.} \bibnamefont{Wiringa}},
  \bibinfo{author}{\bibfnamefont{S.~C.} \bibnamefont{Pieper}},
  \bibinfo{author}{\bibfnamefont{J.}~\bibnamefont{Carlson}}, \bibnamefont{and}
  \bibinfo{author}{\bibfnamefont{V.~R.} \bibnamefont{Pandharipande}},
  \bibinfo{journal}{Phys. Rev. C} \textbf{\bibinfo{volume}{62}},
  \bibinfo{pages}{014001} (\bibinfo{year}{2000}),
  \urlprefix\url{http://link.aps.org/doi/10.1103/PhysRevC.62.014001}.

\bibitem[{\citenamefont{Elliott}(1958)}]{Ell58}
\bibinfo{author}{\bibfnamefont{J.~P.} \bibnamefont{Elliott}},
  \bibinfo{journal}{Proc. Roy. Soc. A} \textbf{\bibinfo{volume}{245}},
  \bibinfo{pages}{128} (\bibinfo{year}{1958}).

\bibitem[{\citenamefont{Gnilozub et~al.}(2006)\citenamefont{Gnilozub, Kurgalin,
  and Tchuvil'sky}}]{GKT06}
\bibinfo{author}{\bibfnamefont{I.~A.} \bibnamefont{Gnilozub}},
  \bibinfo{author}{\bibfnamefont{S.~D.} \bibnamefont{Kurgalin}},
  \bibnamefont{and} \bibinfo{author}{\bibfnamefont{Y.~M.}
  \bibnamefont{Tchuvil'sky}}, \bibinfo{journal}{Phys. At. Nucl.}
  \textbf{\bibinfo{volume}{69}}, \bibinfo{pages}{1014} (\bibinfo{year}{2006}).

\bibitem[{\citenamefont{Gnilozub et~al.}(2007)\citenamefont{Gnilozub, Kurgalin,
  and Tchuvil'sky}}]{GKT07}
\bibinfo{author}{\bibfnamefont{I.~A.} \bibnamefont{Gnilozub}},
  \bibinfo{author}{\bibfnamefont{S.~D.} \bibnamefont{Kurgalin}},
  \bibnamefont{and} \bibinfo{author}{\bibfnamefont{Y.~M.}
  \bibnamefont{Tchuvil'sky}}, \bibinfo{journal}{Nucl. Phys. A}
  \textbf{\bibinfo{volume}{790}}, \bibinfo{pages}{687C} (\bibinfo{year}{2007}).

\bibitem[{\citenamefont{Gnilozub et~al.}(2008)\citenamefont{Gnilozub, Kurgalin,
  and Tchuvil'sky}}]{GKT08}
\bibinfo{author}{\bibfnamefont{I.~A.} \bibnamefont{Gnilozub}},
  \bibinfo{author}{\bibfnamefont{S.~D.} \bibnamefont{Kurgalin}},
  \bibnamefont{and} \bibinfo{author}{\bibfnamefont{Y.~M.}
  \bibnamefont{Tchuvil'sky}}, \bibinfo{journal}{Phys. At. Nucl.}
  \textbf{\bibinfo{volume}{71}}, \bibinfo{pages}{1213} (\bibinfo{year}{2008}).

\bibitem[{\citenamefont{Wolsky et~al.}(2010)\citenamefont{Wolsky, Gnilozub,
  Kurgalin, and Tchuvil'sky}}]{GKT10}
\bibinfo{author}{\bibfnamefont{R.}~\bibnamefont{Wolsky}},
  \bibinfo{author}{\bibfnamefont{I.~A.} \bibnamefont{Gnilozub}},
  \bibinfo{author}{\bibfnamefont{S.~D.} \bibnamefont{Kurgalin}},
  \bibnamefont{and} \bibinfo{author}{\bibfnamefont{Y.~M.}
  \bibnamefont{Tchuvil'sky}}, \bibinfo{journal}{Phys. At. Nucl.}
  \textbf{\bibinfo{volume}{73}}, \bibinfo{pages}{1405} (\bibinfo{year}{2010}).

\bibitem[{\citenamefont{Gnilozub et~al.}(2013)\citenamefont{Gnilozub, Kurgalin,
  and Tchuvil'sky}}]{GKT13}
\bibinfo{author}{\bibfnamefont{I.~A.} \bibnamefont{Gnilozub}},
  \bibinfo{author}{\bibfnamefont{S.~D.} \bibnamefont{Kurgalin}},
  \bibnamefont{and} \bibinfo{author}{\bibfnamefont{Y.~M.}
  \bibnamefont{Tchuvil'sky}}, \bibinfo{journal}{J. Phys.: Conf. Ser.}
  \textbf{\bibinfo{volume}{436}}, \bibinfo{pages}{012034}
  (\bibinfo{year}{2013}).

\bibitem[{\citenamefont{Maris et~al.}(2009)\citenamefont{Maris, Vary, and
  Shirokov}}]{Maris2009}
\bibinfo{author}{\bibfnamefont{P.}~\bibnamefont{Maris}},
  \bibinfo{author}{\bibfnamefont{J.~P.} \bibnamefont{Vary}}, \bibnamefont{and}
  \bibinfo{author}{\bibfnamefont{A.~M.} \bibnamefont{Shirokov}},
  \bibinfo{journal}{Phys. Rev. C} \textbf{\bibinfo{volume}{79}},
  \bibinfo{pages}{014308} (\bibinfo{year}{2009}).

\bibitem[{\citenamefont{Volya and Tchuvil'sky}()}]{CNCIM}
\bibinfo{author}{\bibfnamefont{A.}~\bibnamefont{Volya}} \bibnamefont{and}
  \bibinfo{author}{\bibfnamefont{Y.}~\bibnamefont{Tchuvil'sky}},
  \bibinfo{note}{{IASEN2013} Conference Proceedings, World Scientific (2014)}.

\bibitem[{\citenamefont{Zhao et~al.}(1989)\citenamefont{Zhao, Gai, Lund,
  Rugari, Mikolas, Brown, Nolen, and Samuel}}]{Zhao89}
\bibinfo{author}{\bibfnamefont{Z.}~\bibnamefont{Zhao}},
  \bibinfo{author}{\bibfnamefont{M.}~\bibnamefont{Gai}},
  \bibinfo{author}{\bibfnamefont{B.~J.} \bibnamefont{Lund}},
  \bibinfo{author}{\bibfnamefont{S.~L.} \bibnamefont{Rugari}},
  \bibinfo{author}{\bibfnamefont{D.}~\bibnamefont{Mikolas}},
  \bibinfo{author}{\bibfnamefont{B.~A.} \bibnamefont{Brown}},
  \bibinfo{author}{\bibfnamefont{J.~A.} \bibnamefont{Nolen}}, \bibnamefont{and}
  \bibinfo{author}{\bibfnamefont{M.}~\bibnamefont{Samuel}},
  \bibinfo{journal}{Phys. Rev. C} \textbf{\bibinfo{volume}{39}},
  \bibinfo{pages}{1985} (\bibinfo{year}{1989}),
  \urlprefix\url{http://link.aps.org/doi/10.1103/PhysRevC.39.1985}.

\bibitem[{\citenamefont{Buchmann et~al.}(2007)\citenamefont{Buchmann, D'Auria,
  Dombsky, Giesen, Jackson, McNeely, Powell, and Volya}}]{Buch07}
\bibinfo{author}{\bibfnamefont{L.}~\bibnamefont{Buchmann}},
  \bibinfo{author}{\bibfnamefont{J.}~\bibnamefont{D'Auria}},
  \bibinfo{author}{\bibfnamefont{M.}~\bibnamefont{Dombsky}},
  \bibinfo{author}{\bibfnamefont{U.}~\bibnamefont{Giesen}},
  \bibinfo{author}{\bibfnamefont{K.~P.} \bibnamefont{Jackson}},
  \bibinfo{author}{\bibfnamefont{P.}~\bibnamefont{McNeely}},
  \bibinfo{author}{\bibfnamefont{J.}~\bibnamefont{Powell}}, \bibnamefont{and}
  \bibinfo{author}{\bibfnamefont{A.}~\bibnamefont{Volya}},
  \bibinfo{journal}{Phys. Rev. C} \textbf{\bibinfo{volume}{75}},
  \bibinfo{pages}{012804} (\bibinfo{year}{2007}),
  \urlprefix\url{http://link.aps.org/doi/10.1103/PhysRevC.75.012804}.

\bibitem[{\citenamefont{Fortune and Kurath}(1978)}]{Fort78}
\bibinfo{author}{\bibfnamefont{H.~T.} \bibnamefont{Fortune}} \bibnamefont{and}
  \bibinfo{author}{\bibfnamefont{D.}~\bibnamefont{Kurath}},
  \bibinfo{journal}{Phys. Rev. C} \textbf{\bibinfo{volume}{18}},
  \bibinfo{pages}{236} (\bibinfo{year}{1978}),
  \urlprefix\url{http://link.aps.org/doi/10.1103/PhysRevC.18.236}.

\bibitem[{\citenamefont{von Oertzen et~al.}(2010)\citenamefont{von Oertzen,
  Dorsch, Bohlen, Krücken, Faestermann, Hertenberger, Kokalova, Mahgoub,
  Milin, Wheldon et~al.}}]{Oert10}
\bibinfo{author}{\bibfnamefont{W.}~\bibnamefont{von Oertzen}},
  \bibinfo{author}{\bibfnamefont{T.}~\bibnamefont{Dorsch}},
  \bibinfo{author}{\bibfnamefont{H.~G.} \bibnamefont{Bohlen}},
  \bibinfo{author}{\bibfnamefont{R.}~\bibnamefont{Krücken}},
  \bibinfo{author}{\bibfnamefont{T.}~\bibnamefont{Faestermann}},
  \bibinfo{author}{\bibfnamefont{R.}~\bibnamefont{Hertenberger}},
  \bibinfo{author}{\bibfnamefont{{\relax Tz}.}~\bibnamefont{Kokalova}},
  \bibinfo{author}{\bibfnamefont{M.}~\bibnamefont{Mahgoub}},
  \bibinfo{author}{\bibfnamefont{M.}~\bibnamefont{Milin}},
  \bibinfo{author}{\bibfnamefont{C.}~\bibnamefont{Wheldon}},
  \bibnamefont{et~al.}, \bibinfo{journal}{Eur. Phys. J}
  \textbf{\bibinfo{volume}{43}}, \bibinfo{pages}{17} (\bibinfo{year}{2010}),
  ISSN \bibinfo{issn}{1434-6001},
  \urlprefix\url{http://dx.doi.org/10.1140/epja/i2009-10894-2}.

\bibitem[{\citenamefont{Cunsolo et~al.}(1981)\citenamefont{Cunsolo, Foti,
  Imm\`e, Pappalardo, Raciti, and Saunier}}]{Cuns81}
\bibinfo{author}{\bibfnamefont{A.}~\bibnamefont{Cunsolo}},
  \bibinfo{author}{\bibfnamefont{A.}~\bibnamefont{Foti}},
  \bibinfo{author}{\bibfnamefont{G.}~\bibnamefont{Imm\`e}},
  \bibinfo{author}{\bibfnamefont{G.}~\bibnamefont{Pappalardo}},
  \bibinfo{author}{\bibfnamefont{G.}~\bibnamefont{Raciti}}, \bibnamefont{and}
  \bibinfo{author}{\bibfnamefont{N.}~\bibnamefont{Saunier}},
  \bibinfo{journal}{Phys. Rev. C} \textbf{\bibinfo{volume}{24}},
  \bibinfo{pages}{476} (\bibinfo{year}{1981}),
  \urlprefix\url{http://link.aps.org/doi/10.1103/PhysRevC.24.476}.

\bibitem[{\citenamefont{Cunsolo et~al.}(1982)\citenamefont{Cunsolo, Foti,
  Imm\`e, Pappalardo, and Raciti}}]{Cuns82}
\bibinfo{author}{\bibfnamefont{A.}~\bibnamefont{Cunsolo}},
  \bibinfo{author}{\bibfnamefont{A.}~\bibnamefont{Foti}},
  \bibinfo{author}{\bibfnamefont{G.}~\bibnamefont{Imm\`e}},
  \bibinfo{author}{\bibfnamefont{G.}~\bibnamefont{Pappalardo}},
  \bibnamefont{and} \bibinfo{author}{\bibfnamefont{G.}~\bibnamefont{Raciti}},
  \bibinfo{journal}{Physics Letters B} \textbf{\bibinfo{volume}{112}},
  \bibinfo{pages}{121} (\bibinfo{year}{1982}), ISSN \bibinfo{issn}{0370-2693},
  \urlprefix\url{http://www.sciencedirect.com/science/article/pii/0370269382903100}.

\bibitem[{\citenamefont{Cunsolo et~al.}(1983)\citenamefont{Cunsolo, Foti,
  Imm\`e, Pappalardo, and Raciti}}]{Cuns83}
\bibinfo{author}{\bibfnamefont{A.}~\bibnamefont{Cunsolo}},
  \bibinfo{author}{\bibfnamefont{A.}~\bibnamefont{Foti}},
  \bibinfo{author}{\bibfnamefont{G.}~\bibnamefont{Imm\`e}},
  \bibinfo{author}{\bibfnamefont{G.}~\bibnamefont{Pappalardo}},
  \bibnamefont{and} \bibinfo{author}{\bibfnamefont{G.}~\bibnamefont{Raciti}},
  \bibinfo{journal}{Lett. Nuo. Cim.} \textbf{\bibinfo{volume}{37}},
  \bibinfo{pages}{193} (\bibinfo{year}{1983}).

\bibitem[{\citenamefont{Smithson et~al.}(1986)\citenamefont{Smithson, Watson,
  and Fortune}}]{Smit85}
\bibinfo{author}{\bibfnamefont{M.~J.} \bibnamefont{Smithson}},
  \bibinfo{author}{\bibfnamefont{D.~L.} \bibnamefont{Watson}},
  \bibnamefont{and} \bibinfo{author}{\bibfnamefont{H.~T.}
  \bibnamefont{Fortune}}, \bibinfo{journal}{Phys. Rev. C}
  \textbf{\bibinfo{volume}{33}}, \bibinfo{pages}{509} (\bibinfo{year}{1986}),
  \urlprefix\url{http://link.aps.org/doi/10.1103/PhysRevC.33.509}.

\bibitem[{\citenamefont{Smithson et~al.}(1988)\citenamefont{Smithson, Watson,
  and Fortune}}]{Smit88}
\bibinfo{author}{\bibfnamefont{M.~J.} \bibnamefont{Smithson}},
  \bibinfo{author}{\bibfnamefont{D.~L.} \bibnamefont{Watson}},
  \bibnamefont{and} \bibinfo{author}{\bibfnamefont{H.~T.}
  \bibnamefont{Fortune}}, \bibinfo{journal}{Phys. Rev. C}
  \textbf{\bibinfo{volume}{37}}, \bibinfo{pages}{1036} (\bibinfo{year}{1988}),
  \urlprefix\url{http://link.aps.org/doi/10.1103/PhysRevC.37.1036}.

\bibitem[{\citenamefont{Sanders}(1956)}]{San56}
\bibinfo{author}{\bibfnamefont{R.~M.} \bibnamefont{Sanders}},
  \bibinfo{journal}{Phys. Rev.} \textbf{\bibinfo{volume}{104}},
  \bibinfo{pages}{1434} (\bibinfo{year}{1956}),
  \urlprefix\url{http://link.aps.org/doi/10.1103/PhysRev.104.1434}.

\bibitem[{\citenamefont{Bair et~al.}(1966)\citenamefont{Bair, Ford, and
  Jones}}]{Bair66}
\bibinfo{author}{\bibfnamefont{J.~K.} \bibnamefont{Bair}},
  \bibinfo{author}{\bibfnamefont{J.~L.~C.} \bibnamefont{Ford}},
  \bibnamefont{and} \bibinfo{author}{\bibfnamefont{C.~M.} \bibnamefont{Jones}},
  \bibinfo{journal}{Phys. Rev.} \textbf{\bibinfo{volume}{144}},
  \bibinfo{pages}{799} (\bibinfo{year}{1966}),
  \urlprefix\url{http://link.aps.org/doi/10.1103/PhysRev.144.799}.

\bibitem[{\citenamefont{Wagemans et~al.}(2002)\citenamefont{Wagemans, Wagemans,
  Goeminne, Serot, Loiselet, and Gaelens}}]{Wag02}
\bibinfo{author}{\bibfnamefont{J.}~\bibnamefont{Wagemans}},
  \bibinfo{author}{\bibfnamefont{C.}~\bibnamefont{Wagemans}},
  \bibinfo{author}{\bibfnamefont{G.}~\bibnamefont{Goeminne}},
  \bibinfo{author}{\bibfnamefont{O.}~\bibnamefont{Serot}},
  \bibinfo{author}{\bibfnamefont{M.}~\bibnamefont{Loiselet}}, \bibnamefont{and}
  \bibinfo{author}{\bibfnamefont{M.}~\bibnamefont{Gaelens}},
  \bibinfo{journal}{Phys. Rev. C} \textbf{\bibinfo{volume}{65}},
  \bibinfo{pages}{034614} (\bibinfo{year}{2002}),
  \urlprefix\url{http://link.aps.org/doi/10.1103/PhysRevC.65.034614}.

\bibitem[{\citenamefont{Cobern et~al.}(1981)\citenamefont{Cobern, Bland,
  Fortune, Moore, Mordechai, and Middleton}}]{Cobe80}
\bibinfo{author}{\bibfnamefont{M.~E.} \bibnamefont{Cobern}},
  \bibinfo{author}{\bibfnamefont{L.~C.} \bibnamefont{Bland}},
  \bibinfo{author}{\bibfnamefont{H.~T.} \bibnamefont{Fortune}},
  \bibinfo{author}{\bibfnamefont{G.~E.} \bibnamefont{Moore}},
  \bibinfo{author}{\bibfnamefont{S.}~\bibnamefont{Mordechai}},
  \bibnamefont{and}
  \bibinfo{author}{\bibfnamefont{R.}~\bibnamefont{Middleton}},
  \bibinfo{journal}{Phys. Rev. C} \textbf{\bibinfo{volume}{23}},
  \bibinfo{pages}{2387} (\bibinfo{year}{1981}),
  \urlprefix\url{http://link.aps.org/doi/10.1103/PhysRevC.23.2387}.

\bibitem[{\citenamefont{Yildiz et~al.}(2006)\citenamefont{Yildiz, Freer,
  Soi\ifmmode~\acute{c}\else \'{c}\fi{}, Ahmed, Ashwood, Clarke, Curtis,
  Fulton, Metelko, Novatski et~al.}}]{Yild06}
\bibinfo{author}{\bibfnamefont{S.}~\bibnamefont{Yildiz}},
  \bibinfo{author}{\bibfnamefont{M.}~\bibnamefont{Freer}},
  \bibinfo{author}{\bibfnamefont{N.}~\bibnamefont{Soi\ifmmode~\acute{c}\else
  \'{c}\fi{}}}, \bibinfo{author}{\bibfnamefont{S.}~\bibnamefont{Ahmed}},
  \bibinfo{author}{\bibfnamefont{N.~I.} \bibnamefont{Ashwood}},
  \bibinfo{author}{\bibfnamefont{N.~M.} \bibnamefont{Clarke}},
  \bibinfo{author}{\bibfnamefont{N.}~\bibnamefont{Curtis}},
  \bibinfo{author}{\bibfnamefont{B.~R.} \bibnamefont{Fulton}},
  \bibinfo{author}{\bibfnamefont{C.~J.} \bibnamefont{Metelko}},
  \bibinfo{author}{\bibfnamefont{B.}~\bibnamefont{Novatski}},
  \bibnamefont{et~al.}, \bibinfo{journal}{Phys. Rev. C}
  \textbf{\bibinfo{volume}{73}}, \bibinfo{pages}{034601}
  (\bibinfo{year}{2006}),
  \urlprefix\url{http://link.aps.org/doi/10.1103/PhysRevC.73.034601}.

\bibitem[{\citenamefont{Curtis et~al.}(2002)\citenamefont{Curtis, Caussyn,
  Chandler, Cooper, Fletcher, Laird, and Pavan}}]{Curt02}
\bibinfo{author}{\bibfnamefont{N.}~\bibnamefont{Curtis}},
  \bibinfo{author}{\bibfnamefont{D.~D.} \bibnamefont{Caussyn}},
  \bibinfo{author}{\bibfnamefont{C.}~\bibnamefont{Chandler}},
  \bibinfo{author}{\bibfnamefont{M.~W.} \bibnamefont{Cooper}},
  \bibinfo{author}{\bibfnamefont{N.~R.} \bibnamefont{Fletcher}},
  \bibinfo{author}{\bibfnamefont{R.~W.} \bibnamefont{Laird}}, \bibnamefont{and}
  \bibinfo{author}{\bibfnamefont{J.}~\bibnamefont{Pavan}},
  \bibinfo{journal}{Phys. Rev. C} \textbf{\bibinfo{volume}{66}},
  \bibinfo{pages}{024315} (\bibinfo{year}{2002}),
  \urlprefix\url{http://link.aps.org/doi/10.1103/PhysRevC.66.024315}.

\bibitem[{\citenamefont{Jahn et~al.}(1978)\citenamefont{Jahn, Stahel, Wozniak,
  de~Meijer, and Cerny}}]{Jahn78}
\bibinfo{author}{\bibfnamefont{R.}~\bibnamefont{Jahn}},
  \bibinfo{author}{\bibfnamefont{D.~P.} \bibnamefont{Stahel}},
  \bibinfo{author}{\bibfnamefont{G.~J.} \bibnamefont{Wozniak}},
  \bibinfo{author}{\bibfnamefont{R.~J.} \bibnamefont{de~Meijer}},
  \bibnamefont{and} \bibinfo{author}{\bibfnamefont{J.}~\bibnamefont{Cerny}},
  \bibinfo{journal}{Phys. Rev. C} \textbf{\bibinfo{volume}{18}},
  \bibinfo{pages}{9} (\bibinfo{year}{1978}),
  \urlprefix\url{http://link.aps.org/doi/10.1103/PhysRevC.18.9}.

\bibitem[{\citenamefont{Woodworth et~al.}(1979)\citenamefont{Woodworth,
  McNeill, Jury, Alvarez, Berman, Faul, and Meyer}}]{Wood78}
\bibinfo{author}{\bibfnamefont{J.~G.} \bibnamefont{Woodworth}},
  \bibinfo{author}{\bibfnamefont{K.~G.} \bibnamefont{McNeill}},
  \bibinfo{author}{\bibfnamefont{J.~W.} \bibnamefont{Jury}},
  \bibinfo{author}{\bibfnamefont{R.~A.} \bibnamefont{Alvarez}},
  \bibinfo{author}{\bibfnamefont{B.~L.} \bibnamefont{Berman}},
  \bibinfo{author}{\bibfnamefont{D.~D.} \bibnamefont{Faul}}, \bibnamefont{and}
  \bibinfo{author}{\bibfnamefont{P.}~\bibnamefont{Meyer}},
  \bibinfo{journal}{Phys. Rev. C} \textbf{\bibinfo{volume}{19}},
  \bibinfo{pages}{1667} (\bibinfo{year}{1979}),
  \urlprefix\url{http://link.aps.org/doi/10.1103/PhysRevC.19.1667}.

\bibitem[{\citenamefont{Morgan et~al.}(1970{\natexlab{a}})\citenamefont{Morgan,
  Tilley, Mitchell, Hilko, and Roberson}}]{Morg70}
\bibinfo{author}{\bibfnamefont{G.~L.} \bibnamefont{Morgan}},
  \bibinfo{author}{\bibfnamefont{D.~R.} \bibnamefont{Tilley}},
  \bibinfo{author}{\bibfnamefont{G.~E.} \bibnamefont{Mitchell}},
  \bibinfo{author}{\bibfnamefont{R.~A.} \bibnamefont{Hilko}}, \bibnamefont{and}
  \bibinfo{author}{\bibfnamefont{N.~R.} \bibnamefont{Roberson}},
  \bibinfo{journal}{Nuc. Phys. A} \textbf{\bibinfo{volume}{148}},
  \bibinfo{pages}{480} (\bibinfo{year}{1970}{\natexlab{a}}), ISSN
  \bibinfo{issn}{0375-9474},
  \urlprefix\url{http://www.sciencedirect.com/science/article/pii/037594747090641X}.

\bibitem[{\citenamefont{Weinman and Silverstein}(1958)}]{Wein58}
\bibinfo{author}{\bibfnamefont{J.~A.} \bibnamefont{Weinman}} \bibnamefont{and}
  \bibinfo{author}{\bibfnamefont{E.~A.} \bibnamefont{Silverstein}},
  \bibinfo{journal}{Phys. Rev.} \textbf{\bibinfo{volume}{111}},
  \bibinfo{pages}{277} (\bibinfo{year}{1958}),
  \urlprefix\url{http://link.aps.org/doi/10.1103/PhysRev.111.277}.

\bibitem[{\citenamefont{Goldberg et~al.}(2005)\citenamefont{Goldberg,
  K\"allman, Lönnroth, Manngård, and Skorodumov}}]{Gold04}
\bibinfo{author}{\bibfnamefont{V.~Z.} \bibnamefont{Goldberg}},
  \bibinfo{author}{\bibfnamefont{K.-M.} \bibnamefont{K\"allman}},
  \bibinfo{author}{\bibfnamefont{T.}~\bibnamefont{Lönnroth}},
  \bibinfo{author}{\bibfnamefont{P.}~\bibnamefont{Manngård}},
  \bibnamefont{and} \bibinfo{author}{\bibfnamefont{B.~B.}
  \bibnamefont{Skorodumov}}, \bibinfo{journal}{Physics of Atomic Nuclei}
  \textbf{\bibinfo{volume}{68}}, \bibinfo{pages}{1079} (\bibinfo{year}{2005}),
  ISSN \bibinfo{issn}{1063-7788},
  \urlprefix\url{http://dx.doi.org/10.1134/1.1992561}.

\bibitem[{\citenamefont{Johnson et~al.}(2009)\citenamefont{Johnson, Rogachev,
  Goldberg, Brown, Robson, Crisp, Cottle, Fu, Giles, Green et~al.}}]{John09}
\bibinfo{author}{\bibfnamefont{E.~D.} \bibnamefont{Johnson}},
  \bibinfo{author}{\bibfnamefont{G.~V.} \bibnamefont{Rogachev}},
  \bibinfo{author}{\bibfnamefont{V.~Z.} \bibnamefont{Goldberg}},
  \bibinfo{author}{\bibfnamefont{S.}~\bibnamefont{Brown}},
  \bibinfo{author}{\bibfnamefont{D.}~\bibnamefont{Robson}},
  \bibinfo{author}{\bibfnamefont{A.~M.} \bibnamefont{Crisp}},
  \bibinfo{author}{\bibfnamefont{P.~D.} \bibnamefont{Cottle}},
  \bibinfo{author}{\bibfnamefont{C.}~\bibnamefont{Fu}},
  \bibinfo{author}{\bibfnamefont{J.}~\bibnamefont{Giles}},
  \bibinfo{author}{\bibfnamefont{B.~W.} \bibnamefont{Green}},
  \bibnamefont{et~al.}, \bibinfo{journal}{The European Physical Journal A}
  \textbf{\bibinfo{volume}{42}}, \bibinfo{pages}{135} (\bibinfo{year}{2009}),
  ISSN \bibinfo{issn}{1434-6001},
  \urlprefix\url{http://dx.doi.org/10.1140/epja/i2009-10887-1}.

\bibitem[{\citenamefont{Artemov et~al.}(1990)\citenamefont{Artemov, Belyanin,
  Vetoshkin, Wolski, Golovkov, Goldberg, Madeja, Pankratov, Serikov, Timofeev
  et~al.}}]{Arte90}
\bibinfo{author}{\bibfnamefont{K.~P.} \bibnamefont{Artemov}},
  \bibinfo{author}{\bibfnamefont{O.~P.} \bibnamefont{Belyanin}},
  \bibinfo{author}{\bibfnamefont{A.~L.} \bibnamefont{Vetoshkin}},
  \bibinfo{author}{\bibfnamefont{R.}~\bibnamefont{Wolski}},
  \bibinfo{author}{\bibfnamefont{M.~S.} \bibnamefont{Golovkov}},
  \bibinfo{author}{\bibfnamefont{V.~Z.} \bibnamefont{Goldberg}},
  \bibinfo{author}{\bibfnamefont{M.}~\bibnamefont{Madeja}},
  \bibinfo{author}{\bibfnamefont{V.~V.} \bibnamefont{Pankratov}},
  \bibinfo{author}{\bibfnamefont{I.~N.} \bibnamefont{Serikov}},
  \bibinfo{author}{\bibfnamefont{V.~A.} \bibnamefont{Timofeev}},
  \bibnamefont{et~al.}, \bibinfo{journal}{Sov. J. Nucl. Phys.}
  \textbf{\bibinfo{volume}{52}}, \bibinfo{pages}{634} (\bibinfo{year}{1990}).

\bibitem[{\citenamefont{Goldberg and Pakhomov}(1993)}]{Gold93}
\bibinfo{author}{\bibfnamefont{V.~Z.} \bibnamefont{Goldberg}} \bibnamefont{and}
  \bibinfo{author}{\bibfnamefont{A.~E.} \bibnamefont{Pakhomov}},
  \bibinfo{journal}{Phys. Atomic Nuclei} \textbf{\bibinfo{volume}{56}},
  \bibinfo{pages}{1167} (\bibinfo{year}{1993}).

\bibitem[{\citenamefont{Markenroth et~al.}(2000)\citenamefont{Markenroth,
  Axelsson, Baxter, Borge, Donzaud, Fayans, Fynbo, Goldberg, Grevy,
  Guillemaud-Mueller et~al.}}]{Mark00}
\bibinfo{author}{\bibfnamefont{K.}~\bibnamefont{Markenroth}},
  \bibinfo{author}{\bibfnamefont{L.}~\bibnamefont{Axelsson}},
  \bibinfo{author}{\bibfnamefont{S.}~\bibnamefont{Baxter}},
  \bibinfo{author}{\bibfnamefont{M.}~\bibnamefont{Borge}},
  \bibinfo{author}{\bibfnamefont{C.}~\bibnamefont{Donzaud}},
  \bibinfo{author}{\bibfnamefont{S.}~\bibnamefont{Fayans}},
  \bibinfo{author}{\bibfnamefont{H.}~\bibnamefont{Fynbo}},
  \bibinfo{author}{\bibfnamefont{V.}~\bibnamefont{Goldberg}},
  \bibinfo{author}{\bibfnamefont{S.}~\bibnamefont{Grevy}},
  \bibinfo{author}{\bibfnamefont{D.}~\bibnamefont{Guillemaud-Mueller}},
  \bibnamefont{et~al.}, \bibinfo{journal}{Phys.\ Rev.\ C}
  \textbf{\bibinfo{volume}{62}}, \bibinfo{pages}{034308}
  (\bibinfo{year}{2000}).

\bibitem[{\citenamefont{L\"onnroth et~al.}(2010)\citenamefont{L\"onnroth,
  Norrby, Goldberg, Rogachev, Golovkov, K\"allman, Lattuada, Perov, Romano,
  Skorodumov et~al.}}]{Lonn10}
\bibinfo{author}{\bibfnamefont{T.}~\bibnamefont{L\"onnroth}},
  \bibinfo{author}{\bibfnamefont{M.}~\bibnamefont{Norrby}},
  \bibinfo{author}{\bibfnamefont{V.~Z.} \bibnamefont{Goldberg}},
  \bibinfo{author}{\bibfnamefont{G.~V.} \bibnamefont{Rogachev}},
  \bibinfo{author}{\bibfnamefont{M.~S.} \bibnamefont{Golovkov}},
  \bibinfo{author}{\bibfnamefont{K.-M.} \bibnamefont{K\"allman}},
  \bibinfo{author}{\bibfnamefont{M.}~\bibnamefont{Lattuada}},
  \bibinfo{author}{\bibfnamefont{S.~V.} \bibnamefont{Perov}},
  \bibinfo{author}{\bibfnamefont{S.}~\bibnamefont{Romano}},
  \bibinfo{author}{\bibfnamefont{B.~B.} \bibnamefont{Skorodumov}},
  \bibnamefont{et~al.}, \bibinfo{journal}{The European Physical Journal A}
  \textbf{\bibinfo{volume}{46}}, \bibinfo{pages}{5} (\bibinfo{year}{2010}),
  ISSN \bibinfo{issn}{1434-6001},
  \urlprefix\url{http://dx.doi.org/10.1140/epja/i2010-11021-2}.

\bibitem[{\citenamefont{Rogachev et~al.}(2010)\citenamefont{Rogachev, Johnson,
  Mitchell, Goldberg, Kemper, and Wiedenhöver}}]{SantaTecla10}
\bibinfo{author}{\bibfnamefont{G.~V.} \bibnamefont{Rogachev}},
  \bibinfo{author}{\bibfnamefont{E.~D.} \bibnamefont{Johnson}},
  \bibinfo{author}{\bibfnamefont{J.}~\bibnamefont{Mitchell}},
  \bibinfo{author}{\bibfnamefont{V.~Z.} \bibnamefont{Goldberg}},
  \bibinfo{author}{\bibfnamefont{K.~W.} \bibnamefont{Kemper}},
  \bibnamefont{and}
  \bibinfo{author}{\bibfnamefont{I.}~\bibnamefont{Wiedenhöver}},
  \bibinfo{journal}{AIP Conf. Proc.} \textbf{\bibinfo{volume}{1213}},
  \bibinfo{pages}{137} (\bibinfo{year}{2010}),
  \urlprefix\url{http://dx.doi.org/10.1063/1.3362563}.

\bibitem[{\citenamefont{Lane and Thomas}(1958)}]{Lane58}
\bibinfo{author}{\bibfnamefont{A.~M.} \bibnamefont{Lane}} \bibnamefont{and}
  \bibinfo{author}{\bibfnamefont{R.~G.} \bibnamefont{Thomas}},
  \bibinfo{journal}{Rev. Mod. Phys.} \textbf{\bibinfo{volume}{30}},
  \bibinfo{pages}{257} (\bibinfo{year}{1958}),
  \urlprefix\url{http://link.aps.org/doi/10.1103/RevModPhys.30.257}.

\bibitem[{\citenamefont{Johnson}(2008)}]{EDJ-Thesis}
\bibinfo{author}{\bibfnamefont{E.~D.} \bibnamefont{Johnson}}, Ph.D. thesis,
  \bibinfo{school}{Florida State University} (\bibinfo{year}{2008}),
  \urlprefix\url{http://diginole.lib.fsu.edu/etd/3494/}.

\bibitem[{\citenamefont{Avila}(2013)}]{Avila13}
\bibinfo{author}{\bibfnamefont{M.~L.} \bibnamefont{Avila}}, Ph.D. thesis,
  \bibinfo{school}{Florida State University} (\bibinfo{year}{2013}),
  \urlprefix\url{http://diginole.lib.fsu.edu/etd/8525}.

\bibitem[{\citenamefont{Tilley et~al.}(1995)\citenamefont{Tilley, Weller,
  Cheves, and Chasteler}}]{Till95}
\bibinfo{author}{\bibfnamefont{D.~R.} \bibnamefont{Tilley}},
  \bibinfo{author}{\bibfnamefont{H.~R.} \bibnamefont{Weller}},
  \bibinfo{author}{\bibfnamefont{C.~M.} \bibnamefont{Cheves}},
  \bibnamefont{and} \bibinfo{author}{\bibfnamefont{R.~M.}
  \bibnamefont{Chasteler}}, \bibinfo{journal}{Nucl. Phys. A}
  \textbf{\bibinfo{volume}{595}}, \bibinfo{pages}{1} (\bibinfo{year}{1995}),
  ISSN \bibinfo{issn}{0375-9474},
  \urlprefix\url{http://www.sciencedirect.com/science/article/pii/0375947495003381}.

\bibitem[{\citenamefont{Ashwood et~al.}(2006)\citenamefont{Ashwood, Freer,
  Sakuta, Ahmed, Curtis, MCEwan, Metelko, Novatoski, Soic, Stepanov
  et~al.}}]{Ashw06}
\bibinfo{author}{\bibfnamefont{N.~I.} \bibnamefont{Ashwood}},
  \bibinfo{author}{\bibfnamefont{M.}~\bibnamefont{Freer}},
  \bibinfo{author}{\bibfnamefont{S.}~\bibnamefont{Sakuta}},
  \bibinfo{author}{\bibfnamefont{N.~M.} \bibnamefont{Ahmed}},
  \bibinfo{author}{\bibfnamefont{N.}~\bibnamefont{Curtis}},
  \bibinfo{author}{\bibfnamefont{P.}~\bibnamefont{MCEwan}},
  \bibinfo{author}{\bibfnamefont{C.~J.} \bibnamefont{Metelko}},
  \bibinfo{author}{\bibfnamefont{B.}~\bibnamefont{Novatoski}},
  \bibinfo{author}{\bibfnamefont{N.}~\bibnamefont{Soic}},
  \bibinfo{author}{\bibfnamefont{D.}~\bibnamefont{Stepanov}},
  \bibnamefont{et~al.}, \bibinfo{journal}{J. Phys. G : Nuc. Part. Phys.}
  \textbf{\bibinfo{volume}{32}}, \bibinfo{pages}{463} (\bibinfo{year}{2006}),
  \urlprefix\url{http://stacks.iop.org/0954-3899/32/i=4/a=005}.

\bibitem[{\citenamefont{Sellers et~al.}(1995)\citenamefont{Sellers, Manley,
  Niboh, Weerasundara, Lindgren, Clausen, Farkhondeh, Norum, and
  Berman}}]{Sell95}
\bibinfo{author}{\bibfnamefont{R.~M.} \bibnamefont{Sellers}},
  \bibinfo{author}{\bibfnamefont{D.~M.} \bibnamefont{Manley}},
  \bibinfo{author}{\bibfnamefont{M.~M.} \bibnamefont{Niboh}},
  \bibinfo{author}{\bibfnamefont{D.~S.} \bibnamefont{Weerasundara}},
  \bibinfo{author}{\bibfnamefont{R.~A.} \bibnamefont{Lindgren}},
  \bibinfo{author}{\bibfnamefont{B.~L.} \bibnamefont{Clausen}},
  \bibinfo{author}{\bibfnamefont{M.}~\bibnamefont{Farkhondeh}},
  \bibinfo{author}{\bibfnamefont{B.~E.} \bibnamefont{Norum}}, \bibnamefont{and}
  \bibinfo{author}{\bibfnamefont{B.~L.} \bibnamefont{Berman}},
  \bibinfo{journal}{Phys. Rev. C} \textbf{\bibinfo{volume}{51}},
  \bibinfo{pages}{1926} (\bibinfo{year}{1995}),
  \urlprefix\url{http://link.aps.org/doi/10.1103/PhysRevC.51.1926}.

\bibitem[{\citenamefont{Fortune et~al.}(1985)\citenamefont{Fortune, Bland, and
  Rae}}]{Fortune1985}
\bibinfo{author}{\bibfnamefont{H.~T.} \bibnamefont{Fortune}},
  \bibinfo{author}{\bibfnamefont{L.~C.} \bibnamefont{Bland}}, \bibnamefont{and}
  \bibinfo{author}{\bibfnamefont{W.~D.~M.} \bibnamefont{Rae}},
  \bibinfo{journal}{Journal of Physics G: Nuclear Physics}
  \textbf{\bibinfo{volume}{11}}, \bibinfo{pages}{1175} (\bibinfo{year}{1985}),
  \urlprefix\url{http://stacks.iop.org/0305-4616/11/i=10/a=013}.

\bibitem[{\citenamefont{Manley et~al.}(1990)\citenamefont{Manley, Millener,
  Berman, Bertozzi, Buti, Finn, Hersman, Hyde-Wright, Hynes, Kelly
  et~al.}}]{Man90}
\bibinfo{author}{\bibfnamefont{D.~M.} \bibnamefont{Manley}},
  \bibinfo{author}{\bibfnamefont{D.~J.} \bibnamefont{Millener}},
  \bibinfo{author}{\bibfnamefont{B.~L.} \bibnamefont{Berman}},
  \bibinfo{author}{\bibfnamefont{W.}~\bibnamefont{Bertozzi}},
  \bibinfo{author}{\bibfnamefont{T.~N.} \bibnamefont{Buti}},
  \bibinfo{author}{\bibfnamefont{J.~M.} \bibnamefont{Finn}},
  \bibinfo{author}{\bibfnamefont{F.~W.} \bibnamefont{Hersman}},
  \bibinfo{author}{\bibfnamefont{C.~E.} \bibnamefont{Hyde-Wright}},
  \bibinfo{author}{\bibfnamefont{M.~V.} \bibnamefont{Hynes}},
  \bibinfo{author}{\bibfnamefont{J.~J.} \bibnamefont{Kelly}},
  \bibnamefont{et~al.}, \bibinfo{journal}{Phys. Rev. C}
  \textbf{\bibinfo{volume}{41}}, \bibinfo{pages}{448} (\bibinfo{year}{1990}),
  \urlprefix\url{http://link.aps.org/doi/10.1103/PhysRevC.41.448}.

\bibitem[{\citenamefont{Rae and Bhowmik}(1984)}]{Rae84}
\bibinfo{author}{\bibfnamefont{W.}~\bibnamefont{Rae}} \bibnamefont{and}
  \bibinfo{author}{\bibfnamefont{R.}~\bibnamefont{Bhowmik}},
  \bibinfo{journal}{Nuc. Phys. A} \textbf{\bibinfo{volume}{427}},
  \bibinfo{pages}{142} (\bibinfo{year}{1984}), ISSN \bibinfo{issn}{0375-9474},
  \urlprefix\url{http://www.sciencedirect.com/science/article/pii/0375947484901428}.

\bibitem[{\citenamefont{Norrby et~al.}(2011{\natexlab{a}})\citenamefont{Norrby,
  Lonnroth, Goldberg, Rogachev et~al.}}]{Norrby2011a}
\bibinfo{author}{\bibfnamefont{M.}~\bibnamefont{Norrby}},
  \bibinfo{author}{\bibfnamefont{T.}~\bibnamefont{Lonnroth}},
  \bibinfo{author}{\bibfnamefont{V.~Z.} \bibnamefont{Goldberg}},
  \bibinfo{author}{\bibfnamefont{G.~V.} \bibnamefont{Rogachev}},
  \bibnamefont{et~al.}, \bibinfo{journal}{Eur. Phys. J A}
  \textbf{\bibinfo{volume}{47}}, \bibinfo{pages}{73}
  (\bibinfo{year}{2011}{\natexlab{a}}).

\bibitem[{\citenamefont{Norrby et~al.}(2011{\natexlab{b}})\citenamefont{Norrby,
  Lonnroth, Goldberg, Rogachev et~al.}}]{Norrby2011b}
\bibinfo{author}{\bibfnamefont{M.}~\bibnamefont{Norrby}},
  \bibinfo{author}{\bibfnamefont{T.}~\bibnamefont{Lonnroth}},
  \bibinfo{author}{\bibfnamefont{V.~Z.} \bibnamefont{Goldberg}},
  \bibinfo{author}{\bibfnamefont{G.~V.} \bibnamefont{Rogachev}},
  \bibnamefont{et~al.}, \bibinfo{journal}{Eur. Phys. J A}
  \textbf{\bibinfo{volume}{47}}, \bibinfo{pages}{96}
  (\bibinfo{year}{2011}{\natexlab{b}}).

\bibitem[{\citenamefont{Descouvemont and Baye}(1985)}]{Desc85}
\bibinfo{author}{\bibfnamefont{P.}~\bibnamefont{Descouvemont}}
  \bibnamefont{and} \bibinfo{author}{\bibfnamefont{D.}~\bibnamefont{Baye}},
  \bibinfo{journal}{Phys. Rev. C} \textbf{\bibinfo{volume}{31}},
  \bibinfo{pages}{2274} (\bibinfo{year}{1985}),
  \urlprefix\url{http://link.aps.org/doi/10.1103/PhysRevC.31.2274}.

\bibitem[{\citenamefont{Furutachi et~al.}(2008)\citenamefont{Furutachi, Kimura,
  Doté, Kanada-En'yo, and Oryu}}]{Furu08}
\bibinfo{author}{\bibfnamefont{N.}~\bibnamefont{Furutachi}},
  \bibinfo{author}{\bibfnamefont{M.}~\bibnamefont{Kimura}},
  \bibinfo{author}{\bibfnamefont{A.}~\bibnamefont{Doté}},
  \bibinfo{author}{\bibfnamefont{Y.}~\bibnamefont{Kanada-En'yo}},
  \bibnamefont{and} \bibinfo{author}{\bibfnamefont{S.}~\bibnamefont{Oryu}},
  \bibinfo{journal}{Progress of Theoretical Physics}
  \textbf{\bibinfo{volume}{119}}, \bibinfo{pages}{403} (\bibinfo{year}{2008}),
  \urlprefix\url{http://ptp.oxfordjournals.org/content/119/3/403.abstract}.

\bibitem[{\citenamefont{Utsuno and Chiba}(2011)}]{Utsu11}
\bibinfo{author}{\bibfnamefont{Y.}~\bibnamefont{Utsuno}} \bibnamefont{and}
  \bibinfo{author}{\bibfnamefont{S.}~\bibnamefont{Chiba}},
  \bibinfo{journal}{Phys. Rev. C} \textbf{\bibinfo{volume}{83}},
  \bibinfo{pages}{021301} (\bibinfo{year}{2011}),
  \urlprefix\url{http://link.aps.org/doi/10.1103/PhysRevC.83.021301}.

\bibitem[{\citenamefont{Smirnov and Yu.M.~Tchuvilsky}(1977)}]{Sm77}
\bibinfo{author}{\bibfnamefont{Y.~F.} \bibnamefont{Smirnov}} \bibnamefont{and}
  \bibinfo{author}{\bibfnamefont{Y.~M.} \bibnamefont{Yu.M.~Tchuvilsky}},
  \bibinfo{journal}{Phys. Rev. C} \textbf{\bibinfo{volume}{15}},
  \bibinfo{pages}{84} (\bibinfo{year}{1977}).

\bibitem[{\citenamefont{Morgan et~al.}(1970{\natexlab{b}})\citenamefont{Morgan,
  Tilley, Mitchell, Hilko, and Roberson}}]{Morgan1970}
\bibinfo{author}{\bibfnamefont{G.~L.} \bibnamefont{Morgan}},
  \bibinfo{author}{\bibfnamefont{D.~R.} \bibnamefont{Tilley}},
  \bibinfo{author}{\bibfnamefont{G.~E.} \bibnamefont{Mitchell}},
  \bibinfo{author}{\bibfnamefont{R.~A.} \bibnamefont{Hilko}}, \bibnamefont{and}
  \bibinfo{author}{\bibfnamefont{N.~R.} \bibnamefont{Roberson}},
  \bibinfo{journal}{Phys. Lett. B} \textbf{\bibinfo{volume}{32}},
  \bibinfo{pages}{353} (\bibinfo{year}{1970}{\natexlab{b}}).

\bibitem[{\citenamefont{Auerbach and Zelevinsky}(2011)}]{Auerbach:2011}
\bibinfo{author}{\bibfnamefont{N.}~\bibnamefont{Auerbach}} \bibnamefont{and}
  \bibinfo{author}{\bibfnamefont{V.}~\bibnamefont{Zelevinsky}},
  \bibinfo{journal}{Rep.Prog.Phys.} \textbf{\bibinfo{volume}{74}},
  \bibinfo{pages}{106301} (\bibinfo{year}{2011}).

\bibitem[{\citenamefont{Volya and Zelevinsky}(2003)}]{Volya:2003PRC}
\bibinfo{author}{\bibfnamefont{A.}~\bibnamefont{Volya}} \bibnamefont{and}
  \bibinfo{author}{\bibfnamefont{V.}~\bibnamefont{Zelevinsky}},
  \bibinfo{journal}{Phys. Rev. C} \textbf{\bibinfo{volume}{67}},
  \bibinfo{pages}{054322} (\bibinfo{year}{2003}).

\bibitem[{\citenamefont{Volya}(2009)}]{Volya:2009}
\bibinfo{author}{\bibfnamefont{A.}~\bibnamefont{Volya}},
  \bibinfo{journal}{Phys. Rev. C} \textbf{\bibinfo{volume}{79}},
  \bibinfo{pages}{044308} (\bibinfo{year}{2009}).

\bibitem[{\citenamefont{Fynbo et~al.}(2005)}]{Fynbo2005}
\bibinfo{author}{\bibfnamefont{H.~O.~U.} \bibnamefont{Fynbo}}
  \bibnamefont{et~al.}, \bibinfo{journal}{Nature}
  \textbf{\bibinfo{volume}{433}}, \bibinfo{pages}{136} (\bibinfo{year}{2005}).

\bibitem[{\citenamefont{Kuchera et~al.}(2011)\citenamefont{Kuchera, Rogachev,
  Goldberg, Johnson et~al.}}]{Kuchera2011}
\bibinfo{author}{\bibfnamefont{A.~N.} \bibnamefont{Kuchera}},
  \bibinfo{author}{\bibfnamefont{G.~V.} \bibnamefont{Rogachev}},
  \bibinfo{author}{\bibfnamefont{V.~Z.} \bibnamefont{Goldberg}},
  \bibinfo{author}{\bibfnamefont{E.~D.} \bibnamefont{Johnson}},
  \bibnamefont{et~al.}, \bibinfo{journal}{Phys. Rev. C}
  \textbf{\bibinfo{volume}{84}}, \bibinfo{pages}{054615}
  (\bibinfo{year}{2011}).

\bibitem[{\citenamefont{Goldberg et~al.}(1974)\citenamefont{Goldberg, Rudakov,
  and Timofeev}}]{Goldberg1974}
\bibinfo{author}{\bibfnamefont{V.~Z.} \bibnamefont{Goldberg}},
  \bibinfo{author}{\bibfnamefont{V.~P.} \bibnamefont{Rudakov}},
  \bibnamefont{and} \bibinfo{author}{\bibfnamefont{V.~A.}
  \bibnamefont{Timofeev}}, \bibinfo{journal}{Sov. J. Nucl. Phys.}
  \textbf{\bibinfo{volume}{19}}, \bibinfo{pages}{253} (\bibinfo{year}{1974}).

\bibitem[{\citenamefont{Buck et~al.}(1975)\citenamefont{Buck, Dover, and
  Vary}}]{Buck1975}
\bibinfo{author}{\bibfnamefont{B.}~\bibnamefont{Buck}},
  \bibinfo{author}{\bibfnamefont{C.~B.} \bibnamefont{Dover}}, \bibnamefont{and}
  \bibinfo{author}{\bibfnamefont{J.~P.} \bibnamefont{Vary}},
  \bibinfo{journal}{Phys. Rev. C} \textbf{\bibinfo{volume}{11}},
  \bibinfo{pages}{1803} (\bibinfo{year}{1975}).

\bibitem[{\citenamefont{Fortune}(2012)}]{Fort12}
\bibinfo{author}{\bibfnamefont{H.~T.} \bibnamefont{Fortune}},
  \bibinfo{journal}{The European Physical Journal A}
  \textbf{\bibinfo{volume}{48}}, \bibinfo{pages}{1} (\bibinfo{year}{2012}),
  ISSN \bibinfo{issn}{1434-6001},
  \urlprefix\url{http://dx.doi.org/10.1140/epja/i2012-12063-0}.

\end{thebibliography}

\end{document}